\documentclass[12pt]{iopart}

\usepackage{iopams}
\usepackage{graphics}
\usepackage[bookmarks=false]{hyperref}
\usepackage{graphicx}
\usepackage{epsfig}
\usepackage{rotating}
\usepackage[latin1]{inputenc}
\usepackage{hyperref}
\usepackage{color}

\def\thetabold{\mbox{\boldmath $\theta$\unboldmath}}
\newcommand{\signum}{{\rm sgn}}
\newcommand{\Dcross}{D_\times}
\newcommand{\Dcrossa}{{\tilde D}_\times}
\newcommand{\Dcrossij}{D^{\times}_{ij}}

\newcommand{\DcrossNullij}{{\Dcrossij}^{(0)}}

\newcommand{\betacross}{N_s}
\newcommand{\betacrossas}{N_a}
\newcommand{\elss}{Els\"asser}

\begin{document}

\title{Scaling and universality in coupled driven diffusive models}

\author{Abhik Basu} \address{Theoretical Condensed Matter Physics
  Division and Centre for Advanced Research and Education,
Saha Institute of Nuclear Physics, Calcutta 700064, India\cite{byemail}}
\author{Erwin Frey} \address{Arnold Sommerfeld Center for
  Theoretical Physics and Center for NanoScience, Department of
  Physics, Ludwig-Maximilians-Universit\"at M\"unchen, Theresienstrasse
  37, D-80333 M\"unchen, Germany\cite{byemail1}}

\date{\today}

\begin{abstract}
Inspired by the physics of magnetohydrodynamics (MHD) a simplified
coupled Burgers-like model in one dimension ($1d$), a generalization of the Burgers model
to coupled degrees of freedom, is proposed to
describe $1d$MHD. In addition to MHD, this model serves as a $1d$
reduced model for driven binary fluid mixtures. Here we have
performed a comprehensive study of the universal properties of the
generalized  $d$-dimensional version of the reduced model. We employ both
analytical and numerical approaches. In particular, we determine the
scaling exponents and the amplitude-ratios of the relevant two-point
time-dependent correlation functions in the model. We demonstrate that these
quantities vary continuously with the amplitude of the noise
cross-correlation. Further our numerical studies corroborate the
continuous dependence of long wavelength and long time-scale physics of the model on the amplitude
of the noise cross-correlations, as found in our analytical studies.
We construct and simulate {\em lattice-gas} models of coupled
degrees of freedom in $1d$, belonging to the universality class of
our coupled Burgers-like model, which display similar behavior.  We
use a variety of
  numerical (Monte-Carlo and Pseudo-spectral methods) and analytical
  (Dynamic Renormalization Group, Self-Consistent Mode-Coupling Theory
  and Functional Renormalization Group) approaches for our
  work. The results from our different approaches complement one
  another. Possible realizations of our results in various nonequilibrium models are discussed.
\end{abstract}


\maketitle

\section{Introduction}

Physical descriptions of many natural driven systems involve coupled
dynamics of several degrees of freedom. A prominent example is the
dynamics of a driven symmetric mixture of a miscible binary
fluid \cite{ruiz}. Coarse-grained dynamical descriptions of such a
system are in terms of the local velocity field ${\bf v}({\bf r},t)$
and the difference in the local concentrations of the two components
$\psi ({\bf r},t)$. The feature which sets this system apart from a
{\em passively} advected system is that density fluctuations, while
being advected by the flow, create stresses which in turn feed back
on the flow. Thus, this forms an example of {\em active} advection.
Typical experimental measurements include correlation functions of
the dynamical variables and their universal scaling properties
\cite{swiney} in the non-equilibrium  statistical steady states
(NESS). A similar system with a closely related theoretical
structure is magnetohydrodynamics in three dimensions ($3d$MHD). It
deals with the coupled evolution of the velocity ${\bf v} ({\bf r},t)$
and magnetic ${\bf b}({\bf r},t)$ fields in quasi-neutral plasmas
\cite{jackson,biskampbook,arnab}. There the $\bf v$-field advects
and stretches the $\bf b$-field lines; the $\bf b$-field in turn
acts on the $\bf v$-field through the Lorentz force. As in driven
binary fluids, one is typically interested in the correlation
functions and their universal properties characterized by
appropriate scaling exponents in the NESS. Despite the differences
in the microscopic physics of these two systems, they have a great
many commonalities in the physical descriptions at large scales and
long times. In particular, they show similar universal scaling
behavior in the hydrodynamic limit. Often one asks similar questions
concerning scaling behavior in the NESS and the nature of the
dynamical equations these systems follow. Inspired by these
generalities a $1d$ generalized Burgers model \cite{abjkb} was
proposed as a simpler reduced model, which captures many qualitative
features of two real systems we described above. This model consists
of the Burgers equation for the velocity field, containing a coupling term
to the second field (representing feedback), which in turn obeys a
diffusion-advection equation. We generalize this model to
$d$-dimensions, where it has the same continuous symmetries as the
binary fluid mixture model or $3d$MHD model equations, and
investigate its universal properties as a function of the dimension
$d$.

In the vicinity of a critical point, equilibrium systems show
universal scaling properties for thermodynamic functions and
correlation functions. These are characterized by universal scaling
exponents that depend on the spatial dimension $d$ and the symmetry
of the order parameter (e.g., Ising, XY etc.) \cite{fisherrev}, but
not on the material parameters that specify the (bare) Hamiltonian.
Prominent exceptions are the $2d\;XY$ model and its relatives, where
the renormalization group flow is characterized by a fixed line and
as a consequence the scaling exponents exhibit a continuous
dependence on the stiffness parameter. Equilibrium dynamics close to
critical points also show universality though the dynamic scaling
exponents, which characterize the time-dependence of unequal time
correlation functions, now also depend upon the presence or absence
of conservation laws and the non-dissipative terms in the dynamical
equations \cite{halperin}. A different situation arises in driven,
dissipative, nonequilibrium systems with NESS. Significant advances
have been made in classifying the physics of non-equilibrium systems
at long time and large length scales into universality classes. It
has been shown that standard universality classes in critical
dynamics are quite robust to detailed-balance violating
perturbations~\cite{tauber-etal:02}. Novel features are found only
for models with conserved order parameter and spatially anisotropic
noise correlations. In contrast, truly non-equilibrium dynamic
phenomena, whose steady states cannot be described in terms of
Gibbsian distributions, are found to be rather sensitive to all
kinds of perturbations. Prominent examples are driven diffusive
systems~\cite{zia:review}, and fluid- and magnetohydrodynamic-
turbulence~\cite{biskampbook,rahulrev,abthesis}. In contrast to
systems in equilibrium, what characterizes the universality classes
in nonequilibrium systems remains a yet unsolved question.

In this article we examine the particular issue of universality in
the non-equilibrium in the context of {\em driven coupled
generalized Burgers model} \cite{abjkb} in $d$-dimensions. We show
that its NESS depends sensitively on the parameters of the model. A
brief account of the physics of this model has been reported in
Ref.\cite{rapcom}. Here we extend the results of Ref.\cite{rapcom},
and present a comprehensive study of the universal properties of the
model. In order to study our model systematically, we have employed a variety of analytical
and numerical techniques, all of which together bring out a
coherent and consistent picture: We find that the universal properties of the model characterized by
dimensionless amplitude ratios and scaling exponents of various
correlation functions in the model depend explicitly on the noise
crosscorrelations. Our results provide valuable insight on the issue
of universality in coupled nonequilibrium systems and, in
particular, what might characterize universality classes in a simple coupled model discussed here.
We use analytical methods including dynamic renormalization
group (DRG), self-consistent mode coupling methods (SCMC) and
functional renormalization group (FRG) to calculate the relevant
scaling exponents and amplitude-ratios of the correlation functions.
Furthermore, to complement our analytical results we use
pseudo-spectral methods to numerically solve the stochastically
driven model equations in one ($1d$) and two ($2d$) space dimensions
and use them to calculate scaling exponents and amplitude ratios.
We, in addition, construct lattice-gas models in one space dimension
belonging to the same universality class as the model equations of
\cite{rapcom}, following closely the lattice-gas models for the
Kardar-Parisi-Zhang (KPZ) equation for growing surfaces \cite{kpz}.
We perform Monte-Carlo simulations on this coupled lattice-gas model
and calculate the scaling exponents and the amplitude ratios. Our
analytical and numerical results agree well with each other. The
organization of the rest of the paper is as follows: In Section
\ref{model} we consider the model equations and the noise
correlations which we use in our work here. In Sec. \ref{noisecorr}
we describe the noise correlations which we use in our analytical
work. Then in the Section~\ref{coup-lat} we describe our
constructions of the lattice-gas models in $1d$ belonging to the
universality class of the continuum model equations. Then in
Sections~ \ref{results:field_theory} and \ref{numrec} we present our
analytical and numerical results respectively, concerning the
scaling exponents and the amplitude-ratios. We finally summarize our
work in Sec.\ref{conclu}.

\section{The stochastically driven model equations}
\subsection{Construction of the model equations} \label{model}

Let us begin by recalling the general principles which go in setting
up the one-dimensional ($1d$) Burgers \cite{fns} as a reduced model
for the $3d$ Navier-Stokes (NS) equation for the velocity field $\bf
v$. The NS equation is given by
\begin{equation}
\frac{\partial {\bf v}}{\partial t} + {\bf v\cdot \nabla v}= -\nabla
p/\rho + \nu_v\nabla^2 {\bf v} + {\bf f}_v, \label{ns1}
\end{equation}
where $p$ is the pressure, $\rho$ is the density, $\nu_v$ is the
kinematic viscosity and $\bf f_v$ is an external force
(deterministic or random). Equation (\ref{ns}) has two constants of
motion in the inviscid, unforced  limit ($\nu_v=0,\;{\bf f}_v=0$): (i)
The kinetic energy $\int d^dx\, v^2$ and (ii) helicity $\int d^dx\,
{\bf v\cdot \nabla\times v}$. Further equation (\ref{ns1}) is
invariant under the Galilean transformation and, as a consequence,
is of the conservation law form. The $1d$ simplest non-linear
equation which obeys the conservation of kinetic energy (helicity
 cannot be defined in $1d$ properly, since it involves curl of a vector) and also is Galilean invariant
is the famous Burgers equation \cite{fns}
\begin{equation}
\frac{\partial v}{\partial t} + \frac{1}{2}\partial_x v^2 =\nu
\partial_{xx} v + f.
\label{burgers}
\end{equation}
Here $v$ is the $1d$ Burgers velocity field - it is a pressureless
fluid since the pressure has been dropped, $\nu$ is the Burgers
viscosity and $f$ is an external force.

Fluid systems having coupled degrees of freedom require descriptions
more general than pure one-component fluid systems.
Two such well-known examples are

\paragraph{(i) Symmetric (50-50) binary fluid mixture:-}
The coarse-grained dynamics of a driven symmetric binary fluid mixture is
described by two continuum variables: a velocity field ${\bf v}({\bf
r},t)$ and a concentration-gradient field ${\bf b}({\bf r},t)=\nabla
\psi ({\bf r},t)$ where $\psi$ is the relative difference of local
densities of the two components \cite{ruiz}. The coupled equations
of motion, in the incompressible limit of the dynamics, consist of
the generalized Navier-Stokes (which includes the stresses from the
concentration gradients) for $\bf v$ and a diffusion-advection
equation for $\bf b$
\begin{eqnarray}
\frac{\partial {\bf v}}{\partial t} + {\bf v\cdot \nabla
v}&=&-\frac{\nabla p}{\rho} - \alpha {\bf b} {\bf \nabla \cdot b}
+ \nu \nabla^2 {\bf v} + {\bf f}_v,\label{gen-ns-binary}\\
\frac{\partial{\bf b}}{\partial t} + \nabla ({\bf v\cdot b})&=&\mu
\nabla^2 {\bf b} +{\bf f}_b,\label{advec-bin}
\end{eqnarray}
together with the incompressibility condition $\nabla \cdot {\bf
v}=0$. We also have $\alpha > 0$ for thermodynamic stability; $\mu$
is diffusivity of the concentration gradient. Functions ${\bf f}_v$
and ${\bf f}_b$ are stochastic forces required to maintain a
statistical steady-state. Force ${\bf f}_b$ is curl-free. Equations
(\ref{gen-ns-binary}) and (\ref{advec-bin}) are Galilean invariant
and admit the following conservation laws in the inviscid, unforced
limit \cite{ruiz}:
\begin{itemize}
\item Spatial integral of the square of the
concentration $\psi$: $C=\frac{1}{2} \int d^3 r [\psi ({\bf
r},t]^2$,
\item Total energy $E=\frac{1}{2}\int d^3r [ v({\bf r},t)|^2
+ \alpha b ({\bf r},t)^2]$.
\end{itemize}
We now ask for a
$1d$ model, whose relation with  equations
(\ref{gen-ns-binary}) and (\ref{advec-bin}) is same as that between
the Burgers equation and the Navier-Stokes equation. The simplest of
such $1d$ equations, to the leading orders in the gradients and
bilinear order in the fields $\bf u$ and $\bf b$ are \cite{abjkb}
\begin{eqnarray}
\frac{\partial u}{\partial t} + \frac{1}{2}\partial_x u^2 +
\frac{1}{2} \partial_x b^2 &=& \nu \partial_{xx} u + f, \label{binburg1}\\
\frac{\partial b}{\partial t} + \partial_x (ub) &=& \mu\partial_{xx} b
+ g. \label{binburg2}
\end{eqnarray}
Here, $u$ and $b$ are the $1d$ Burgers velocity and concentration
gradient fields, $\nu$ and $\mu$ are diffusion coefficients for $u$ and $b$,
and $f$ and $g$ are external sources. Equations (\ref{binburg1}) and
(\ref{binburg2}) the $1d$ analogs of $E$ and $C$ as defined above in
the inviscid ($\nu=\mu=0$), unforced ($f=g=0$) limit obey the two
conservation laws mentioned above. Note that equations (\ref{binburg1}) and
(\ref{binburg2}) are invariant under
the Galilean transformation.

\paragraph{Magnetohydrodynamics in three dimensions:-}
The subject of Magnetohydrodynamics (MHD)
\cite{jackson,biskampbook,arnab} deals with the dynamics of a
quasi-neutral plasma in terms of coupled evolution of velocity ${\bf
v}({\bf r},t)$ and magnetic ${\bf b}({\bf r},t)$ fields. the
equations of motion consist of the generalized Navier-Stokes
equation containing the stresses from the magnetic fields for the
$\bf v$ field and the Induction equation for the $\bf b$ field:
\begin{eqnarray}
\frac{\partial {\bf v}}{\partial t} + {\bf v\cdot\nabla v} =
-\frac{\nabla p}{\rho} + \frac{(\nabla \times \bf b)\times
b}{4\pi\rho} +\nu\nabla^2 {\bf v} + {\bf f}_v,\label{mhdv}\\
\frac{\partial {\bf b}}{\partial t} + \nabla \times ({\bf v\times
b}) =\mu \nabla^2 {\bf b} + {\bf f}_b.\label{mhdm}
\end{eqnarray}
These are to be supplemented by $\nabla\cdot \bf b=0$, and, in case
of an incompressible fluid, $\bf \nabla\cdot v=0$. Here, $\nu,\,
\mu$ are kinetic and magnetic viscosities and ${\bf f}_v,\,{\bf
f}_b$ are the external sources, with ${\bf f}_b$ being solenoidal.
Equations (\ref{mhdv}) and (\ref{mhdm}) are Galilean invariant and
admit the following conserved quantities in the inviscid and
unforced limit:
(i) Total energy (the sum of the kinetic and magnetic energies)
$E=\frac{1}{2}\int d^3\, r[v^2 +b^2]$, and (ii) Crosshelicity
$C=\int d^3r\, {\bf v\cdot b}$ 
As in the previous example of binary fluid mixture we what the $1d$
model should be whose relation with the $3d$MHD equation above is
same as that of Burgers with Navier-Stokes. Such $1d$ equations are
the same as (\ref{binburg1}) and (\ref{binburg2}), which conserves
the $1d$ analog of the total crosshelicity as well \footnote{The
third invariant of $3d$MHD, i.e., the total magnetic helicity cannot
be properly defined in $1d$, since it involves curl of a vector.
Hence it is not considered in writing down the $1d$ model.} Further,
Eqs. of motion (\ref{mhdv}) and (\ref{mhdm}) are invariant under the
Galilean transformation. Note also that under parity reversal $\bf
r\rightarrow -r$, $\bf v$ and $\bf b$ transform differently: $\bf
v\rightarrow -v$ and $\bf b\rightarrow b$.

The $d$-dimensional generalization \cite{rapcom} of the
$1d$-model equations (\ref{binburg1}) and (\ref{binburg2})
are,
\begin{equation}
\frac{\partial {\bf u}}{\partial t} + \frac{\lambda_1}{2}{\nabla}{\bf u}^2
 + \frac{\lambda_2}{2} {\nabla}{\bf b}^2= \nu_0{\nabla^2}{\bf u} +{\bf f},
\label{eq:burgers_1}
\end{equation}
\begin{equation}
\frac{\partial {\bf b}}{\partial t} + \lambda_3{\nabla}({\bf u}\cdot{\bf b}) =
\mu_0{\nabla^2\bf b} +{\bf g}. \label{eq:burgers_2}
\end{equation}
We refer to these equations as the {\em Generalized Burgers Model}
(henceforth GBM). Here, $\bf u$ and $\bf b$ are $d$-dimensional {\em
Burgers} fields, respectively.  Parameters $\lambda_1,\,\lambda_2$
and $\lambda_3$ are actually identical to unity but kept for formal
book keeping purposes in the renormalization group and mode coupling
approaches below. Parameters $\nu_0$ and $\mu_0$ are bare
(unrenormalized; see below) viscosities. Functions $\bf f$ and $\bf
g$ are external noise sources required to maintain a statistical
steady state. The noise $\bf g$, in general, may have both a
non-zero divergence and a curl (see the discussion just before
Eq.~(\ref{flow7a}). The quantities of interests are the correlation
functions as functions of Fourier wavevector $\bf k$ and frequency
$\omega$
\begin{eqnarray}
C_{ij}^u ({\bf k},\omega)\equiv \langle u_i({\bf k},\omega)u_j({\bf
-k},-\omega)\rangle,\\
C_{ij}^b({\bf k},\omega)\equiv \langle b_i ({\bf k},\omega)b_j ({\bf
-k},-\omega)\rangle,\\
C_{ij}^\times ({\bf k},\omega) \equiv \langle u_i ({\bf k},\omega)
b_j(-{\bf k},-\omega)\rangle. \label{corr-def}
\end{eqnarray}
From the properties of the fields $\bf u$ and $\bf b$ under the
reversal of parity as discussed above $C^u_{ij}$ and $C^b_{ij}$ are
even functions of $\bf k$ and $C^\times_{ij}$ is an odd function of
$\bf k$.

Upon introducing a new set of fields, ${\bf u} = \nabla h$ and ${\bf
b} = \nabla \phi$, the GBM \cite{abjkb} maps onto a model for
drifting lines introduced by Erta\'s and Kardar \cite{ek}
\begin{equation}
\frac{\partial h}{\partial t}+\frac{\lambda_1}{2}(\nabla h)^2
+\frac{\lambda_2}{2} (\nabla \phi)^2=\nu_0 \nabla^2 h +\theta_1,
\label{eq2kpz1}
\end{equation}
\begin{equation}
\frac{\partial \phi}{\partial t}+\lambda_3 (\nabla h)\cdot (\nabla
\phi)= \mu_0\nabla^2 \phi +\theta_2, \label{eq2kpz2}
\end{equation}
where fields $h$ and $\phi$ are the local, instantaneous
longitudinal and transverse fluctuations, respectively, around the
mean position of the drifting line. The functions $\theta_1$ and
$\theta_2$ are given by ${\bf f}=\nabla \theta_1$ and ${\bf
g}=\nabla \theta_2$.
%
%
If the fluid and magnetic viscosities are equal, $\nu_0 = \mu_0$,
the Burgers-like MHD model, Eqs. (\ref{eq:burgers_1}) and
(\ref{eq:burgers_2}), can also be mapped onto a model of coupled
growing surfaces. Further by introducing Els\"asser variables, $\bf
z^{\pm}=u\pm b$ one finds (setting
$\lambda_1=\lambda_2=\lambda_3=1$), for $\eta_0=\nu_0=\mu_0$,
\begin{equation}
   \frac{\partial {\bf z^{\pm}}}{\partial t}
 + \frac{1}{2}\nabla {z^{\pm}}^2 =
    \eta_0\nabla^2 {\bf z^{\pm}}+{\thetabold^{\pm}} \, ,
    \label{els}
\end{equation}
where $\theta^{\pm}_i=(f_i\pm g_i)/2$.

Each of the Els\"asser variables obeys a stochastically driven
Burgers equation, where the coupling between the two fields $\bf
z^\pm$ arises solely due to the cross-correlations in the external
(stochastic) driving forces. Physically, this model describes the
growth of surfaces on two interpenetrating sub-lattices (say, A and
B), where the growth of each of the surfaces follows a KPZ dynamics.
The dynamics becomes coupled since depositions of particles on the A
and B sub-lattices are correlated with each other. 
Such a coupled surface growth problem can be mapped to a related
equilibrium problem of a pair of two Directed Polymers (DP) in
random medium. A DP in $d+1$ dimension is just a directed string
stretched along one particular direction with free fluctuations in
all other $d$ transverse directions. When placed in a random medium
competitions between the elastic energy of the string and the random
potential of the medium lead to phase transitions between smooth and
rough phases \cite{halpin}. The DP phases and phase transition can
be mapped exactly to the phases and phase transitions in the KPZ
surface growth model. In the present case, the pair of variables
$\bf z^{\pm}$ model the free energies of the two DPs in a random
medium. The variable $\bf x$ refers to the directions transverse to
the DP and $t$ is the longitudinal dimension. However, the external
noise sources $ \thetabold^{\pm}$ are in general cross-correlated
leading to interesting phase diagram of the coupled system
\cite{abpol} (see also below).

\subsection{Noise distributions}
\label{noisecorr}

The noise sources $\bf f$ and $\bf g$ or alternatively $
\thetabold^{\pm}$ are chosen to be Gaussian-distributed with zero
mean and specified variances. It should be noted that, in
nonequilibrium situations there are no restrictions on the noise
variances as there are no detailed balance conditions relating the
diffusivities and the noise-variances, unlike for systems in
equilibrium \cite{fdt}. Furthermore, for analytical conveniences we
assume them to be conserved noise sources which suffices for our
purposes in this article. In particular we make the following
choices:
\begin{eqnarray}
  \langle f_i ({\bf k},t) f_j ({-{\bf k},0}) \rangle
          &=&  2 k_i k_j D_u^{(0)} ({\bf k}) \delta (t),
\label{noise1}\\
  \langle g_i ({\bf k},t) g_j ({-{\bf k},0}) \rangle
          &=&  2 k_i k_j D_b^{(0)} ({\bf k}) \delta (t),
\label{noise2}\\
  \langle f_i ({\bf k},t) g_j ({-{\bf k},0}) \rangle
          &=&  2 i {\DcrossNullij} ({\bf k}) \delta (t),
\label{noise3}
\end{eqnarray}
where the functions $D_{u,b}^{(0)} ({\bf k})$ are even and
${\DcrossNullij} ({\bf k})$ is odd in $\bf k$, respectively. A
superscript $0$ refers to the bare (unrenormalized) quantities. The
above structures of the auto-correlators of $f_i$ and $g_i$ are the
simplest choices consistent with their tensorial structures and the
conservation law form of the equations of motion
(\ref{eq:burgers_1}) and (\ref{eq:burgers_2}). Equations
(\ref{noise1}) and (\ref{noise2}) are invariant under inversion,
rotation and exchange of $i$ with $j$. We take the noise
cross-correlation, equation (\ref{noise3}) to be invariant under
spatial inversion, but we allow it to break (i) rotational
invariance, and (separately) (ii) symmetry with respect to an
interchange of the cartesian indices $i$ and $j$. The matrix
${\DcrossNullij} ({\bf k})$ in general has a symmetric part
${\DcrossNullij} ({\bf k})_s$ and an antisymmetric part
${\DcrossNullij} ({\bf k})_a$ with respect to interchanges between
$i$ and $j$. In this article most of our results involve a finite
${\DcrossNullij} ({\bf k})_s$ but zero ${\DcrossNullij} ({\bf
k})_a$, although some effects of ${\DcrossNullij}_a$ are also
discussed. Effects of the noise cross-correlations of the forms
given in Eq.(\ref{noise3}) on the universal properties of the model
Eqs. (\ref{eq:burgers_1}) and (\ref{eq:burgers_2}) are discussed
briefly in Ref.\cite{rapcom}. The properties of noise correlators
described above use the explicit symmetries of the GBM model and the
fields $\bf u$ and $\bf b$. Such choices, however, leave the
functional forms (as functions of $\bf k$) of $D_u^{(0)}({ \bf
k}),\,D_b^{(0)} ({\bf k})$ and ${\DcrossNullij} ({\bf k})$, which
are the amplitudes of the noise variances, arbitrary. In
order to define the model completely by making specific choices, one
further defines {\em amplitudes} of the symmetric and antisymmetric
parts of the cross-correlations by the relations ${\DcrossNullij}
({\bf k})_s = 2D_{s}^0({\bf k})k_ik_j,\,|D_{s}^0({\bf k})|=D_{s}^0$
and ${\DcrossNullij} ({\bf k})_a {\DcrossNullij} ({\bf k})_a =
4{D_{a}^0}^2 k^4$ where $D_{s}^0$ and $D_{a}^0$ are amplitudes of
the respective parts. Further, we choose $D_u^{(0)}({\bf k})=D_u^0$
and $D_b^{(0)} ({\bf k}) =D_b^0$.


\section{Coupled lattice-gas models in one dimension}
\label{coup-lat}

Studies of lattice-gas models to understand the long-time,
large-scale properties of continuum model equations in
non-equilibrium statistical mechanics have a long history. For
example, several lattice-gas model for the KPZ equation have been
proposed and studied (see, e.g., \cite{rsos,latt-kpz}). Often
numerical simulations are used to investigate the scaling
properties. Such studies have several advantages. The main advantage
is: Due to the analytically intractable nature of such models it is
often much easier to numerically simulate a lattice-gas model than
to obtain the numerical solution of the corresponding noise-driven
continuum model equation. In addition such studies allow us to
explore and understand the applicability of the concept of {\em
universality classes}, originally developed in the context of
critical phenomena and equilibrium critical dynamics
\cite{fisherrev}, in physical situations out of equilibrium,  by
comparing the lattice model results with the results from the
continuum model.

In this section we propose a lattice-gas model for the
one-dimensional ($1d$) GBM, Eqs.  (\ref{eq:burgers_1}) and
(\ref{eq:burgers_2}). Our starting point is the observation made in
Ref.~\cite{rapcom}, that in the hydrodynamic limit the {\em
effective} ({\rm renormalized}) Prandtl number $P_m=\nu/\mu=1$ for
the model Eqs. (\ref{eq:burgers_1}) and (\ref{eq:burgers_2})
\footnote{Technically, as discussed in Ref.~\cite{rapcom},
renormalized $P_m=1$ is the fixed point of the model.} Henceforth,
since we are interested in the asymptotic scaling properties, we may
set the bare magnetic Prandtl number $P_m^0=\mu_0/\nu_0=1$, i.e.,
take the two bare viscosities as identical.  Upon setting our book
keeping parameters to their physical values,
$\lambda_1=\lambda_2=\lambda_3=1$, and introducing the Els\"asser
variables $z^{\pm}=u\pm b$, one obtains a set of coupled Burgers
equations
\begin{equation}
\partial_t z^{\pm}+ \frac{1}{2} \partial_x (z^{\pm})^2
= \nu_0 \partial_x^2z^{\pm}+\theta_{\pm} \, ,
\label{coupled_burgers_1d}
\end{equation}
where the coupling between $z^+$ and $z^-$ is mediated only via the
cross-correlations in the noise $\theta_{\pm} = \frac{1}{2} (f \pm
g)$ only. With the standard mapping, $z^+=\partial_x
h_1,\;z^-=\partial_x h_2$, we may rewrite
Eq.~(\ref{coupled_burgers_1d}) as a set of coupled KPZ equations for
the height fields $h_1$ and $h_2$
\begin{eqnarray}
  \partial_t h_1 + \frac{1}{2} (\partial_x h_1)^2
  &=&\nu_0\partial_{xx}h_1+\psi_1,\nonumber \\
  \partial_t h_2 + \frac{1}{2} (\partial_x h_2)^2
  &=&\nu_0\partial_{xx}h_2+\psi_2,
\label{2kpz}
\end{eqnarray}
where the noises $\partial_x\psi_{1,2}= \theta_\pm$.  Further the
auto-correlations of the fields $h_{1,2}$ are simply related to
those of $h$ and $\phi$: $h_{1,2}=h\pm \phi$. We write them in the
real space
\begin{eqnarray}
\langle h({ x},t)h({ x'},t')\rangle &=& \frac{1}{2}[\langle
h_1(x,t)h_1(x',t')\rangle
+\langle h_2(x,t)h_2(x',t')\rangle\nonumber\\ &+& \langle h_1(x,t)h_2(x,t)\rangle + \langle h_1 (x',t')h_2(x,t)],\\
\langle \phi(x,t)\phi(x',t')\rangle &=& \frac{1}{2}[\langle
h_1(x,t)h_1(x',t')\rangle + \langle h_2 (x,t) h_2 (x',t')\rangle \nonumber \\ &-&
\langle h_1(x,t)h_2(x',t')\rangle - \langle h_1 (x',t')h_2(x,t)\rangle].
\label{lattmod-corr}
\end{eqnarray}

Constructions of lattice models for a single KPZ equation is
well-documented in the literature \cite{rsos,latt-kpz}. In such
models the underlying space (substrate) is taken to be discrete.
Particles are deposited randomly over the discrete lattice and they
settle on different lattice points according to certain
model-dependent local growth rules \cite{rsos,latt-kpz}. In our case
since each of the Eqs.~(\ref{2kpz}) has the structure of a KPZ
equation, such growth rules can be used to represent each of
Eqs.~(\ref{2kpz}), representing two growing surfaces over two
sub-lattices. In such lattice models stochasticity enters into the
model through the randomness in the deposition process. In our case
we model noise-crosscorrelations [see Eqs.~(\ref{psi1}-\ref{psi3})
below] by cross-correlating the randomness in depositions in the two
sublattices. In the Fourier space the correlations of the noise
sources $\psi_{1,2}$ have the form
\begin{eqnarray}
\langle\psi_1(k,t)\psi_1(-k,0)\rangle&=&2D_0\delta(t),\label{psi1} \\
\langle\psi_2(k,t)\psi_2(-k,0)\rangle&=&2D_0\delta(t),\label{psi2} \\
\langle\psi_1(k,t)\psi_2(-k,0)\rangle&=&2i\tilde{D}_0k/|k|\delta(t)\label{psi3}.
\end{eqnarray}
Note that though variances in Eqs.~(\ref{psi1}-\ref{psi2}) in real
space are proportional to $\delta (x-x')$, the third one is
proportional to $1/(x-x')$ (see Appendix \ref{cross1d}). Such noises
can be generated by first generating two independent, zero-mean
short range, white in space and time Gaussian noises, and then
taking appropriate linear combinations of them. A numerical scheme
to generate noises with variances (\ref{psi1}-\ref{psi3}) is given
in Appendix (\ref{noisegen}). Eqs.~(\ref{2kpz}) may be viewed as a
coupled growth model describing the dynamics of two growing surfaces
of two different particles on two interpenetrating sub-lattices A
and B where the deposition process (represented by the noise sources
$\psi_{1,2}$) of the particles A and B are correlated but the local
dynamics of the particles on the sublattices are independent of each
other. In particular each sub-lattice has a local KPZ dynamics.
Monte-Carlo simulations of our coupled lattice models yields fields
$h_1$ and $h_2$. The dynamics of the two growing surfaces on the two
sublattices are coupled through the noise sources. The
auto-correlators of $h$ and $\phi$ are then calculated by using
relations (\ref{lattmod-corr}).

We implement the Newman-Bray (hereafter NB) \cite{ns} and
the Restricted Solid on Solid (RSOS) \cite{rsos} algorithms for each
of the KPZ equations.
In the RSOS algorithm \cite{rsos} noise sources $\psi_{1,2}$ with
correlations characterized by Eqs.~(\ref{psi1}-\ref{psi3}) are used
in the random determination of the lattice sites which are to be
updated in a given time-step; see below for more details.  Since
$\psi_{1,2}$ are cross-correlated the sites of sub-lattices A and B
which are updated in a given time-step get cross-correlated, which
in turn induces cross-correlations in the height fields $h_1$ and
$h_2$. We have performed Monte-Carlo simulations on our coupled
lattice model based on the RSOS lattice model for the KPZ dynamics.
The results from the Monte-Carlo simulations of our coupled RSOS
lattice model are discussed in Section~\ref{rsos}. The details of
generation of random numbers obeying correlations (\ref{psi3}) are
discussed in the Appendix (\ref{noisegen}).

In the NB lattice-gas model, the mapping between the KPZ surface
growth model and the equilibrium problem of directed polymer in a
(quenched) random medium is exploited \cite{halpin,ns}. We extend
this idea in our construction of a lattice-gas model for the Eqs.
(\ref{2kpz}) based on the NB algorithm which is equivalent to
considering two directed polymers in a random medium. The two
free-energies of the two polymers would then represent the heights
of the two interpenetrating sub-lattices as described above. In this
coupled lattice-gas model, the noise cross-correlations represent
{\em effective} interactions induced by the randomness of the
embedding medium between the two polymers. The detailed numerical
results from the model are presented in Section \ref{ns}.

\section{Field theory analysis}
\label{results:field_theory}

Our field theoretic analytical studies include one-loop dynamic
renormalization group (DRG), one-loop self-consistent mode coupling
(SCMC) and functional renormalization group (FRG) studies. From
previous studies on the closely related KPZ model for surface growth
one has learned that DRG schemes are well-suited to study the
scaling properties of the rough phase in $1d$ and the
smooth-to-rough phase transition in spatial dimensions
$d=2+\epsilon,\,\epsilon>0$ \cite{freykpz}. In contrast, the SCMC
and the FRG approches are known to yield results on the scaling
properties of the rough phases in $d=1$ dimensions and higher
\cite{halpin,jkb}. In $d=1$ the results from the DRG and the
SCMC/FRG schemes are identical \cite{frey1dburg}.

We are interested in the physics in the scaling limit, i.e., at long
time and length scales.  In that limit the time-dependent two-point
correlation functions are written in terms of the dynamic exponent
$z$ and the two roughness exponents $\chi_h$ and $\chi_\phi$
as\footnote{The roughness exponents $\chi_u$ and $\chi_b$ of the
  fields $\bf u$ and $\bf b$ are related to $\chi_h$ and $\chi_\phi$:
  $\chi_h=\chi_u +1,\,\chi_\phi=\chi_b +1$.}
\begin{eqnarray}
  C_{hh}({\bf x},t) &\equiv& \langle h({\bf x}, t)h(0,0)\rangle
  = x^{2\chi_h}f_h(t/x^z) \, , \\
  C_{\phi\phi}({\bf x},t) &\equiv&\langle \phi({\bf x},t)\phi(0,0)\rangle=
  x^{2\chi_\phi}f_\phi(t/x^z) \, , \\
  C_\times({\bf x},t) &\equiv& \langle h({\bf x},t)\phi(0,0)\rangle = \signum({\bf
    x})x^{\chi_h +\chi_\phi}f_\times (t/x^z) \, .
\end{eqnarray}
Here, angular brackets $\langle\rangle$ means averaging over the
noise distributions. Functions $f_h,\,f_\phi$ and $f_\times$ are
scaling functions of the scaling variable $t/x^z$. Since
$C_\times({\bf x})$ is an odd function of $\bf x$, a signum function
appears, $\signum({\bf x})=-\signum(-{\bf
  x})$. Ward identities resulting from the Galilean invariance of the
model Eqs. (\ref{eq:burgers_1}) and (\ref{eq:burgers_2}) imply that
$\chi_h$ and $\chi_\phi$ are identical [see below; see also Appendix
\ref{ward}]. Henceforth, we write $\chi_u=\chi_h-1=\chi_\phi-1=\chi_b=\chi$. Clearly then
the ratios of the various equal-time correlators are dimensionless
numbers.  One  also defines widths
\begin{eqnarray}
  W_h(t) &=& \sqrt{ \langle [h({\bf x},t) - \langle h({\bf x},
    t)\rangle]^2 \rangle } =\sqrt {[\langle h({\bf x},t)^2\rangle - \langle h({\bf x},t)\rangle^2]}\, \\
  W_{\phi}(t) &=& \sqrt{ \langle [\phi({\bf x}, t) - \langle\phi({\bf
    x}, t)\rangle]^2 \rangle }=\sqrt{[\langle \phi({\bf x},t)^2\rangle - \langle \phi({\bf x},
    t)\rangle^2]} \, ,
\end{eqnarray}
These are related to the two-point correlators measured at the same
space and time. They exhibit growing parts for small $t$ and yield
the ratio $\chi_h/z$: $W_h(t),\,W_{\phi}(t)\sim t^{\chi_h/z}$, and
saturated parts for large $t$ yielding the exponent $\chi_h$:
$W_h(t),\,W_{\phi}(t)\sim L^{\chi_h}$ for large $t$ where $L$ is the
system size. The ratios of the amplitudes of the correlation
functions or the widths in the steady-state yield the amplitude-ratio
$A$ (defined below).

\subsection{Review of the KPZ Equation}
\label{kpzresults}

The KPZ equation for surface growth is one of the simplest
non-linear generalization of the diffusion equation and serves as a
paradigm for phase transitions and scaling in non-equilibrium
systems; see for example Refs.~\cite{halpin,krugrev} for extensive
reviews. Our GBM, Eq.~(\ref{eq2kpz1}), reduces to the KPZ equation
for $\phi=0$
\begin{equation}
  \frac{\partial h}{\partial t} + \frac{\lambda_1}{2} (\nabla h)^2
  = \nu_0 \nabla^2 h \, .
\label{1kpz}
\end{equation}
The field $h({\bf x},t)$ physically represents the height profile of
a growing surface. The KPZ equation has by now been studied by a
broad variety of approaches. These include dynamic RG \cite{natter},
Monte-Carlo simulations of the equivalent lattice-gas models
\cite{latt-kpz}, and by mapping onto the equilibrium problem of a
directed polymer in a random medium \cite{halpin}. The main results
concerning the statistical properties include
\begin{itemize}
\item Due to the Galilean invariance the scaling exponents $\chi_h$
  and $z$ characterizing the rough phase follow an exact relation
  $\chi_h+z=2$ \cite{fns,freykpz}.
  \item The scaling exponents for the rough phase in $1d$ are exactly
  given by $\chi_h=1/2,\,z=3/2$
  \cite{halpin,freykpz,krugrev}.
  \item For spatial dimensions $d=2+\epsilon$ with $\epsilon>0$ there is
  a phase transition from a smooth to a rough phase.  The exponents in
  the smooth phase are exactly given by $\chi_h=(2-d)/2$ and $z=2$
  \cite{freylongrange}.
  The values of the exponents in the rough phase are controversial.  Functional renormalization group studies
  \cite{halpin} and equivalent mode
  coupling analyses in terms of a small-$\chi$ expansion \cite{jkb}
  give $\chi_h= (4-d)/6$ and $z=(8+d)/6$.  However, a recent critical
  analysis of the mode coupling equations by Canet and Moore
  \cite{moore} modifies these values. They obtain $z=2-(4-d)/4 +
  O((4-d)^2),\,\chi_h=2-z$ near $d=4$ and $z=4/3 + d/3 +
  O(d^2),\,\chi_h=2-z$ near $d=0$. These scaling exponents describe the rough
  phase at $d=1$ and $2<d<4$. They further obtain a new set of scaling
  solutions of their mode-coupling equations for $d<2$.
\end{itemize}

The GBM discussed in this work are expected to exhibit much richer
behavior, since in addition to advection and diffusion it contains
a feedback term (the term $\lambda_2\nabla b^2$) in
Eq.~(\ref{eq:burgers_1}). Furthermore, although equations
(\ref{eq:burgers_1}) and (\ref{eq:burgers_2}) are invariant under
parity inversion, since the fields $\bf u$ and $\bf b$ have
different properties under parity inversion, an intriguing
possibility of breaking parity in the statistical steady state by
the presence of a nonzero cross-correlations of $\bf u$ and $\bf b$
(created by suitable choices external forces) exists. We discuss
some of these issues below.

\subsection{Dynamic renormalization group studies in $d$-dimension}
\label{rg1d}

In this section we employ dynamic renormalization group (DRG) methods
to understand the long-time and large-scale physics of the model
equations (\ref{eq:burgers_1}) and (\ref{eq:burgers_2}).  Before going
into the details of our calculations and results we elucidate the
continuous symmetries under which the equations of motion remain
invariant. As shown in appendix~\ref{ward} these allow us to construct
exact relations between different vertex functions which in turn
impose strict conditions on the renormalization of different
parameters in the model. Here we list the symmetries and summarize the
consequences on the renormalization of the model.
\begin{itemize}
\item The model shows Galilean invariance when $\lambda_1=\lambda_3=
  \lambda$, i.e., the equations of motion are invariant under the
  continuous transformations
   \begin{eqnarray}
     {\bf u}'({\bf x},t) &=& {\bf u}({\bf x} +
     \lambda {\bf u_0}t,t) \, ,\nonumber \\
     {\bf b}'({\bf x},t) &=&  {\bf b}({\bf x} +
     \lambda {\bf u_0}t,t) \, .
   \end{eqnarray}
   This invariance implies that the coupling constants $\lambda_1$ and
   $\lambda_3$ do not renormalize in the long wavelength limit.
 \item There is a rescaling invariance of the field
   \begin{equation}
     {\bf b} \rightarrow \Lambda {\bf b},\,\lambda_2
     \rightarrow \lambda_2/{\sqrt \Lambda} \, .
 \end{equation}
 This ensures that also the coupling constant $\lambda_2$ does not
 renormalize in the long wavelength limit. This can be formulated in
 a more formal language (see Appendix \ref{ward}).
\end{itemize}
Summarizing, none of the coupling constants
$\lambda_1,\,\lambda_2,\,\lambda_3$ renormalizes. Hence, in
perturbative renormalization group treatments, if we were to carry
out a renormalization-group transformation by integrating out a
shell of modes $\Lambda e^{-l}<q< \Lambda$, and rescaling
$x_i\rightarrow e^lx_i,\,u_i\rightarrow e^{\chi l}u_i,\,t\rightarrow e^{lz}t,\,
b_i\rightarrow e^{\chi l}b_i$, the couplings $\lambda_1,\,\lambda_2$
and $\lambda_3$ would be affected only by na\"ive rescaling. Thus,
rescaling of space and time can be done in such a way as to keep the
coupling strengths $\lambda_1,\,\lambda_2,\,\lambda_3$ constant.

We perform a one-loop dynamic renormalization group (DRG)
transformation on the model Eqs. (\ref{eq:burgers_1}) and
(\ref{eq:burgers_2}) with the correlations of the noise sources
specified in Eqs.  (\ref{noise1}), (\ref{noise2}) and
(\ref{noise3}).
The cross-correlation function is imaginary and odd in wavevector
${\bf k}$: It is proportional to $sgn ({\bf k})$ and hence is
non-analytic at ${\bf k}=0$. Since perturbative expansions as in DRG
analyses used here are always analytic at ${\bf k}=0$, non-analytic
terms of the form $sgn(\bf k)$ are not generated. Hence there are no
perturbative corrections to the cross-correlations.
%
Furthermore,  the model equations (\ref{eq:burgers_1}) and
(\ref{eq:burgers_2}) and the variances of the noise sources
(\ref{noise1}), (\ref{noise2}) are invariant under inversion
(reversal of parity), rotation and the interchange of Cartesian
co-ordinates $i$ and $j$. The last two invariances, under rotation
and interchange between $i$ and $j$ respectively, are broken only by
the choice of the noise cross-correlations (\ref{noise3}), of the
external stochastic forces which can be controlled from outside
separately. This is reflected in the fact the presence of the
symmetric (anti-symmetric) noise cross-correlations does not lead to
the generation of the anti-symmetric (symmetric) noise
cross-correlations. This allows us to explore the effects of
symmetric and anti-symmetric noise cross-correlations separately. It
should be noted that the calculations presented here are done at a
fixed dimension $d$, instead of as an expansion about any critical
dimension.

\paragraph{Symmetric cross-correlations.}
We first consider the case when the (bare) noise cross-correlations
are fully symmetric (finite $D_s^0$), i.e., no anti-symmetric
cross-correlations are present ($D_a^0=0$). We perform a
renormalization group transformation as outlined above. The resulting
RG flow equations are presented in terms of the renormalized and
rescaled variables
\begin{equation}
  (\nu,\,\mu)       \rightarrow  (\nu,\,\mu)e^{(z-2)l} \, , \nonumber \\
  (D_u,\,D_b,\,D_s) \rightarrow  (D_u,\,D_b,\,D_s)e^{(z-d-2\chi_h)l}.
\end{equation}
We obtain the following differential flow equations for the {\em
  running} parameters $\nu (l),\,\mu (l),\,D_u(l),\,D_b(l)$ (with
$\lambda_1=\lambda_2=\lambda_3=\lambda$)
\begin{eqnarray}
  \frac{d\nu}{dl}
   &=&\nu \left[z-2+\frac{2-d}{d}\frac{G}{4}
       \left( 1+{\it A} P_m^2 \right) \right],
  \label{flow1} \\
  \frac{d\mu}{dl}
   &=&\mu \left[z-2+ \frac{G}{4}\frac{2-d}{d}\left(\frac{3+P_m}{(1+P_m)^2} +\frac{{\it A}(1+3 P_m)}{(1+P_m)^2}\right)
   \right] \label{flow2} \\
  \frac{dD_u}{dl}
   &=&D_u \left[z-2\chi_h-d+\frac{G}{4}
      \left( 1+{{\it A}^2 P_m^3}+\frac{4N_s P_m^2}{1+P_m} \right) \right],\label{flow3} \\
  \frac{dD_b}{dl}
   &=&D_b \left[z-2\chi_h-d+ G \left(\frac{P_m^2}{1+P_m}-\frac{4P_m^3}{(1+P_m)^3} \frac{N_s}{\it A}
   \right)\right],\label{flow4} \\
  \frac{d\lambda}{dl}
   &=&\lambda[\chi_h +z-2].\label{flow5}
\end{eqnarray}
The parameter $D_s$ does not receive any fluctuation corrections and
is affected only by na\"ive rescaling. Here, we have introduced an
effective coupling constant $G=\frac{\lambda^2D_u}{\nu^3}$ and two
amplitude ratios, ${\it A}= \frac{D_b}{D_u}$ and
$\betacross=(\frac{D_s}{D_u})^2$, characterizing the relative
magnitude of the noise amplitudes for the magnetic field and the
(symmetric) cross-correlations with respect to the noise amplitude
of the velocity field, respectively. Further, $P_m=\nu/\mu$ is the
{\em renormalized} Prandtl number. We find, from the flow equations
(\ref{flow1}) and (\ref{flow2}), at the RG fixed point $\nu=\mu$,
i.e., we have $P_m=1$ at the RG fixed point, regardless of the
values of the bare viscosities. Henceforth, we put $\nu=\mu$ in our
calculations below. Flow equations for the effective coupling
constant $G$ and the amplitude-ratio ${\it A}=D_b/D_u$ may be
obtained from its definition above and by using the flow equations
(\ref{flow1}-\ref{flow5}). They are
\begin{eqnarray}
\frac{d\it A}{dl}
   &=&{\it A} \left[(1+{\it A}^2+2\betacross)-2(1-N_s/{\it A}) \right]
   \frac{G}{4}, \label{flow6}\\
\frac{dG}{dl}&=& G\left[2-d +
2G\frac{2d-3}{2d}(1-\betacross)\right]. \label{flow7}
\end{eqnarray}

At the RG fixed point renormalized parameters are scale invariant
(i.e., do not receive fluctuation corrections anymore under further
mode eliminations); we then set the LHS of Eqs.
(\ref{flow1}-\ref{flow5}) to zero. These yield (a $^*$ denotes fixed
point values),
\begin{eqnarray}
  G^* \in \left\{ 0, \,
    \frac{2d\epsilon}{2d-3} \left( 1+\betacross \right) \right\} \, ,
  \label{paramfix}
\end{eqnarray}
to the lowest order in $\betacross$ with $\epsilon=d-2$.  When
$G^*=0$, ${\it A}$ is undetermined and when
$G^*=\frac{2d\epsilon}{2d-3} \left( 1+\betacross \right)$ we find
\begin{equation}
A^*=1-2\betacross. \label{paramfix1}
\end{equation}
Note that the fixed point value of the effective coupling constants
$G$ and the amplitude ratio ${\it A}$ explicitly depend on the
strength of the noise cross-correlations $\betacross$. We show below
that the parameter $\betacross$ is marginal at the RG fixed point,
i.e., $\betacross$ can have variable values at the fixed point. In
the rough phase at $d=1$, we find $G^*=2(1+\betacross)$ as the
stable fixed point and an amplitude-ratio ${\it A}^* = D_b/D_u = 1 -
2\betacross$. This implies that the non-linearities are relevant and
the asymptotic scaling properties of the correlation functions are
different from those for the corresponding linear model. For $d = 2
+ \epsilon$ with $\epsilon
> 0$, we obtain $G^*=0$ as a stable and $G^* = 4\epsilon (1+N_s)$ as
an unstable fixed point indicating a smooth-to-rough transition. In
the smooth phase ($G^*=0$) the nonlinearities are irrelevant, the
scaling properties are determined by the corresponding linear
equations and we find $d{\it A}/dl = 0$, i.e., $\it A$ does not
change under mode elimination and is simply given by
$D_{b}^0/D_{u}^0$, the bare amplitude ratio.  At the roughening
transition, $G^*=4\epsilon (1+\betacross)$, the amplitude ratio
becomes $A^*=1-2\betacross$. Also note that the value of the
coupling constant $G^*$ at the critical point for $d= 2+\epsilon$
increases with increasing $\betacross$, i.e., with increasing
symmetric noise cross-correlations. These, therefore, suggest that
the presence of symmetric noise cross-correlations helps to
stabilize the smooth surface against roughening perturbations. For
$d>2$ beyond the critical point ({\em roughening transition
  point}) there is presumably a rough phase which is not accessible by
perturbative RG. This is reminiscent of the analogous problem in the
KPZ Equation \cite{freykpz}. Further we obtain $\chi_h=1/2,\,z=3/2$
in the rough phase at $d=1$ and
$\chi=-O(\epsilon^2),\,z=2+O(\epsilon)^2$ at the roughening
transition for $d=2+\epsilon$. Scaling exponents for the rough phase
cannot be obtained by perturbative RG.

\paragraph{Anti-symmetric cross-correlations.}

Having discussed the effects of symmetric cross-correlations, we now
proceed to analyze the effects of the anti-symmetric
cross-correlations in a DRG framework. Therefore, we now have a
finite $D_a^0$ in the bare noise cross-correlations with $D_s^0$
being set to zero. In this situation, fields $\bf u$ and $\bf b$,
and forces $\bf f$ and $\bf g$ are no longer expressible as
gradients of scalars. Hence, Equations (\ref{eq:burgers_1}) and
(\ref{eq:burgers_2}) cannot be reduced to (\ref{eq2kpz1}) and
(\ref{eq2kpz2}).  Therefore, we work with Equations
(\ref{eq:burgers_1}) and (\ref{eq:burgers_2}) directly. We follow
the same scheme of calculations as above. The resulting RG flow
equations are
\begin{eqnarray}
\frac{d\nu}{dl}
   &=&\nu \left[z-2+\frac{2-d}{d}\frac{G_a}{4}
       \left( 1+{A P_m^2} \right) \right],
  \label{flow7a} \\
  \frac{d\mu}{dl}
   &=&\mu \left[z-2+ \frac{G_a}{4}\frac{2-d}{d}(\frac{3+P_m}{(1+P_m)^2} +\frac{A (1+ P_m)}{(1+P_m)^2})
   \right] \label{flow8} \\
\frac{d\lambda}{dl}&=& \lambda [\chi + z-1],\label{flow9}\\
\frac{dD_u}{dl}&=&D_u\left[z-2\chi -2 -d + G_a(1+ A^2P_m^3 +
4\frac{\betacrossas P_m^2}{1+P_m})\right] ,\label{flow10}\\
\frac{dD_b}{dl}&=&D_b\left[z-2\chi -2
-d+G_a\left(\frac{P_m^2}{1+P_m} + \frac{4P_m^3}{(1+P_m)^3}
\frac{N_a}{A}
   \right)
  \right] \label{flow11} .
\end{eqnarray}
The parameter $D_a$, like $D_s$ above, does not receive any
fluctuation correction and is affected only by na\"ive rescaling.
Note that the flow Eqs. (\ref{flow7a}) and (\ref{flow8})  have the
same form as Eqs. (\ref{flow1}) and (\ref{flow2}), the corresponding
flow equations in the symmetric cross-correlations case, except that
we now denote the new dimensionless coupling constant by
$G_a=\lambda^2 D_u/\nu^3$. Here,  $\betacrossas =(D_a/D_u)^2$.
Further, as in the symmetric cross-correlations case, $P_m=1$ at the
RG fixed point. However, the flow equations (\ref{flow10}) and
(\ref{flow11}) are not identical to their counterparts (\ref{flow3})
and (\ref{flow4}) in our discussions on symmetric cross-correlations
above: the fluctuation corrections  contributing to $D_u$ and $D_b$
arising from the anti-symmetric cross-correlations have the same
sign in this case unlike the case with symmetric cross-correlations.
We find that the parameter $A=D_b/D_u$ is unity for all values of
$D_a$, i.e., $A=1$ at the RG fixed point. Further, one can obtain a
flow equation for the the coupling constant $G_a=\lambda^2D_u/\nu^3$
with the help of Eqs. (\ref{flow7a}- \ref{flow9}). It is
\begin{equation}
\frac{dG_a}{dl}=G_a\left[2-d + 2G_a \left(\frac{2d-3}{2d}
+\betacrossas\right)\right].\label{flowga}
\end{equation}
At the RG fixed point $dG_a/dl=0$ yielding $G_a=0$ or
$G_a=\epsilon\left[\frac{2d}{2d -3} + O(\betacrossas)\right]$. The
value $G_a=0$ corresponds to the smooth phase, as in the KPZ case,
whereas $G_a^*=\epsilon\left[\frac{2d}{2d -3} +
O(\betacrossas)\right]$ is an unstable fixed point indicating a
smooth-to-rough phase transition. In the smooth phase the scaling
exponents $z=2,\,\chi=(2-d)/2$ which are identical to their KPZ
counterparts. Further at the phase transition point the exponents
are independent of $\betacrossas$:
$z=2+O(\epsilon)^2,\,\chi=O(\epsilon)^2$; again the statistical
properties of the rough phase cannot be explored by perturbative RG.

Note that our above conclusions on obtaining continuously varying
amplitude-ratios  in the rough phase in $1d$ and at the
smooth-to-rough transition for $d\geq 2$ for finite symmetric
cross-correlations  rest on the requirement that $\betacross$ and
hence $D_s$ can have variable values at the RG fixed point. This can
happen if $D_s$  is {\em marginal} at the RG fixed point. Therefore,
to complete our analysis we now proceed to demonstrate that the
parameter $\betacross$ is strictly marginal at the RG fixed point,
even beyond linearized RG. In our notations $D_s({\bf k})/\nu$ is
the amplitude of $C_{ij}^s ({\bf k},t)$, the symmetric part of the
cross-correlation function matrix, with $D_s({\bf k})D_s({\bf
k})=4D_s^2$. Since $C_{ij}^s({\bf k},t)$ is an odd function of the
wavevector $\bf k$, we must have, for consistency, $C_{ij}^s ({\bf
  k}=0,t)=0$. Therefore, there must be a length scale $l_c$ (which
itself diverges in the thermodynamic limit) such that
\begin{eqnarray}
C_{ij}^s({\bf k},t=0)&=&2iD^{s}_{ij}({\bf k})k^{-d-2\chi}/\nu\,\,
 {\rm for}\ {\bf k}
\rightarrow 0
\end{eqnarray}
up to a scale $l,\,
 1/k\sim l\lesssim l_c \rightarrow \infty$, and
\begin{equation}
C_{ij}^s({\bf k},t=0)= 0 ,\,{\rm at}\;\; k=0 \,\,(l_c\leq l=\infty).
\end{equation}
Under rescaling we have $D_u(l)=D_ul^{z-2\chi-3}$ as
$l\rightarrow\infty$. In contrast, under the same rescaling
cross-correlation $C_{ij}^s ({\bf k},0) =2iD_s({\bf
  k})k^{-d-2\chi}k_ik_j/\nu$ if $k'=k/b \geq l_c^{-1}$ but is zero if
$k'<l_c^{-1}$. Thus the true scaling regime is $l<l_c \rightarrow
\infty$ and at $l\lesssim l_c$, $D_s(l)\sim D_s
l_c^{z-2\chi-d-2}$. The latter does not receive any fluctuation
corrections under mode elimination, as we argued above, and hence is arbitrary because it
depends on $D_s=D_s(l=0)$ and therefore marginal.  Therefore,
$\betacross=(\frac{D_s}{D_u})^2$ is also marginal. The marginality of
$\betacrossas =(\frac{D_a}{D_u})^2$ can be argued similarly. We close
this Section by summarizing our results obtained from the DRG scheme:
\begin{itemize}
\item We obtain the amplitude-ratio $D_b/D_u=1-2N_s$ in the presence
  of finite symmetric cross-correlations $N_s$ in the rough phase at
  $d=1$ and smooth-to-rough transition at $d=2+\epsilon$. The values
  of the scaling exponents are unchanged from their values for the KPZ
  equation.
\item In the presence of the anti-symmetric cross-correlations, the
  amplitude-ratio $D_b/D_u=1$ at $d=2+\epsilon$. The scaling exponents
  are unchanged from their KPZ values.
\end{itemize}
Clearly, the DRG scheme, as in the KPZ equation, fails to yield any
result concerning the strong coupling rough phase at
$d=2+\epsilon$. We now resort to the SCMC and the FRG schemes to
obtain results for the rough phases at $d=2+\epsilon$.

\subsection{Self-consistent mode-coupling analysis in $d$-dimensions}
\label{modecoupling}

In a self-consistent mode-coupling (SCMC) scheme perturbation theories
are formulated in terms of the response and correlation functions of
the fields ${\bf u}({\bf x},t)$ and ${\bf b}({\bf x},t)$. They are
conveniently expressed in terms of self-energies and generalized
kinetic coefficients. As before, without any loss of generality we
assume $\nu=\mu$, i.e., the magnetic Prandtl number $P_m=1$. This
guarantees that there is only one response function and it can be
written as
\begin{equation}
G^{-1}_{u,b}({\bf k},\omega)=-i\omega -\Sigma({\bf k},\omega),
\label{selfener}
\end{equation}
The correlation functions are of the form
\begin{eqnarray}
C^{u,b}_{ij}({\bf k},\omega)&=&2D_{u,b}({\bf k},\omega)k_ik_j
|G({\bf k},\omega)|^2,
\nonumber \\
C^{\times}_{ij}({\bf k},\omega)&=&2iD_{\times}({\bf k},\omega){_ij}
|G({\bf k},\omega)|^2, \label{corrfunc1}
\end{eqnarray}
where $C^{\times}_{ij}({\bf k},\omega)$ stands for the cross
correlation functions of $\bf u$ and $\bf b$. In the scaling limit, in
terms of wavevector $\bf k$ and frequency $\omega$ the self-energy and
the correlation functions exhibit scaling forms characterized by the
scaling exponents $z$ and $\chi$ and appropriate scaling functions:
\begin{eqnarray}
C^{u,b}_{ij}({\bf k},\omega)&=&2D_{u,b}k_ik_jk^{-2\chi-2-d-z}f_{u,b}(\omega
/k^z),\nonumber \\
C^{\times}_{ij}({\bf k},\omega)&=&2iD_{\times}({\bf k})k_ik_j k^{-2\chi-2-d-z}
f_{\times}(\omega /k^z).
\end{eqnarray}
In diagrammatic language a lowest order mode-coupling theory is
equivalent to a self-consistent one-loop theory. The ensuing coupled
set of integral equations are compatible with the scaling forms
above. To solve this set of coupled integral equations we follow
Ref.\cite{jkb} and employ a small-$\chi$ expansion. This essentially
requires matching of correlation functions and the self-energy at zero
frequency with their respective one-loop expressions. We consider the
following two cases separately:

\paragraph{Symmetric cross-correlations:-}
We consider the case when $C^{\times}_{ij}({\bf k},\omega)=
C^{\times}_{ji}({\bf k},\omega)$. This implies that the fields $\bf
u$ and $\bf b$ can be expressed as gradients of scalars: ${\bf
u}=\nabla h$ and ${\bf b} =\nabla \phi$; note that $\phi$ is
actually a pseudo scalar. Such choices imply that the fields $\bf u$
and $\bf b$ are irrotational vectors. The corresponding equations of
motion in terms of the fields $h$ and $\phi$ are given by
(\ref{eq2kpz1}) and (\ref{eq2kpz2}).


In this case the zero frequency expressions for the correlators and the
response function become
\begin{eqnarray}
\Sigma({\bf k},0)&=&\Gamma k^z,\nonumber \\
C^{u,b}_{ij}({\bf k},0)&=&2\frac{D^{u,b}_{ij}}{\Gamma}k_ik_jk^{-d-2\chi-z},\nonumber \\
C^{s}_{ij}({\bf k},0)&=&2i\frac{D_s({\bf k})}{\Gamma}k_ik_jk^{-d-2\chi-z},
\end{eqnarray}
where $C^{s}_{ij}({\bf k},0)$ is the symmetric part of the cross
correlation function of $\bf u$ and $\bf b$. We also define $D_s({\bf
  k})D_s({\bf k}) ={(D_s)}^2$. In an SCMC approach vertex corrections
are neglected which are exact statements for the present problem in
the zero-wavevector limit. Lack of vertex renormalizations in the
zero-wavevector limit yield the exact relation between the scaling
exponents $\chi$ and $z$, as in the case of the noisy
Burgers/Kardar-Parisi-Zhang equation \cite{freykpz}. In the context
of the Burgers equation in 1+1 dimension Frey {\em et al} showed
\cite{frey1dburg}, by using nonrenormalization of the advective
nonlinearity and second order perturbation theories that the effects
of the vertex corrections at finite wavevectors on the correlation
functions are small. Presumably the same conclusion regarding the
effects of vertex renormalization at finite wavevectors follows for
this model in the present problem also. However, a rigorous
calculation is still lacking. With these definitions the one-loop
self-consistent equations yield the following relations between the
amplitudes. For the self-energy we obtain (without any loss of
generality we set $\lambda_1=\lambda_2=\lambda_3=\lambda$)
\begin{equation}
\frac{\Gamma^2}{D_u\lambda^2}=\frac{S_d}{(2\pi)^d}\frac{1}{2d}(1+\frac{D_u}{D_b}),
\label{selfmode}
\end{equation}
and for the one-loop correlation functions
\begin{eqnarray}
\frac{\Gamma^2}{D_u\lambda^2}&=&\frac{1}{4}\frac{S_d}{(2\pi)^d}\frac{1}{d+3\chi
-2}
\left[1+\left(\frac{D_b}{D_u}\right)^2+2\left(\frac{D_s}{D_u}\right)^2\right],\nonumber
\\
\frac{\Gamma^2}{D_b\lambda^2}&=&\frac{1}{2}\frac{S_d}{(2\pi)^d}\frac{1}{d+3\chi
-2} \left[\frac{D_u}{D_b}-\left(\frac{D_s}{D_b}\right)^2\right],
\label{corrmode}
\end{eqnarray}
Here $S_d$ is the surface of a $d$-dimensional sphere. From
Eqs.(\ref{selfmode}) and (\ref{corrmode}) we obtain
\begin{equation}
\left(\frac{D_b}{D_u}\right)^2+2N_s\left(\frac{D_u}{D_b}+1\right)-1=0.
\label{Ns}
\end{equation}
where $\betacross\equiv (D_s/D_u)^2$ is a dimensionless ratio as
defined above. Since the ratio $D_b/D_u$ is positive semi-definite,
in Eq.(\ref{Ns}) the range of $N_s$ is determined by the range of
positive values for $D_b/D_u$ starting from unity (obtained when
$N_s=0$). Thus for small $\betacross$ we can expand around zero and
look for solutions of the form $A\equiv D_b/D_u=1+aN_s$, such that
for $N_s=0$ we recover $D_u=D_b$ (the result of Ref.\cite{ek}). We
obtain $a=-2$, i.e., $D_b /D_u=1-2N_s$, implying that within this
leading order calculation $N_s$ cannot exceed 1/2, i.e., $D_s\leq
D_u/\sqrt 2$. An important consequence of this calculation is that
the amplitude ratio $D_b /D_u$ is no longer fixed to unity but can
vary continuously with the strength of the noise cross-correlation
(renormalized) amplitude $D_s$. Our results from this section  are
in agreement and complementary to the those obtained in a DRG
framework above (see Sec. \ref{rg1d}). These results are already
confirmed by a one-loop DRG calculation for the rough phase at $d=1$
(see Sec.\ref{rg1d}). In addition, the application of one-loop DRG
demonstrates that the above results are valid at the roughening
transitions to lowest order in $N_s$ in a $d=2+\epsilon$ expansion
as well.


In contrast, the scaling exponents $\chi$ and $z$ are not affected
by the presence of cross correlations. From the above one-loop
mode-coupling Eqs. (\ref{selfmode}) and ({\ref{corrmode}) we obtain
$\chi=-1/2$ and $z =3/2$ in $d=1$ dimensions from our SCMC which are
same as obtained by DRG calculations. Further equations
(\ref{selfmode}) and (\ref{corrmode})
  yield the following values for the scaling exponents in the strong
  coupling regime which are unaffected by the presence of symmetric
  cross-correlations.
\begin{equation}
\chi=-\frac{1}{3}-\frac{d}{6};\;z=\frac{4}{3}+\frac{d}{6}.\label{scstrong}
\end{equation}
These are identical to those obtained by Bhattacharjee in a
small-$\chi$ expansion \cite{jkb} and it is still controversial
whether these values for the exponents actually correspond to the
usual strong coupling case. Recently, Canet {\em et al} performed a
more critical analysis of the self-consistent mode-coupling
equations for the KPZ Equation \cite{moore} and showed the
corresponding mode-coupling equations have two branches (or
universality classes) of the solutions: the $F$ branch having the
upper critical dimension $d_c=4$ and the $S$ solution with $d_c=2$.
The $F$ solution is believed to correspond to the usual rough phase
and the $S$ solution has been discussed in some calculations on the
directed polymer problem \cite{s-soln}. Our solutions or rather
Bhattacharjee's small $\chi$ expansion yields $d_c=4$ as the $F$
solution and agrees with the $F$ solution at $d=0$ and $d=1$ as
well. At other dimensions there are small quantitative differences
between the values for $z$.

\paragraph{Antisymmetric cross-correlations:}
So far we have restricted ourselves to the case where the vector
fields $\bf u$ and $\bf b$ are {\em irrotational}. If however the
fields $\bf a=u,b$ are rotational and have the form
\begin{equation}
{\bf a}={\bf\nabla\times V}_a+\nabla S_a, \label{choiceanti}
\end{equation}
with vectors ${\bf V}_a$ being cross-correlated but the scalars
$S_a$ uncorrelated then the variance $D_{ij}^{\times}({\bf k})$
satisfies
\begin{equation}
D_{ij}^{\times}({\bf k})=-D_{ij}^{\times}({\bf -k})=D_{ji}({\bf
k})=-[D_{ij}^{\times}({\bf k})]^*.
\end{equation}
This is the antisymmetric part of the cross-correlations. Choices
(\ref{choiceanti}) ensures that the vectors $\bf u$ and $\bf b$ are
no longer irrotational. The corresponding (renormalized) noise
strength $D_a$ is formally defined through the relation
\begin{equation}
D_{ij}^{\times}({\bf k})D_{ij}^{\times}({\bf -k})=4D_a^2k^4.
\end{equation}
Similar to the previous case, in the scaling limit (zero frequency
limit) the self energy reads $\Sigma(k,\omega=0)= \Gamma k^z$, the
correlation functions are $C^u_{ij} (k,\omega=0) = k_i k_j D_u
k^{-d-2\chi-z}$, $C^b_{ij} (k,\omega=0) = k_i k_j D_b k^{-d-2\chi-z}$,
and the antisymmetric part of the cross-correlation function reads
$C_{ij}^a (k,\omega=0) = D_{ij}^a ({\bf k})k^{-2\chi-z-d}$.

Following the method outlined above we obtain

\begin{equation}
 \frac{\Gamma^2}{D_u\lambda^2} = \frac{S_d}{(2\pi)^d}
 \frac{1}{2d} \left(1+\frac{D_b}{D_u}\right) \, ,
\label{selfena}
\end{equation}
\begin{eqnarray}
 \frac{\Gamma^2}{D_u \lambda^2}
 &=& \frac14 \frac{S_d}{(2\pi)^d}
     \frac{1}{d-2+3\chi}
     \left[ 1 + \left(\frac{D_b}{D_u}\right)^2 +
               2\left( \frac{\Dcrossa}{D_u}\right)^2
     \right], \nonumber \\
 \frac{\Gamma^2}{D_b \lambda^2}
 &=& \frac12 \frac{S_d}{(2\pi)^d} \frac{1}{d-2+3\chi}
     \left[ \frac{D_u}{D_b} +
            \left( \frac{{\Dcrossa}}{D_b}\right)^2
     \right].
\label{selfcra}
\end{eqnarray}
Equations (\ref{selfena}) and (\ref{selfcra}) give $D_u/D_b=1$ at the
fixed point for arbitrary values of $N_a=({D_a / D_u})^2$. Hence no
restrictions on $N_a$ arises from that.  In contrast to the effects of
the symmetric cross-correlations, the exponents now depend
continuously on $N_a$. To obtain the scaling exponents we use that
$D_b/D_u=1$ and equate Eqs.(\ref{selfena}) and (\ref{selfcra}). To
leading order, we get
\begin{equation}
\chi= -\frac{1}{3}- \frac{d}{6}+ \frac{\betacrossas d}{6} \, , \quad
z=\frac{4}{3}+\frac{d}{6}-\frac{\betacrossas d}{6}. \label{expo}
\end{equation}
These exponents presumably describe the rough phase above $d>2$,
with the same caveats as above~\cite{freylongrange}.  With
increasing $D_a$ the exponent $\chi$ grows (and $z$ decreases).
Obviously this cannot happen indefinitely. We estimate the upper
limit of $N_a$ in the following way: Notice that the
Eqs.(\ref{eq:burgers_1}) and (\ref{eq:burgers_2}) along with the
prescribed noise correlations (i.e., equivalently the dynamic
generating functional) are of conservation law form, i.e.\/ they
vanish as $\bf k\rightarrow 0$. Thus there is no information of any
infrared cut off in the dynamic generating functional.  Moreover, we
know the solutions of the equations {\em exactly} if we drop the
non-linear terms (and hence, the exponents: $\chi=1-d/2,\,z=2$).
Therefore, physically relevant quantities like the total  {\em
energies} of the fields $\bf u$ and $\bf b$ fields \footnote{ These
are kinetic and magnetic energies when $\bf u$ and $\bf b$ are
interpreted as {\em Burgers velocity} and {\em Burgers magnetic}
fields.}, $\int_k\langle {\bf u} ({\bf k},t) {\bf u} (-{\bf
k},t)\rangle$ and $\int_k\langle {\bf b} ({\bf k},t) {\bf b} (-{\bf
k},t) \rangle$, remain {\em finite} as the system size diverges, and
are thus independent of the system size: In particular
\begin{eqnarray}
\int_k\langle {\bf u} ({\bf k},t) {\bf u} (-{\bf k},t)\rangle &&\sim \int
d^dk\; k^{2-d+2\chi},\nonumber \\
\int_k\langle {\bf b} ({\bf k},t) {\bf b} (-{\bf k},t)\rangle &&\sim
\int d^dk \;k^{2-d+2\chi}
\end{eqnarray}
which, for $\chi=-d/2$ (the {\em exact} value of $\chi$ without the
nonlinear terms), are finite in the infinite system size limit. Since
the non-linear terms are of the conservation law form, inclusion of
them {\em cannot } bring a system size dependence on the values of the
total energies. However, if $\chi$ continues to
increase with $D_a$ at some stage these energies would start to depend
on the system size which is unphysical \cite{foot1}:
\begin{eqnarray}
\int_k\langle {\bf u} ({\bf k},t) {\bf u} (-{\bf k},t)\rangle\sim
\int d^d k \,k^{2-4/3-
d\betacrossas} \nonumber \\
\int_k\langle {\bf b} ({\bf k},t) {\bf b} (-{\bf k},t)\rangle\sim
\int d^d k \,k^{2-4/3-d\betacrossas}.
\end{eqnarray}
Therefore, in order to make our model meaningful in the presence of
anti-symmetric cross-correlations, we have to restrict $D_a$ to
values smaller than the maximum value for which these energy
integrals are just system-size independent: This gives
$\betacrossas^{\rm max}=\frac{2}{d}(d/2+1)$.  Note that the limits
on $\betacross$ and $\betacrossas$ impose consistency conditions on
the ratios of the amplitudes of the measured correlation functions
but not on the bare noise correlators. We can use the values of the
dynamic exponent to estimate the upper critical dimension $d_c$ of
the model in the presence of the antisymmetric cross-correlations.
From our expressions (\ref{expo}) the dynamic exponent $z$ increases
with the spatial dimension $d$ for a given strength of the
antisymmetric cross-correlation $\betacrossas$. The value of $d_c$
is given by the dimension in which the dynamic exponent $z$ attains
a value 2 equal to its value {\em without} the nonlinear term.
Clearly, from expressions (\ref{expo}) $z=2$ yields
$d_c=4/(1-\betacrossas)$. Therefore, the antisymmetric
cross-correlations have the effects of increasing the upper critical
dimension of the model.


Antisymmetric cross-correlations stabilize the short-range fixed
point with respect to perturbations by long-range noise with
correlations of the form (in the Fourier space) $k^{-y},\, y>0$.
This can easily be seen: In presence of noise correlations
sufficiently singular in the infra-red limit, i.e.\/ large enough
$y$, the dynamic exponent is  known exactly
\cite{freylongrange,jkbamit}: For a sufficiently large $y$ the
one-loop corrections to the correlators scale same as the bare
correlators for zero external frequency; the one-loop diagrams are
finite for finite external frequencies. Thus they are neglected and
this, together with the Ward identities discussed above yield
$z_{\rm lr} = \frac{2+d}{3} - \frac{y}{3}$. The short range fixed
point remains stable as long as $z_{\rm sr}<z_{\rm lr}$ which gives
$y<-2+(1+\betacrossas)d/2$. Hence we conclude that in the presence
of antisymmetric cross-correlations a long range noise must be {\em
more
  singular} for the short range noise fixed point to loose its
stability or in other words, antisymmetric cross-correlations
increases the stability of the short range noise fixed point with
respect to perturbations from long range noise sources.

We close this section by summarizing our results obtained from the
SCMC calculations:
\begin{itemize}
\item The SCMC method yields results about the rough phase at $d=1$
  and $d>2$.
\item We find that the amplitude $D_b/D_u$ decreases monotonically
  from unity as the symmetric cross-correlations, parametrized by
  $N_s$ increases from zero.  Our result here is in agreement with
  that obtained from the DRG method for $d=1$. The ratio $D_b/D_u$ is
  unaffected by the antisymmetric cross-correlations.
\item Our SCMC calculations yield for the scaling exponent also. In
  the presence of the antisymmetric cross-correlations parameterized
  by $N_a$ they are: $\chi= -\frac{1}{3}- \frac{d}{6}+
  \frac{\betacrossas d}{6} \, , \quad
  z=\frac{4}{3}+\frac{d}{6}-\frac{\betacrossas d}{6}$. The scaling
  exponents are unaffected by the symmetric cross-correlations.
\item The maximum value of the parameter $N_s$ is obtained by setting
  $D_b/D_u$ to zero, while the maximum value of $N_a$ is obtained by
  setting $\chi$ to zero in any dimension. The minimum values for
  both of them are zero.
\end{itemize}

\subsection{Functional renormalization group analyses on the model}
\label{funcrg}

In this Section we study the model Eqs. (\ref{eq:burgers_1}) and
(\ref{eq:burgers_2}) in a functional renormalization group (FRG)
framework. This study is complementary to our DRG and SCMC studies
above. In Section \ref{coup-lat} it has been discussed that the
model Eqs. (\ref{eq:burgers_1}) and (\ref{eq:burgers_2}), for the
bare Prandtl number $P_m^o=\nu_0/\mu_0=1$, reduces to two KPZ
equations [see, Eqs. (\ref{2kpz}) for their $1d$ representations]
representing two growing surfaces $h_1({\bf x},t)$ and $h_2({\bf
x},t)$ which are coupled by noise sources [see, Eqs. (\ref{psi3})
for the noise sources in $1d$]. Such equations in general
$d$-dimensions are

\begin{eqnarray}
\partial_t h_1+1/2(\nabla h_1)^2&=&\nu_0\nabla^2 h_1+\psi_1,\nonumber \\
\partial_t h_2+1/2(\nabla h_2)^2&=&\nu_0\nabla^2 h_2+\psi_2.
\label{2kpzalld}
\end{eqnarray}
Here, the noise correlations in arbitrary dimension $d$ are given by
\begin{eqnarray}
\langle\psi_1({\bf k},t)\psi_1(-{\bf k},0)\rangle&=&2D_0\delta(t),\nonumber \\
\langle\psi_2({\bf k},t)\psi_2(-{\bf k},0)\rangle&=&2D_0\delta(t),\nonumber \\
\langle\psi_1({\bf k},t)\psi_2(-{\bf k},0)\rangle&=&2\hat D_0 \delta
(t) +2i\tilde D_0({\bf k})\delta (t), \label{noisealld}
\end{eqnarray}
omitting a formally divergent factor $\delta(k=0)$ in each of the variances above.
Note that in the third equation of (\ref{noisealld}) the
cross-correlation of the noise sources $\psi_1$ and $\psi_2$ has a
real and an imaginary parts, where as in its one-dimensional
version, used to introduce our lattice-gas models, given by
Eqs.~(\ref{psi1}-\ref{psi3}) the cross-correlation has no real part.
This is because in the bare theory even if the real part of the
noise cross-correlation is zero it would be rendered non-zero
self-consistently in the presence of the imaginary part. In other
words, the imaginary part gives rise to the real part in the
(one-loop) self-consistent theory. The real part, however, remains
zero self-consistently if there in no imaginary part in the bare
noise cross-correlations.

We begin by applying the well-known Cole-Hopf transformation
\cite{cole} to the Eqs. (\ref{2kpzalld}) : $h_{1,2}({\bf
  x},t)=(2\nu/\lambda)\ln Z_{1,2}$.  These transformations reduce the
Eqs. (\ref{2kpzalld}) to
\begin{eqnarray}
\partial_t Z_1=\nu_0\nabla^2 Z_1 + V_1 Z_1,\nonumber \\
\partial_t Z_2=\nu_0\nabla^2 Z_2 + V_2 Z_2,
\label{colehopfeq}
\end{eqnarray}
where $V_1=\frac{\psi_1}{2\nu_0},\,V_2=\frac{\psi_2}{2\nu_0}$.
Equations (\ref{colehopfeq}) can be interpreted as the equations for
the partition functions $Z_1$ and $Z_2$ for two identical directed
polymers (DP), each having $d$ transverse components, in a random
medium whose combined Hamiltonian is given by
\begin{equation}
H=\int dt \left[\frac{\nu_0}{2}\left(\frac{d{\bf x_1}}{dt}\right)^2
+\frac{\nu_0}{2} \left(\frac{d{\bf x_2}}{dt}\right)^2 +V_1({\bf
x_1},t)+V_2({\bf x_2},t) \right]. \label{hamil}
\end{equation}
Functions $V_1({\bf x_1},t)$ and $V_2({\bf x_2},t)$, for this
generalized DP problem, are to be interpreted as quenched random
potentials experienced by the two DPs embedded in them. Clearly, the
potentials $V_1$ and $V_2$ are Gaussian distributed with zero-mean
and variances given by Eqs. (\ref{noisealld}). The coordinate $t$ in
the Hamiltonian $H$ in Eq. (\ref{hamil}), which denotes the physical
{\em time} in the coupled surface growth problem, now becomes the
arc-length of the DPs; ${\bf x}_1(t)$ and ${\bf x}_2(t)$ are the
transverse spatial coordinates of the two DPs. In the Hamiltonian
(\ref{hamil}) the first two terms are the energies of the two DPs
due to transverse fluctuations (elastic energies) which are
minimized if the DPs are straight, and $V_1,\,V_2$ are the potential
energies due to the quenched disorder which can be minimized  if the
DPs follow the minima of the potential landscapes (and hence they
will not be straight). Thus there will be competition between the
two opposite tendencies and there will be different phases depending
upon which one wins in the thermodynamic limit. Due to the structure
of the noise correlations given by the Eqs. (\ref{noisealld}) it is
clear that the cross-correlations of the quenched random potentials
$V_1$ and $V_2$ have a part odd in wavevector $\bf k$, suggesting
that the disorder distribution of the embedding disordered medium
lacks reflection symmetry, i.e., it has a chiral nature. The phase
diagram of two DPs in a reflection-symmetric random environment has
been discussed in Ref.\cite{abpol}. In the present work we include
the effects of chirality and discuss its consequence on the
statistical properties of the two DPs. Thus, our studies of the Eqs.
(\ref{eq:burgers_1}) and (\ref{eq:burgers_2}) can equivalently, in
terms of the Directed Polymer (DP) language, be considered as
investigating the phase diagram of the following toy model: Let us
assume that the two DPs A and B are embedded in a random medium
which has two kinds of pins A and B which pin polymers A and B
respectively. Both the pins are distributed randomly with specified
distributions. Furthermore, the pins A and B may have some
correlations in their distributions, or may not have. If they do
have, then, the effects of such correlations should be modeled by
the cross-correlations of the type we are discussing here.
Physically, if there is a pin A somewhere, then a positive
correlation between the distribution of pins A and B would indicate
that a pin B is likely to be found nearby. Since pins are the places
where polymers are likely to get stuck, then according to the above,
if polymer A is stuck somewhere, then polymer B is also likely to
get stuck nearby with a probability which is higher than if pins A
and B had no correlations among their distributions. This, in some
sense, creates an {\em effective} interaction between polymers A and
B (since they are more likely to get pinned at nearby places). We
elucidate the resulting effects in an FRG framework.

In the present problem, since the two DPs are not interacting with
each other directly, the total partition function $Z$ of the two
DPs, for a given realization of the pinning potentials, is then
given by the product of the individual partition functions:
\begin{equation}
Z=Z_1Z_2=\int {\mathcal D{\bf x}_1}{\mathcal D{\bf x}_2}\exp
[-\frac{1}{T}\int dt\; H]. \label{fullpart}
\end{equation}
Here, $T\equiv 1/\nu_0$ is the temperature. Therefore, following the
standard replica method \cite{replica} the free energy of the
system, after averaging over the distribution of the random
potentials $V_1$ and $V_2$, is given by
\begin{equation}
\langle F\rangle \equiv \langle \ln Z\rangle = lim_{N\rightarrow
0}\frac {\langle Z^N\rangle -1}{N},
\end{equation}
In order to facilitate the usage of the standard FRG method we
generalize the correlations of the random potentials
(\ref{noisealld}), similar to the corresponding FRG treatment for
the single DP problem \cite{halpin}, in the following way:
\begin{eqnarray}
  \langle V_1({\bf x},t)V_1(0,0)\rangle&=&R_1({\bf x})\delta(t),\nonumber \\
  \langle V_2({\bf x},t)V_2(0,0)\rangle&=&R_2({\bf x})\delta(t),\nonumber \\
  \langle V_1({\bf x},t)V_2(0,0)\rangle&=&\hat R ({\bf x})\delta (t)
  +R_{\times}({\bf x})\delta (t). \label{potalld}
\end{eqnarray}
Here, following Ref. \cite{halpin}, the spatial $\delta$-functions
in (\ref{noisealld}) have been replaced by short range functions
$R_1,\,R_2,\,\hat R$ for convenience. The function $R_{\times} ({\bf
  x})$ is an odd function of $\bf x$ representing the imaginary part
of the cross-correlations $\langle\psi_1({\bf k},t)\psi_2(-{\bf
k},0)\rangle$. Hence, $R_{\times}(0)=0$. Note that in the
DP-language functions $R_1({\bf
  x}),\,R_2({\bf x}),\,\hat R ({\bf x})$ and $R_{\times} ({\bf x})$
are proportional to the noise variances in Eqs. (\ref{noisealld}) and
hence to the corresponding correlators. With the above definitions and
notations, we have
\begin{eqnarray}
\langle Z^N\rangle&=&\int \Pi_{\mu,\nu}{\mathcal D}{\bf x}_{1\mu}
{\mathcal D} {\bf x}_{2\nu} \exp\left[-\frac{1}{2T}\int dt
\Sigma_\alpha \left\{\left(\frac{d{\bf x}_{1\alpha}}{dt}\right)^2
+\Sigma_\alpha  \left(\frac{d{\bf x}_{2\alpha
}}{dt}\right)^2\right\}\right] \nonumber
\\&&\times \exp\left[ \frac{1}{T^2}\Sigma_{\alpha,\beta} R_1({\bf x}_{1\alpha} - {\bf
x}_{1\beta})+\frac{1}{T^2}\Sigma_{\alpha,\beta} R_2({\bf x}_{2\beta}
- {\bf x}_{2\beta}) \right] \nonumber
\\&& \times \exp\left[\frac{1}{T^2}\Sigma_{\alpha,\beta} \hat R({\bf x}_{1\alpha} - {\bf
x}_{2\beta}) +\frac{1}{T^2}\Sigma_{\alpha\beta} R_{\times}({\bf
x}_{1\alpha} - {\bf x}_{2\beta})\right]. \label{reppart}
\end{eqnarray}
Here, indices $\alpha,\beta$ correspond to the replica indices
arising out of the replica method used above, representing identical
copies of the same system. Clearly, averaging over the distribution
of the potentials lead to generation of terms with mixed replica
indices - systems having different replica indices now interact with
each other.

In order to set up the functional renormalization group (FRG)
calculation for establishing the long wavelength forms of the
disorder correlators we rescale $\bf x_{1,2}\rightarrow e^l
x_{1,2}$ and $t\rightarrow e^{l\zeta t}$ such that $t\sim |{\bf
x_{1,2}}|^\zeta$, relating longitudinal and transverse fluctuations.
Clearly, the exponent $\zeta$ is the inverse of the dynamic exponent
$z$: $\zeta=1/z$. Such a rescaling yields for temperature
$T\rightarrow e^{(1-2\zeta)l}T$. The
differential flow equation for $T$ then reads
\begin{equation}
\frac{dT}{dl}=(1-2\zeta)T. \label{Tnor}
\end{equation}
Hence, if the disorder induced roughening dominates over thermal
roughening ($\zeta >1/2$) we have $T\rightarrow 0$ under
renormalization and the long wavelength physics is governed by a
zero-temperature fixed point \cite{halpin}. In a functional
renormalization group (FRG) analysis one splits the degrees of
freedom (here $\bf x_1$ and $\bf x_2$) into their long and short
wavelength parts:
\begin{equation}
{\bf x_1}_{\alpha}={\bf x_1}_{\alpha}^<+{\bf x_1}_{\alpha}^>,\, {\bf
x_2}_{\beta}={\bf x_2}_{\beta}^<+{\bf x_2}_{\beta}^>. \label{split}
\end{equation}
We write the degrees of freedoms $\bf x_1$ and $\bf x_2$ as in
(\ref{split}) and  consider the part of $H$ quadratic in the short
wavelength parts of the degrees of freedom ($H^>$). We expand the
disorder potential terms containing $\bf x_1^>,\;x_2^>$ in the
exponential of (\ref{reppart}) up to the second order \cite{halpin},
average over the $\bf x_1^>,\;x_2^>$, and neglect terms containing
three-replica indices due to their irrelevance \cite{halpin} to
obtain
\begin{equation}
H^>=T_1 + T_2 + T_3,\label{hamilgr}
\end{equation}
where
\begin{eqnarray}
T_1&\equiv&\frac{1}{4T^2}\int
\frac{dq}{q^4}[\Sigma_{\alpha,\mu}R_1^{\prime\prime}({\bf
x_1}_{\alpha}^<- {\bf x_1}_{\mu}^<)^2
-2\Sigma_{\alpha,\mu}R_1^{\prime\prime} (0) R_1^{\prime\prime}({\bf
x_1}_{\alpha}^<- {\bf x_1}_{\mu}^<)\nonumber \\&+&
\Sigma_{\alpha,\mu} \frac{n-1}{2}\left[\frac{R_1^{\prime}({\bf
x_{1\alpha}^< - x_{1\mu}}^<)}{ |\bf x_{1\alpha}^< -
x_{1\mu}^<|}\right]^2  - \Sigma_{\alpha\mu}(n-1) \frac{R_1({\bf
x_{1\alpha}^< - x_{1\mu}^<})}{|\bf
x_{1\alpha}^<-x_{1\mu}^<|}R_1^{\prime\prime}(0)],\nonumber \\
T_2&\equiv& \frac{1}{4T^2}\int
\frac{dq}{q^4}[\Sigma_{\alpha,\mu}R_2^{\prime\prime}({\bf
x_2}_{\alpha}^<- {\bf x_2}_{\mu}^<)^2
-2\Sigma_{\alpha,\mu}R_2^{\prime\prime} (0) R_2^{\prime\prime}({\bf
x_2}_{\alpha}^<- {\bf x_2}_{\mu}^<)\nonumber \\&+&
\Sigma_{\alpha,\mu} \frac{n-1}{2}\left[\frac{R_2^{\prime}({\bf
x_{2\alpha}^< - x_{1\mu}}^<)}{ |\bf x_{2\alpha}^< -
x_{2\mu}^<|}\right]^2  - \Sigma_{\alpha\mu}(n-1) \frac{R_2({\bf
x_{2\alpha}^< - x_{2\mu}^<})}{|\bf
x_{2\alpha}^<-x_{1\mu}^<|}R_2^{\prime\prime}(0)],\nonumber \\
T_3&\equiv&\frac{1}{4T^2}\int
\frac{dq}{q^4}[\Sigma_{\alpha,\mu}\hat{R}^{\prime\prime}({\bf
x_1}_{\alpha}^<- {\bf x_2}_{\mu}^<)^2
-2\Sigma_{\alpha,\mu}\hat{R}^{\prime\prime}
(0)R_2^{\prime\prime}({\bf x_1}_{\alpha}^<- {\bf x_2}_{\mu}^<)
\nonumber \\ &+&\Sigma_{\alpha,\mu}R_{\times}({\bf x_1}_{\alpha}^<-
{\bf x_2}_{\mu}^<)^2 + \Sigma_{\alpha,\mu}
\frac{n-1}{2}\left[\frac{\hat R^{\prime}({\bf x_{1\alpha}^< -
x_{2\mu}^<})}{ |\bf x_{1\alpha} - x_{2\mu}|}\right]^2\nonumber
\\ &-&\Sigma_{\alpha\mu}(n-1) \frac{\hat R({\bf x_{1\alpha}^< -
x_{2\mu}^<})}{|\bf x_{1\alpha}^<-x_{2\mu}^<|}\hat
R^{\prime\prime}(0) + \Sigma_{\alpha,\mu} \frac{n-1}{2}\left[\frac{
R_{\times}^{\prime}({\bf x_{1\alpha}^< - x_{2\mu}^<})}{ |\bf
x_{1\alpha} - x_{2\mu}|}\right]^2]
\end{eqnarray}
with $q$ being the Fourier conjugate variable of $t$.  Note that
arguments of $T_1,\,T_2$ and $T_3$ are, pure ${\bf x_{1\alpha}
-x_{1\mu}}$, $\bf x_{2\alpha} - x_{2\mu}$ and mixed $\bf
x_{1\alpha}- x_{2\mu}$ respectively. Comparing with the existing
(bare) terms in $H^<$ we find that $T_1$, $T_2$ and $T_3$,
renormalize, respectively $R_1^<({\bf x_{1\alpha} - x_{1\mu}}),\,
R_2^<({\bf x_{2\alpha} -x_{2\mu}})$ and $\hat R^< ({\bf x_{1\alpha}
- x_{2\mu}})$. Since the all of $T_1,\,T_2$ and $T_3$ are even under
inversion of their arguments, there are no corrections to
$R_\times$. Corrections $T_1$ and $T_2$, to $R_1$ and $R_2$
respectively, have identical forms and are same with the
corresponding corrections in the single DP case \cite{halpin}. This
is expected, since before disorder averaging, the free energies of
each of the DPs, like the free energy of the single DP problem,
follow the usual KPZ equation. In the expansion of the disorder
correlation terms in the exponential of (\ref{reppart}) the terms in
the first order of the expansion do not contribute. This is because
the above expansion is essentially perturbative in $T$ as in the
single DP case \cite{halpin}: $T$ flows to zero under
renormalization [see eq.~(\ref{Tnor})]. Hence the first order terms
having an uncompensated power of $T$ flow to zero and are irrelevant
(in an RG sense). This feature is same as in the single DP case
\cite{halpin}. Note that we have made use of the fact that
$R_{\times}(0)=0$ while arriving at the expression (\ref{hamilgr}).
Different terms in (\ref{hamilgr}) contributes to the fluctuation
corrections to $R_1^<, \,R_2^<,\,\hat R^<$ which can be identified
by their arguments. Note that there are no corrections to
$R_{\times}^<$ which is reminiscent of the lack of fluctuation
corrections to the noise cross-correlations in the Eqs.
(\ref{eq:burgers_1}) and (\ref{eq:burgers_2}) in the long wavelength
limit.

In the next step, we argue that  {\em all} of the functions
$R_1,\,R_2,\,\hat R,\,R_{\times}$ are characterized by the same
scaling behavior in the long wavelength limit. This is because all
of them are proportional to various noise variances in the model
given by Equations (\ref{noise1}), (\ref{noise2}) and
(\ref{noise3}). Now all the correlation functions in the model Eqs.
(\ref{eq:burgers_1}) and (\ref{eq:burgers_2}), in stochastic
Langevin descriptions, are proportional to the noise variances
(\ref{noise1}), (\ref{noise2}) and (\ref{noise3}). Further, these
correlation functions, due to the symmetries of the GBM model, have
the same scaling behavior in the hydrodynamic limit, characterized
by a single roughness ($\chi$) and dynamic ($z$) exponents. Hence,
the effective noise variances, and therefore, the functions
$R_1,\,R_2,\,\tilde R$ and $\hat R$ must have the same scaling
behavior in the long wavelength limit. With the rescaling of $t$ and
$\bf x$ mentioned above and the $R$-functions scale as
\begin{equation}
R\rightarrow [1+ (3-4\zeta)\delta l]R,
\end{equation}
where $R$ in the above stands for all of $R_1,\,R_2,\,\hat R,\,\tilde
R$. These then yield the following differential flow equations:
\begin{eqnarray}
\frac{\partial R_1}{\partial l}&=&(3 -4\zeta)R_1 +\zeta x R_1
^{\prime}
+\frac{1}{2}(R_1^{\prime\prime})^2-R_1^{\prime\prime}(0)R_1^{\prime\prime}
+ \frac{n-1}{2}\left[\frac{R_1'}{x}\right]^2 - \nonumber \\ &&(n-1)
R_1''(0)\frac{R_1''(x)}{x},
\nonumber \\
\frac{\partial R_2}{\partial l}&=&(3 -4\zeta)R_2 +\zeta x R_2
^{\prime}
+\frac{1}{2}(R_2^{\prime\prime})^2-R_2^{\prime\prime}(0)R_2^{\prime\prime}
+ \frac{n-1}{2}\left[\frac{R_2'}{x}\right]^2 - \nonumber \\ &&(n-1)
R_2''(0)\frac{R_2''(x)}{x},
\nonumber \\
\frac{\partial \hat R}{\partial l}&=&(3 -4\zeta)\hat R +\zeta x \hat
R ^{\prime}  +\frac{1}{2}(\hat R^{\prime\prime})^2- \hat
R^{\prime\prime}(0)\hat R^{\prime\prime}+\frac{1}{2}(
R_{\times}^{\prime\prime})^2 \nonumber \\&+&
\frac{n-1}{2}\left[\frac{\hat R'}{x}\right]^2 - (n-1) \hat
R''(0)\frac{\hat R''(x)}{x} + \frac{n-1}{2}\left[\frac{\tilde
R}{x}\right]^2,
\nonumber \\
\frac{\partial R_{\times}}{\partial l}&=&(3 -4\zeta) R_{\times} + \zeta x
 R_{\times} ^{\prime}.
\label{flowfrg}
\end{eqnarray}
In the Eqs. (\ref{flowfrg}) above "$^{\prime}$" denotes a derivative
with respect to $x$, the argument of the functions $R_1$ etc. Note
that the functional flow equations for the functions $R_1$ and $R_2$
are identical to each other which is expected on the ground of
symmetry between the equations of motion (\ref{2kpzalld}) or
(\ref{colehopfeq}). At the RG fixed point all the partial
derivatives with respect to the scale factor $l$ is zero yielding
\begin{eqnarray}
(3 -4\zeta)R_1 &+&\zeta x R_1 ^{\prime}
+\frac{1}{2}(R_1^{\prime\prime})^2-R_1^{\prime\prime}(0)R_1^{\prime\prime}
+ \frac{n-1}{2}\left[\frac{R_1'}{x}\right]^2  \nonumber \\ &-&(n-1)
R_1''(0)\frac{R_1''(x)}{x}=0,\nonumber \\
(3 -4\zeta)R_2 &+&\zeta x R_2 ^{\prime}
+\frac{1}{2}(R_2^{\prime\prime})^2-R_2^{\prime\prime}(0)R_2^{\prime\prime}
+ \frac{n-1}{2}\left[\frac{R_2'}{x}\right]^2  \nonumber \\ &-&(n-1)
R_2''(0)\frac{R_2''(x)}{x}=0,\nonumber \\
(3 -4\zeta)\hat R &+&\zeta x \hat R ^{\prime}  +\frac{1}{2}(\hat
R^{\prime\prime})^2- \hat R^{\prime\prime}(0)\hat
R^{\prime\prime}+\frac{1}{2}( R_{\times}^{\prime\prime})^2 \nonumber
\\&+& \frac{n-1}{2}\left[\frac{\hat R'}{x}\right]^2 - (n-1) \hat
R''(0)\frac{\hat R''(x)}{x} + \frac{n-1}{2}\left[\frac{\tilde
R}{x}\right]^2=0,\nonumber \\
(3 -4\zeta) R_{\times} &+&\zeta x R_{\times} ^{\prime}=0.
\label{fixedfrg}
\end{eqnarray}
In order to proceed further, we make the following choice without
any loss of generality: $\hat R(x) =\gamma R_1(x)=\gamma R_2(x)$ in
the long wavelength limit, where $\gamma$ is a numerical constant.
We further choose $R_{\times}(x)^2 = \Gamma_0 R_1(x)^2=\Gamma_0
R_2(x)^2$ in the long wavelength limit, where $\Gamma_0$ ia another
numerical constant. These parametrizations are consequences of the
symmetries of the GBM model, which ensure, as we have argued above,
functions $R_1,\,R_2,\,\tilde R$ and $\hat R$ are proportional in
the thermodynamic limit. We determine below a relation between
$\gamma$ and $\Gamma_0$.In terms of the parameters $\gamma$ and
$\Gamma_0$, then, the first and the third in the Eqs.
(\ref{flowfrg})  at the fixed point reduce to
\begin{eqnarray}
(3 -4\zeta)R_1+R_1+\zeta xR_1^{\prime}+&&
\frac{1}{2}(R_1^{\prime\prime})
^2-R_1^{\prime\prime}=0,\nonumber \\
(3 -4\zeta)R_1+R_1+\zeta
xR_1^{\prime}+&&\left[\frac{\gamma^2+\Gamma_0}{2\gamma}\right]\left(
(R_1^{\prime\prime})^2-R_1^{\prime\prime}+ \frac{n-1}{2} [\frac{
R_1'}{x}]^2\right) \nonumber \\ &-& (n-1) \frac{R_1}{x}=0.
\label{frgfinal}
\end{eqnarray}
Since  the equations in (\ref{frgfinal}) are identically same, for
consistency we must have
\begin{equation}
\gamma^2 +\Gamma_0=\gamma \Rightarrow \gamma=\frac{1\pm \sqrt
{1-4\Gamma_0}}{2} \simeq \{1-\Gamma_0, \Gamma_0\}, \label{solfrg}
\end{equation}
to the lowest order in $\Gamma_0$. In order to find out the
physically relevant solution from the above two solutions in
(\ref{solfrg}) we argue in the following way: The functions
$R_1,\,R_2$ etc are proportional to respective noise correlators
(\ref{noisealld}) in the coupled-KPZ equations (\ref{2kpzalld}).
Further, in terms of the original field variables $\bf u$ and $\bf
b$ or $h$ and $\phi$, if there is no cross-correlations, i.e., for
$\Gamma_0=0$ the amplitude-ratio of the autocorrelation functions of
$\bf u$ and $\bf b$ (or $h$ and $\phi$), $A$ is unity. Since
$A=(1-\gamma)/(1+\gamma)$ we then have $\gamma=0$ when $A=1$, i.e.,
in the absence of any cross-correlations. Thus we pick up that
relation between $\gamma$ and $\Gamma_0$ which goes to zero in the
limit $\Gamma_0$ goes to zero. Thus we write,
\begin{eqnarray}
\gamma=\Gamma_0 \Rightarrow
A=\frac{1-\Gamma_0}{1+\Gamma_0}=1-2\Gamma_0, \label{frgres}
\end{eqnarray}
to the lowest order in $O(\Gamma_0)$. Further, from its definition,
$\Gamma_0=(R_\times/R_1)^2 = N_s$ in the lowest order. Hence, we
obtain $A=1-2N_s$. Therefore, the relation (\ref{frgres}) agrees
with what we find before from our DRG or SCMC calculations.

The scaling exponents in the present coupled chain problem is
identical to the single-DP problem; this is due to the identical
nature of the functional flow equations of $R_1$ with the
corresponding single DP problem. Therefore, we obtain
$\zeta=6/(8+d)=1/z$ (see also \cite{halpin}) as obtained in our SCMC
calculations before.

As before in the DRG analyses of the problem, to complete our
analysis here, it is required to demonstrate that the parameter
$\Gamma_0$ is marginal in the scaling limit and can have arbitrary
values. We begin by considering the flow equation for
$R_{\times}(x)$ at the RG fixed point. Noting that due to the odd
parity nature $R_{\times} ({\bf x})$ does not receive any
fluctuation corrections we write
\begin{equation}
\frac{\partial R_{\times}}{\partial x}=-\frac{1}{\zeta x}(3 -4\zeta)
R_{\times}.
\end{equation}
This yields, near the fixed point,
\begin{equation}
R_{\times}(x)=\frac{C_0}{x^{3-4\zeta/\zeta}}. \label{rcrossform}
\end{equation}
Here, $C_0$ is a constant of integration which is the value of
$R_\times ({\bf x})$ at small scale. The function $R_\times ({\bf
x})$ is odd under $\bf x\rightarrow -x$ and is non-analytic at $\bf
x=0$. As a result, within perturbative calculations, it does not
receive any fluctuation corrections in the long wavelength limit.
Hence even in that limit the value of $R_\times (x)$ depends upon
$C_0$, its value at the small scale. In contrast, the values of
$R_1(x)$ and $R_2(x)$ in the hydrodynamic limit are independent of
their values at small scales, since fluctuation corrections dominate
over their bare values at large spatial scales. Therefore, the ratio
$\Gamma_0=[R_\times (x)/R_1(x)]^2$ at the large spatial scales,
i.e., in the scaling regime, depends on $C_0$. The constant $C_0$
has no fixed magnitude; it depends upon realizations of the disorder
at small scales and hence can have arbitrary values. Therefore, the
ratio $[R_\times (x)/R_1(x)]^2$ also can have arbitrary values in
the hydrodynamic limit. This completes our FRG analysis. Our FRG
approach to the problem, therefore, yields scaling exponents
$\chi=-\frac{1}{3}-\frac{d}{6};\;z=\frac{4}{3}+\frac{d}{6}$ for
$d=1$ and in the strong coupling phase at $d>2$. It further yields
the amplitude-ratio $D_b/D_u=1-2N_s$ for $d=1$ and in the strong
coupling phase at $d>2$. These results are in agreement with those
from DRG and SCMC approaches above.


\section{Numerical analysis: direct approaches and lattice models}
\label{numrec}

\subsection{Direct Numerical Solutions (DNS) of the model equations}
\label{dns} Having obtained several new results by the applications
of three different analytical perturbative approaches on our model
we now resort to numerical methods to supplement our understanding
of the underlying physics from the above analytical approaches. In
particular, we numerically solve (hereafter referred to as DNS) the
model Eqs. (\ref{eq:burgers_1}) and (\ref{eq:burgers_2}) in one and
two dimensions by using pseudo-spectral methods with the
Adams-Bashforth time evolution scheme [see Appendix (\ref{pseudo})].
Here, we consider only symmetric noise cross-correlations. We
elucidate the scaling properties of the following equal-time
correlation functions of $h ({\bf k},t)$ and $\phi ({\bf k},t)$:
$C_{hh}({ k},0)\equiv \langle h({\bf k},t) h({\bf -k},t)\rangle,\;
C_{\phi\phi}({ k},0)\equiv \langle \phi({\bf k},t)\phi ({\bf -k},t)
\rangle$ in $d=1,\;2$. Since, as discussed before, the scaling
exponents of the fields $h$ and $\phi$ are identical to each other,
the ratio of the equal-time correlation functions
$A=C_{\phi\phi}({\bf k},0)/C_{hh}({\bf k},0)$ is a dimensionless
number which is nothing but the amplitude ratio defined above. We
examine the dependence of $A$ on the parameter $N_s$. We
further consider the time dependence of the widths $W_{h}(t)$ and
$W_{\phi} (t)$ as defined above. In the statistical steady
$W_h(t)^2$ and $W_\phi(t)^2$ approach the equal time steady state
correlation functions $C_{hh}({\bf x=0},0)$ and $C_{\phi\phi}({\bf
x=0},0)$. Therefore, $W_h(t)/W_\phi (t)=\sqrt A$ in the large time
limit (i.e., in the statistical steady state).

Before presenting our numerical results below we discuss a technical
matter, namely, the measurement of the parameter $N_s$. In our
analytical work above, the parameter $N_s\equiv (D_s/D_u)^2$
involves the ratios of the amplitudes of the cross-correlation
function and the velocity auto-correlation function. Therefore,
corresponding numerical works require measurements of the
cross-correlation function as well, in addition to measuring the
auto-correlation functions of $\bf u$ and $\bf b$. It, however, is
much more difficult to obtain data with sufficient statistics for
the cross-correlation function amplitude since it is {\em not
positive definite}. In view of this difficulty, instead of measuring
$N_s$ we use its {\em bare value} as obtained from the amplitude of
the noise cross-correlations in most of our analyses below. We
denote this by $N_s^0$. We would like to mention that the comparison
of our numerical data with the already obtained analytical results
will be largely qualitative, due to the reasons mentioned above. For
our DNS studies the noises $\theta_1$ and $\theta_2$ in
Eqs.~(\ref{eq2kpz1}) and (\ref{eq2kpz2}) are chosen to have
correlations of the form
\begin{eqnarray}
\langle \theta_1 ({\bf k},t)\theta_1 (-{\bf k},0)\rangle &=& \langle
\theta_2 ({\bf k},t)\theta_2 ({\bf -k},0)\rangle = 2D^0 \delta
(t),\nonumber \\
\langle \theta_1 ({\bf k},t)\theta_2(-{\bf k},0)\rangle &=& 2i D_s
({\bf k})\delta (t), \label{noisesimu}
\end{eqnarray}
such that $D_s({\bf k})=-D_s ({\bf -k})$ and $D_s ({\bf k})=D_s^0$
for ${\bf k}>0$. The resultant noise correlation matrix has
eigenvalues $D^0\pm D_s^0$. The fact that the noise correlation
matrix should be positive semi-definite ensures that the upper limit
of $N_s^0\equiv (D_s^0/D^0)^2$ is unity. Although $N_s$ depends
monotonically on $N_s^0$, due to the highly nonlinear nature of the
equations of motion the dependence is not linear. We are able to
measure $N_s$ only in $1d$DNS below. Those measurements indeed show
the monotonic dependences of $N_s$ on $N_s^0$. We present our
results in details below.

\subsubsection{Results in one dimension:}
\label{1ddns}

In this section we present our numerical results from the DNS of the
continuum model Eqs. (\ref{eq2kpz1}) and (\ref{eq2kpz2}) together
with the noise variances (\ref{noisesimu}) in $1d$. We have already
found, from our analytical studies above, that the model Eqs.
(\ref{eq:burgers_1}) and (\ref{eq:burgers_2}) together with the
noise variances (\ref{noise1}), (\ref{noise2}) and (\ref{noise3}) in
$1d$ yields scaling exponents $\chi=-1/2, i.e., \chi_h=1/2$ and
$z=3/2$. In addition, the amplitude-ratio $D_b/D_u$ decreases
monotonically with $N_s$. Our results here confirm our analytical
results as we describe below.

We perform pseudo-spectral simulations of the model Equations in
$1d$ to examine the scaling behavior of the equal time correlation
functions $C_{hh}(k,0),\, C_{\phi\phi}(k,0)$, where $k$ is a Fourier
wavevector, in the statistical steady states. The system sizes $L$
chosen are $L=4096$, $L=2048$ and $L=6144$ where $L$ is the number
of points in a one-dimensional lattice in real space. We present a
log-log plot of the correlation functions versus $k$ in Fig.
\ref{fig6144} for $L=6144$ (left) and $L=4096$ (right); results from
our runs with $L=2048$ have similar behavior. In all the plots the
red point corresponds to the correlation function $C_{hh}$ and the
green points correspond to the correlation function $C_{\phi\phi}$.
The blue line with slope of -2 provides a guide to the eye for
scaling regime in the plots with slope $\approx -2$ which
corresponds to the roughness exponents $ \chi_h$ for the fields $h$
and $\phi$ being 1/2. These values are exact results in the absence
of cross-correlations and obtained in our one-loop DRG and SCMC
above ($1d$) for finite cross-correlations. Our numerical results
clearly yield a value of the roughness exponent $\chi_h$ which is
very close to the analytically calculated value. For our $1d$DNS
studies we estimate the parameter $N_s$ defined above by calculating
the equal-time cross-correlation function for Fourier modes $k$ in
the scaling regimes and taking its ratio with $C_{hh}(k,0)$. In Fig.
(\ref{fig6144}), for a given system size $L$, the amplitude
differences between the scaling regimes of the correlators $C_{hh}$
and $C_{\phi\phi}$, which is same as the parameter $A$ in Section
\ref{rg1d}, increases monotonically with $N_s^0$ (or with $N_s$), a
feature which is in qualitative agreements with our analytical
results above.
\begin{figure}
\includegraphics[width=8cm]{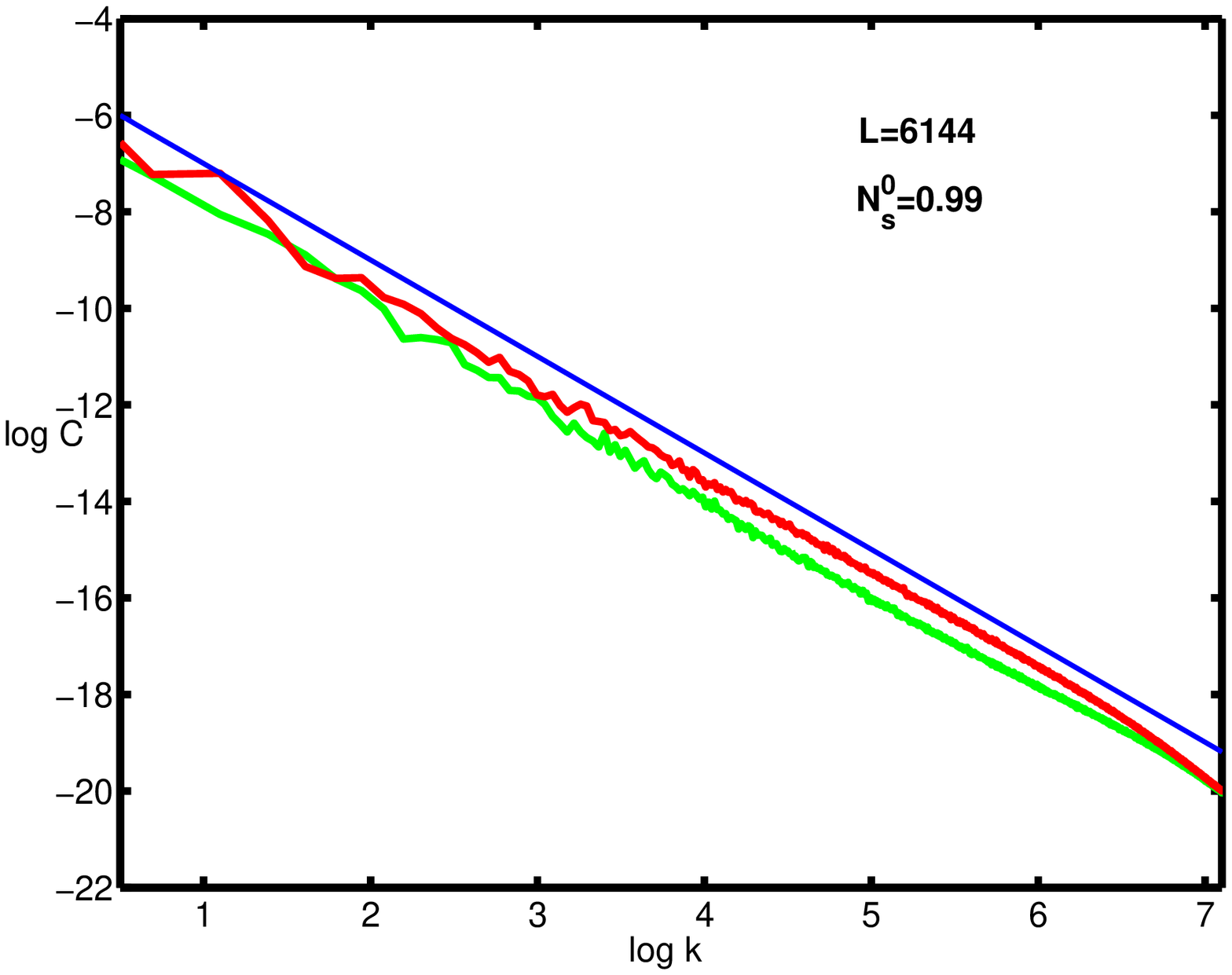}\hfill
\includegraphics[width=8cm]{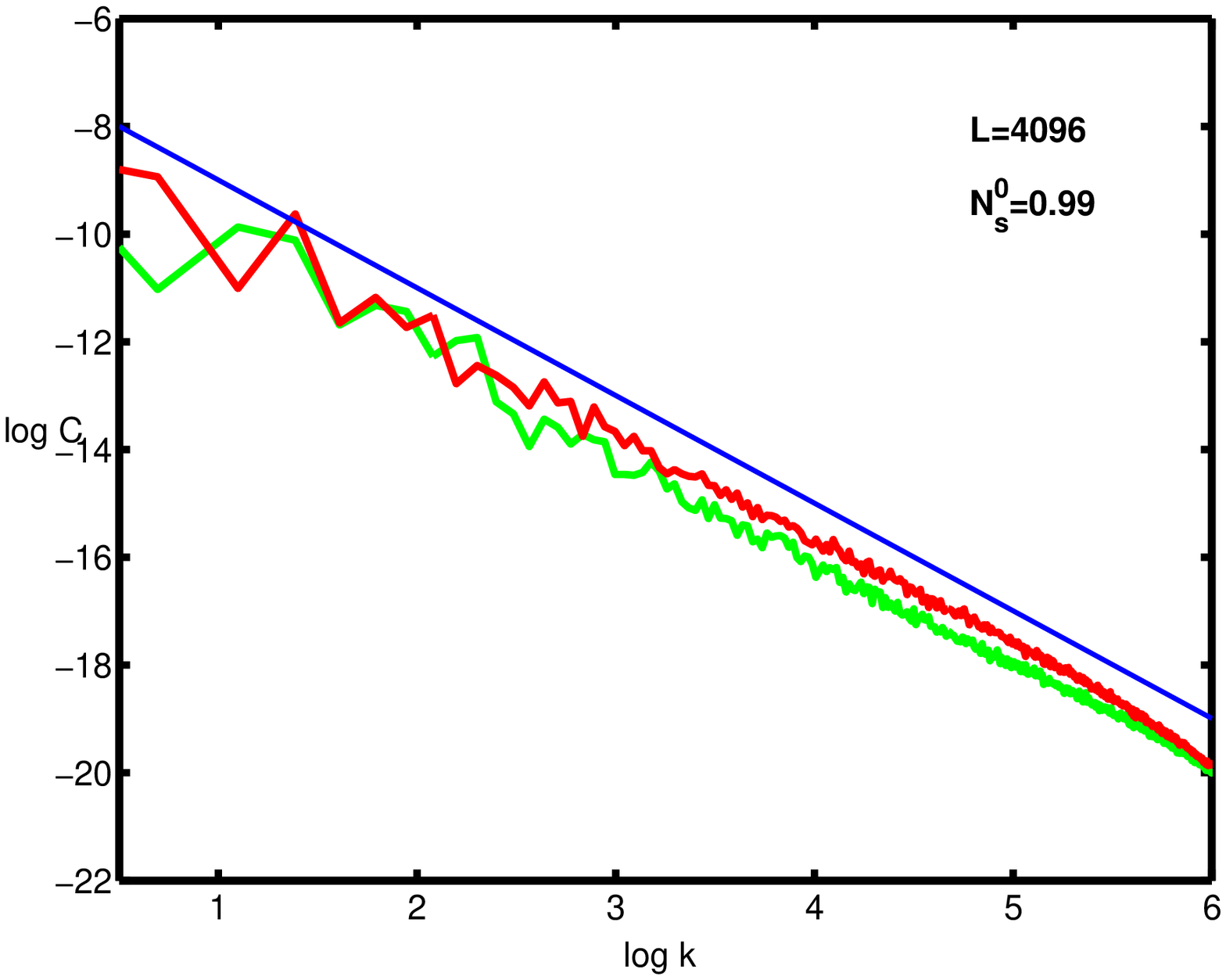}\\
\includegraphics[width=8cm]{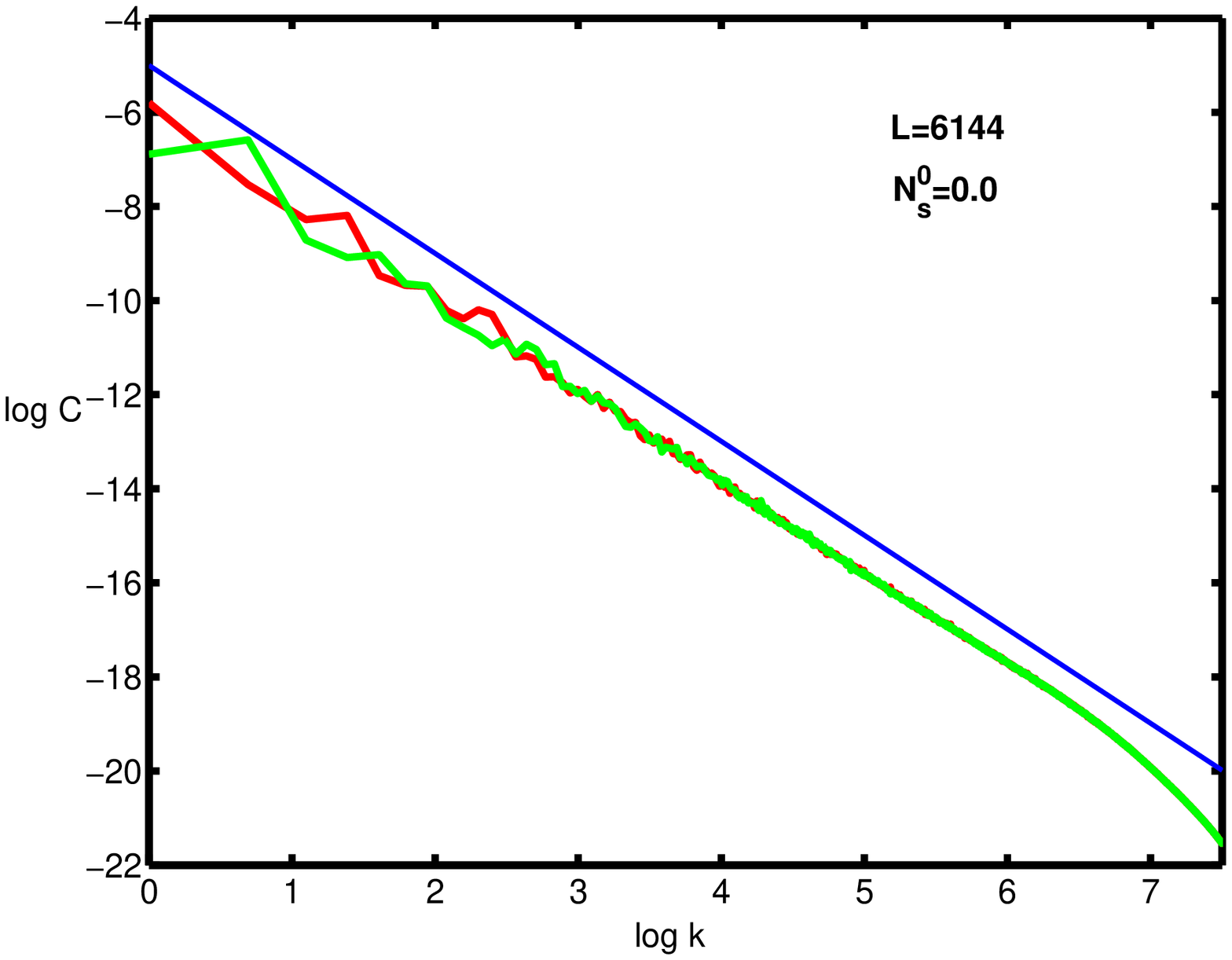}\hfill
\includegraphics[width=8cm]{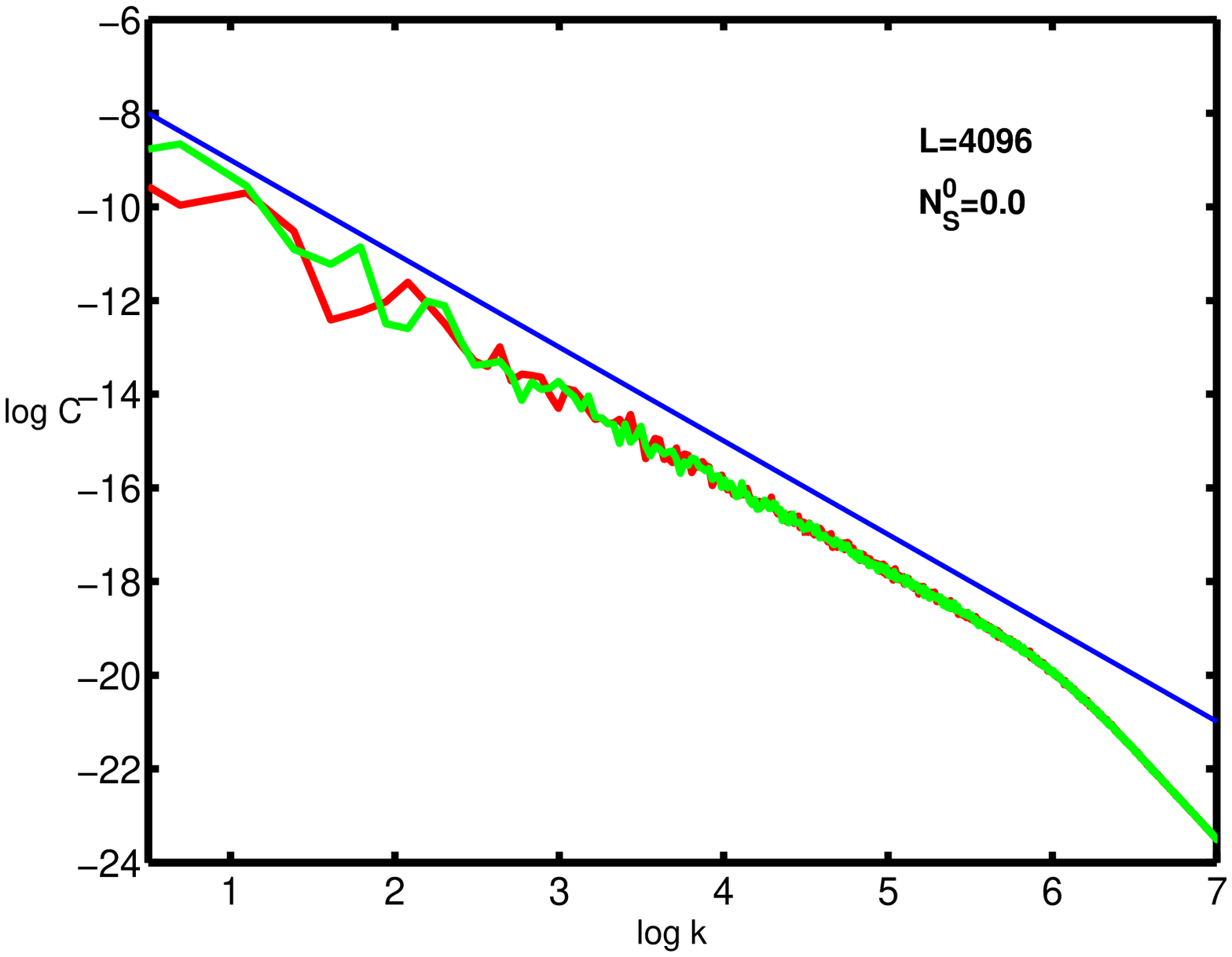}
\caption{ Log-log plots of $C_{hh}(k)$ (red) and $C_{\phi\phi}(k)$
(green) versus $k$ for $L=6144$ with $N_s^0$=0.9 ($N_s=0.82$) (top
left) and $N_s=N_s^0=0.0$ (bottom left), and for $L=4096$ with
$N_s^0$=0.9 ($N_s=0.81$) (top right) and $N_s=N_s^0=0.0$ (bottom
right). The blue line, drawn as a guide to the eye, indicates a
slope of -2, corresponding to the value $\chi_h=1/2$ for the
roughness exponent for the fields $h$ and $\phi$. Note that at
finite $N_s$ there are amplitude differences between the scaling
regimes of $C_{hh}(k)$ and $C_{\phi\phi}(k)$ (top plots) where as
for $N_s^0=0.0$ the amplitude differences disappear (bottom plots)
(see text).} \label{fig6144}
\end{figure}

We now show the time-dependence of the widths $W_{h}(t)$ and
$W_{\phi}(t)$ in Fig. (\ref{t4096}) for system size $L=4096$. We
notice that the saturated amplitude difference between $W_h(t)$ and
$W_{\phi}(t)$ increases as $N_s^0$ increases from 0.0 to 0.9. Since
the ratio of the saturated amplitudes of $W_h(t)$ and $W_{\phi}(t)$
yields the ratio $A$ ($W_\phi/W_h=\sqrt A$), we find that $A$
decreases as $N_s^0$ increases. This is in accordance with the
results as presented in Fig. (\ref{fig6144}).
\begin{figure}[h]
\includegraphics[width=3.0in]{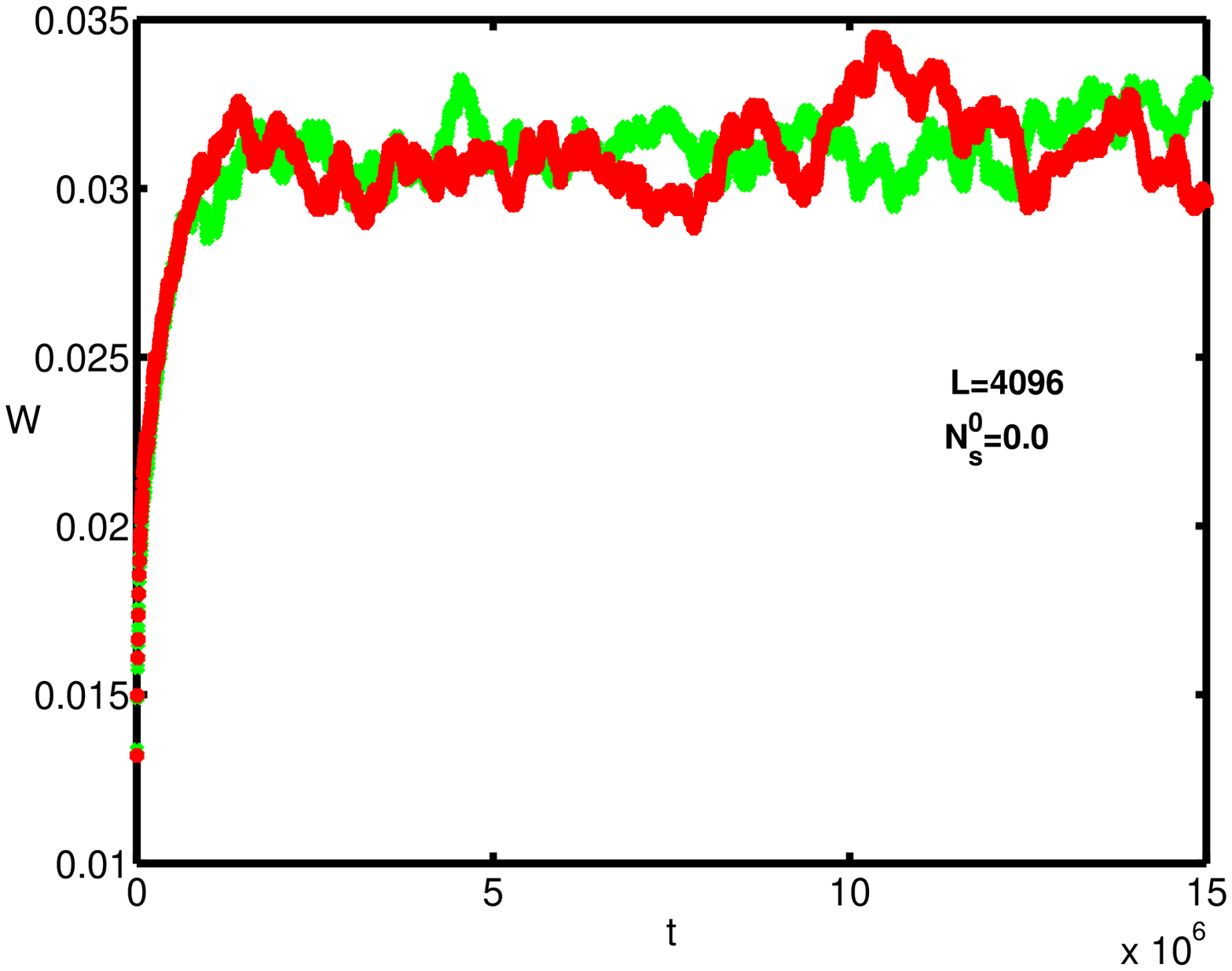}\hfill
\includegraphics[width=3.0in]{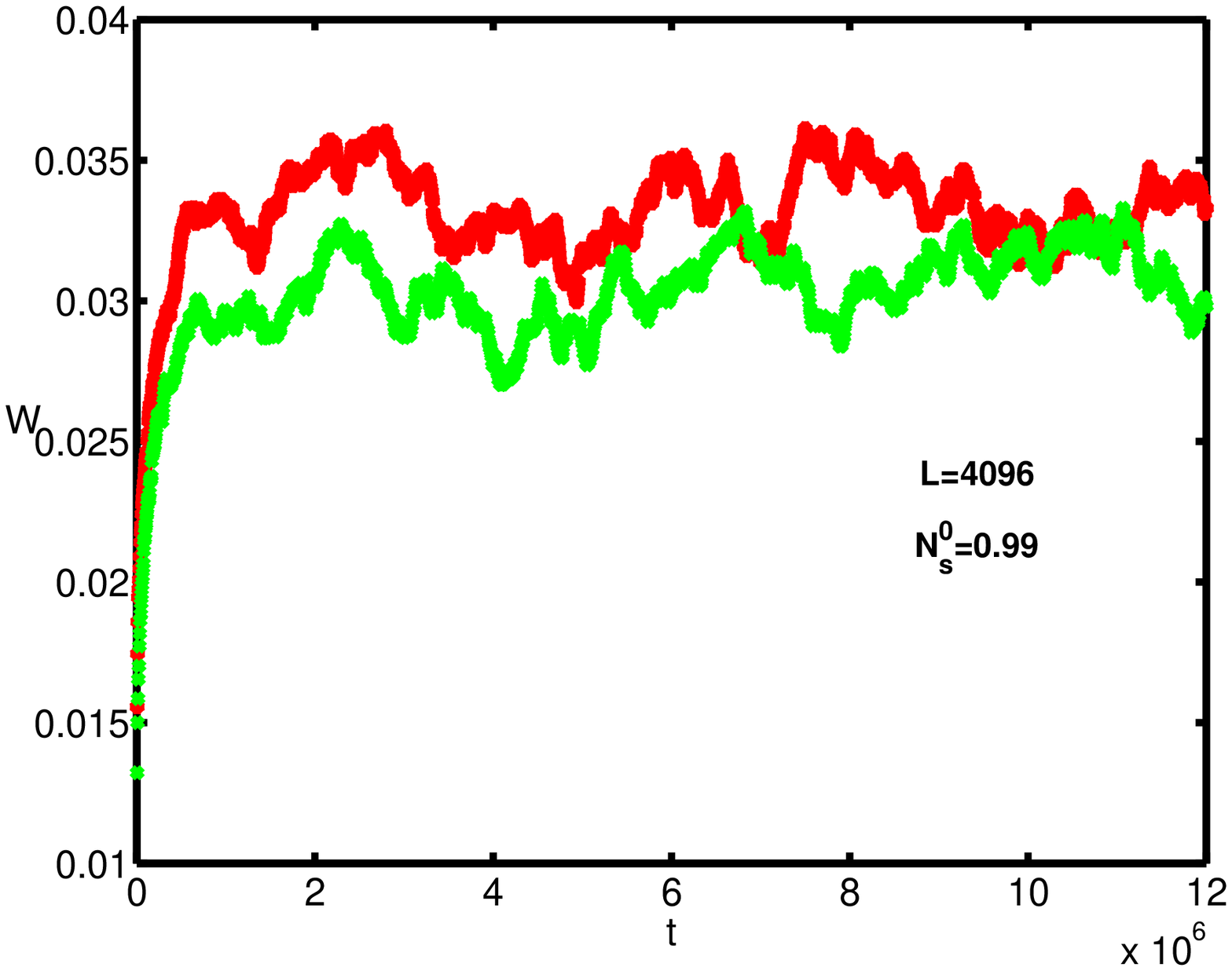}
\caption{A plot of the widths $W_h(t)$ (red) and $W_\phi(t)$ (green)
versus time $t$ for system size $L=4096$ with $N_{s}^0=0.0,\,0.9$
from our $1d$ DNS studies. Clearly the saturated amplitude
difference increases as $N_s^0$ increases.} \label{t4096}
\end{figure}
Figure~(\ref{1dns}) shows the dependence of the parameter $A$
obtained from the plots above on $N_s^0$ for system sizes $L=2048,
4096, 6144$. We find that the amplitude-ratio $A$ decreases
monotonically as $N_{s}^0$ increases in agreement with the behavior
of our mode coupling analyses. We do not observe any noticeable
systematic dependence of this behavior on system sizes. Therefore,
results from our $1d$DNS studies agree with those from the
analytical studies mentioned above.

\begin{figure}[h]
\includegraphics[width=5in]{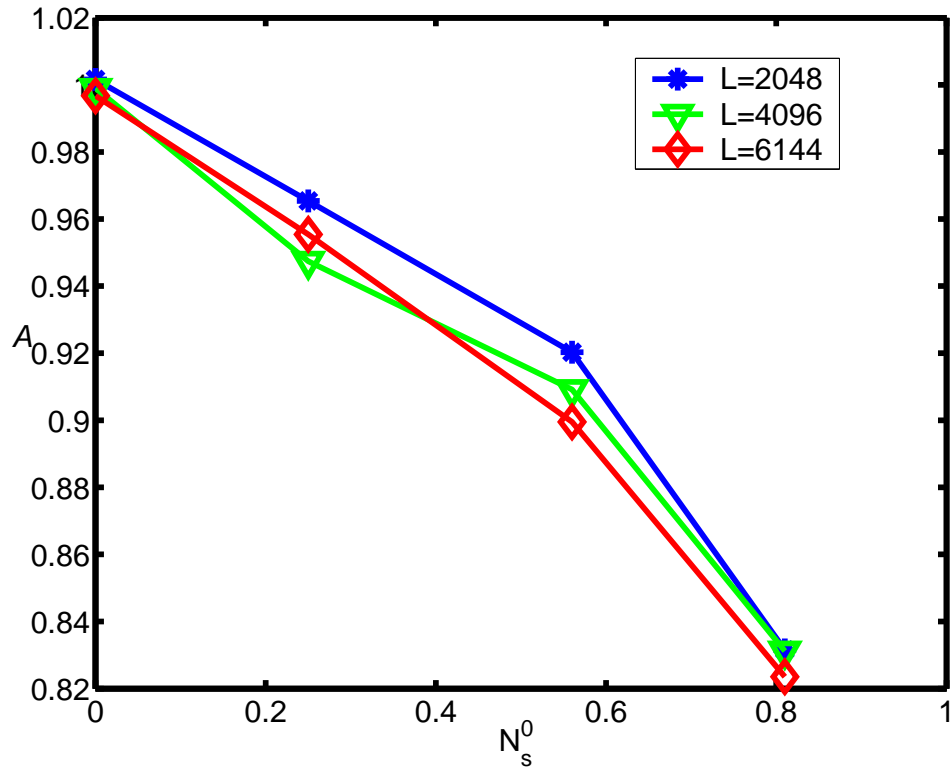}
\caption{A plot of the parameter $A$ versus $N_{s}^0$ for our $1d$
DNS studies for system sizes $L=2048, 4096, 6144$. The general trend
that $A$ decreases with an increase in $N_{s}^0$ is observed for the
all system sizes we worked with.} \label{1dns}
\end{figure}

\subsubsection{Results in two dimensions:}
\label{2ddns}

In this Section we present our results from our $2d$ DNS of the
model Eqs. (\ref{eq2kpz1}) and (\ref{eq2kpz2}) and compare with the
analytical results already obtained. In our DNS studies (using
pseudo-spectral methods) in $2d$ the system sizes we work with are
$96^2$, $128^2$ and $160^2$. Unlike in $1d$, the system exhibits a
{\em non-equilibrium phase transition} from a smooth phase,
characterized by {\em logarithmic roughness} ($\chi_h=0$) and
$\alpha$ independent of $N_s^0$, to a rough phase characterized by
{\em algebraic roughness} ($\chi_h\ge 0$) and a decreasing $A$ as
$N_{s}^0$ increases. In the smooth phase, $A$ is fully determined by
the bare ratio $D_b^0/D_u^0$. Therefore, a simple way of
ascertaining which phase the system is in, is by measuring $A$.

In our $2d$DNS studies there two tuning parameters to reach the
rough phase of the system. They are $D_u^0$ and $N_s^0$. The
crossover from the smooth-to-rough phases is formally determined by
the value of the   {\em coupling constant} $G= \lambda D_u^2/\nu^3$
(see above). Here the renormalized parameter $D_u$ has a monotonic
dependence on the bare amplitude $D_u^0$. For small $D_u^0$,
renomalized $G$ is zero and the system is in its smooth phase. With
increasing $D_u^0$ the system eventually crosses over to the rough
phase where various values of $\alpha$ have been obtained by tuning
$N_s^0$. For each system size we obtain the widths $W_h(t)$ and
$W_{\phi}(t)$ as a function of time $t$.  As for the height field in
the well-known KPZ equation plots of $W_h(t)$, as a function of
time, have a growing part and a saturated part. The system size
dependences of the saturation values of the widths $W_h(t)$ and
$W_\phi (t)$ yield the values of the roughness exponents of the
corresponding fields.  Since we are interested in the statistical
properties of the rough phase, in our $2d$DNS studies we access this
phase by sufficiently large $D_{u}^0$ for reasons as explained
above. As shown below, we find $A <1$ for non-zero $N_{s}^0$. This
ensures that we are indeed able to access the rough phase in our DNS
runs. We present our results from system sizes $96^2$ and $128^2$ in
Fig. (\ref{2d96}) for two values of the parameter $N_s^0=0.0,\,0.9$
for each system size. The data from $160^2$ runs show similar
behavior. We determine $A$ and the ratio $\chi_h/z$. Note that for
each system size the amplitude differences differences between the
saturation values of the widths $W_h(t)$ and $W_{\phi}(t)$ (i.e.,
the parameter $A$ in Sections \ref{rg1d} and \ref{modecoupling})
increase monotonically with $N_{s}^0$, in agreement with  our SCMC
and FRG analyses. Such nontrivial dependence of $A$ on $N_{s}^0$ is
a key signature of the rough phase.
\begin{figure}[h]
\includegraphics[width=3.5in]{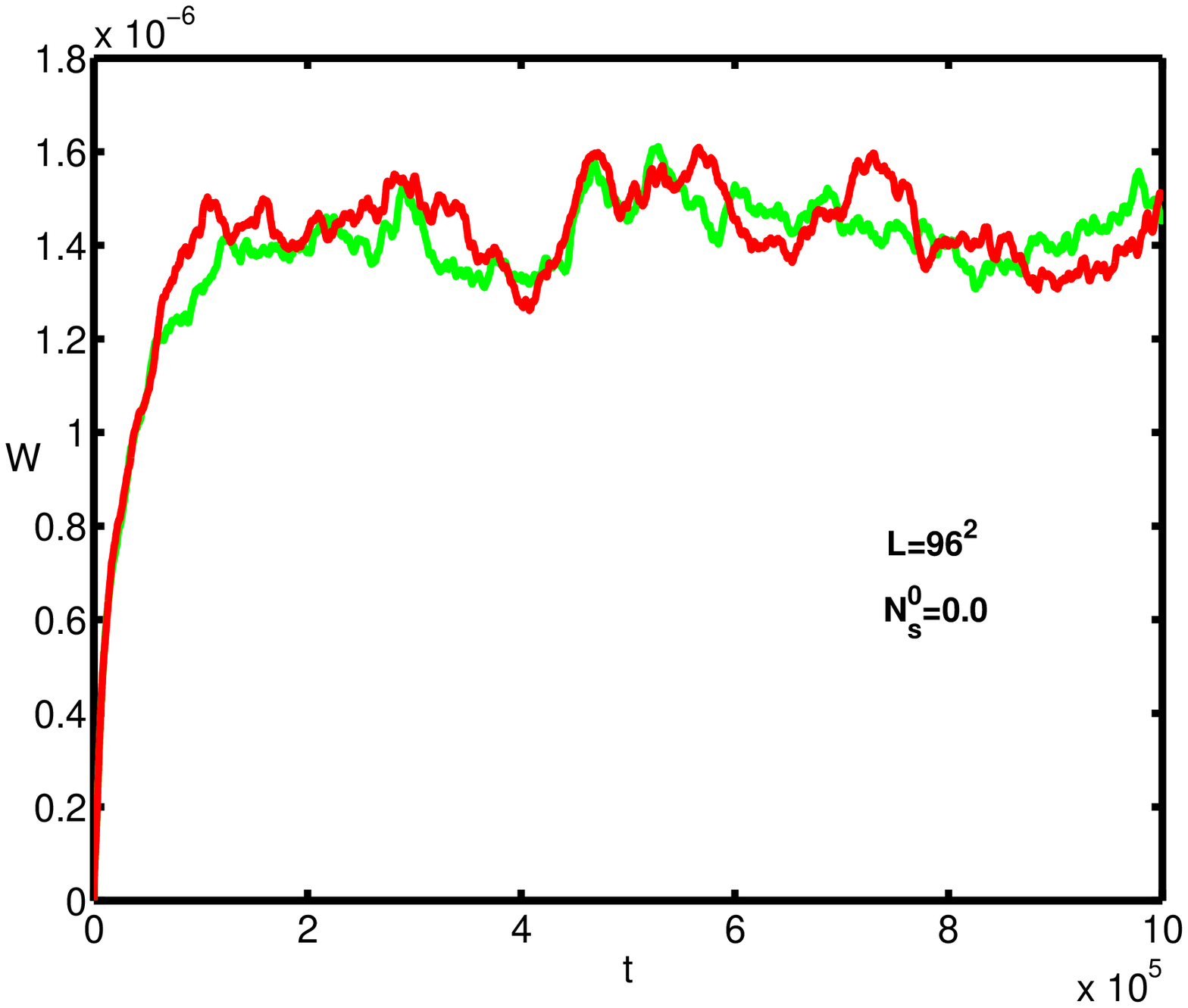}\hfill
\includegraphics[width=3.5in]{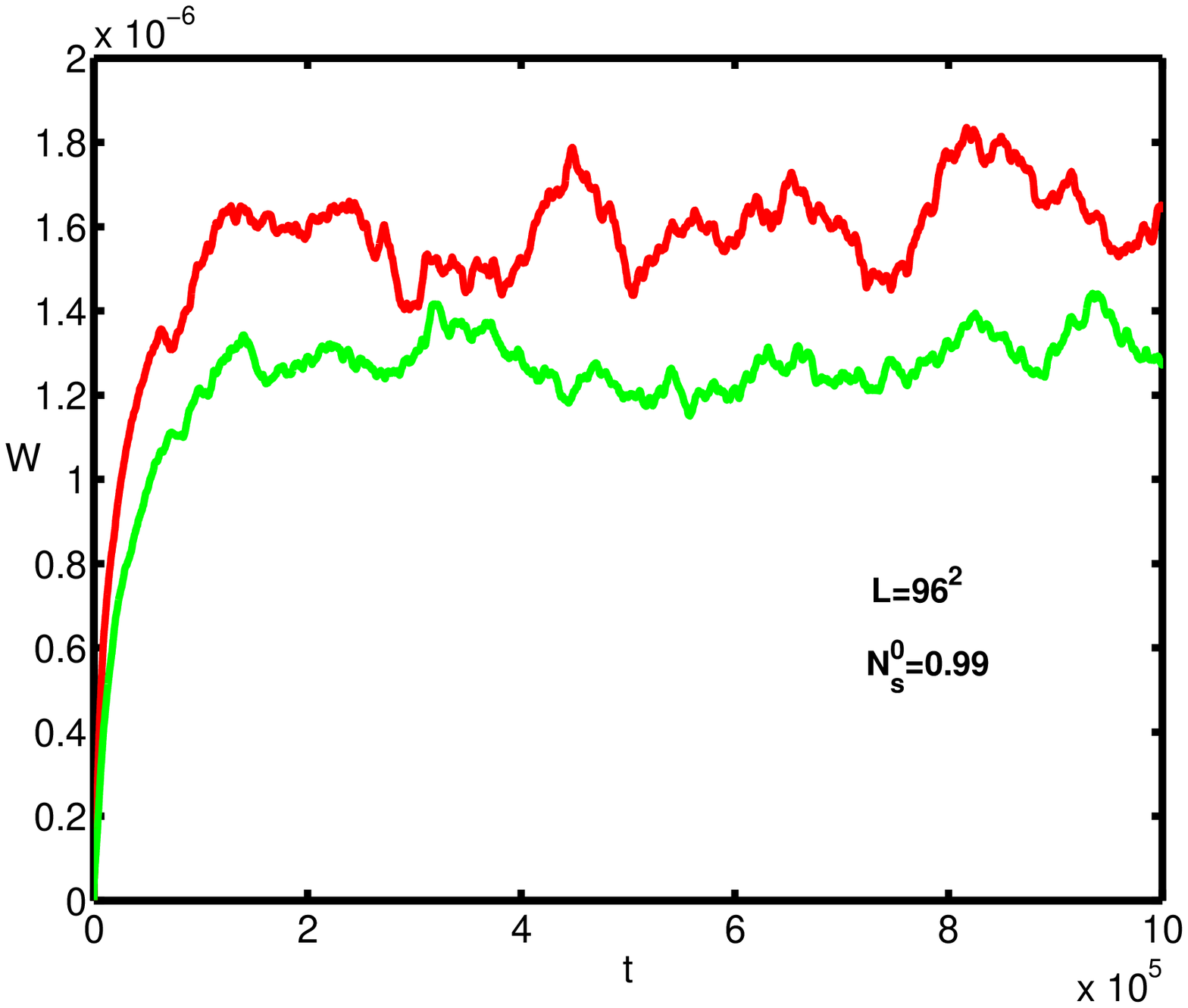}\\
\includegraphics[width=3.5in]{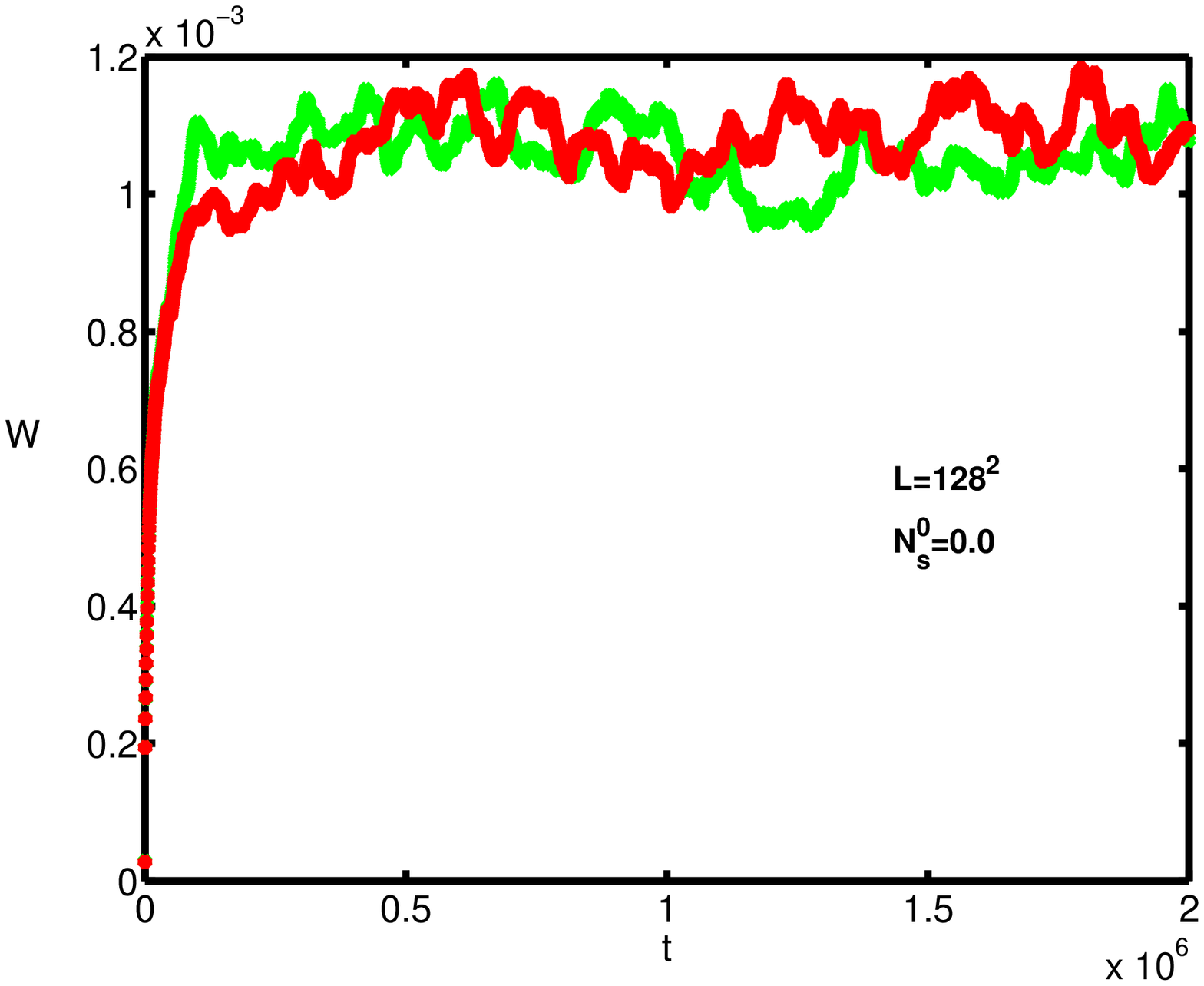}\hfill
\includegraphics[width=3.5in]{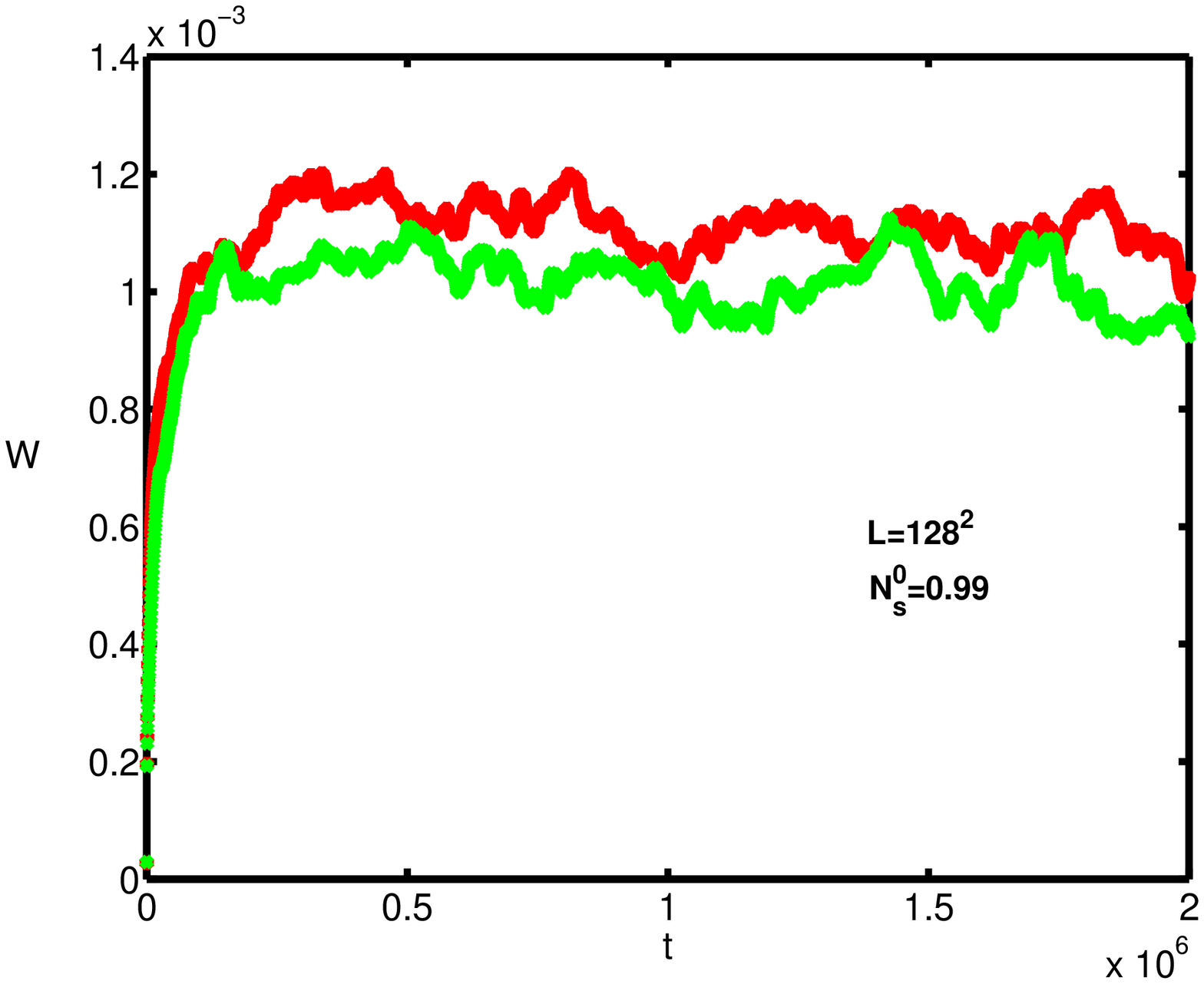}
\caption{Plots of $W_h(t)$ (red) and $W_{\phi}(t)$ (green) versus
$t$ in the rough phase for the system sizes $96^2$ (top) and $128^2$
(bottom) for $N_s^0=0.0,\,0.9$. Note the increasing differences
between the saturation values of $W_h$ and $W_{\phi}$ with
increasing $N_s^0$.} \label{2d96}
\end{figure}
From our DNS studies with system sizes $96^2,\,128^2,\,160^2$ we
calculate the parameter $A$. We show its dependence on $N_{s}^0$ in
the Fig. (\ref{2dcfin}). As in $1d$ we see that $A$ decreases
monotonically as $N_{s}^0$ increases from zero.
\begin{figure}[h]
\includegraphics[width=5in]{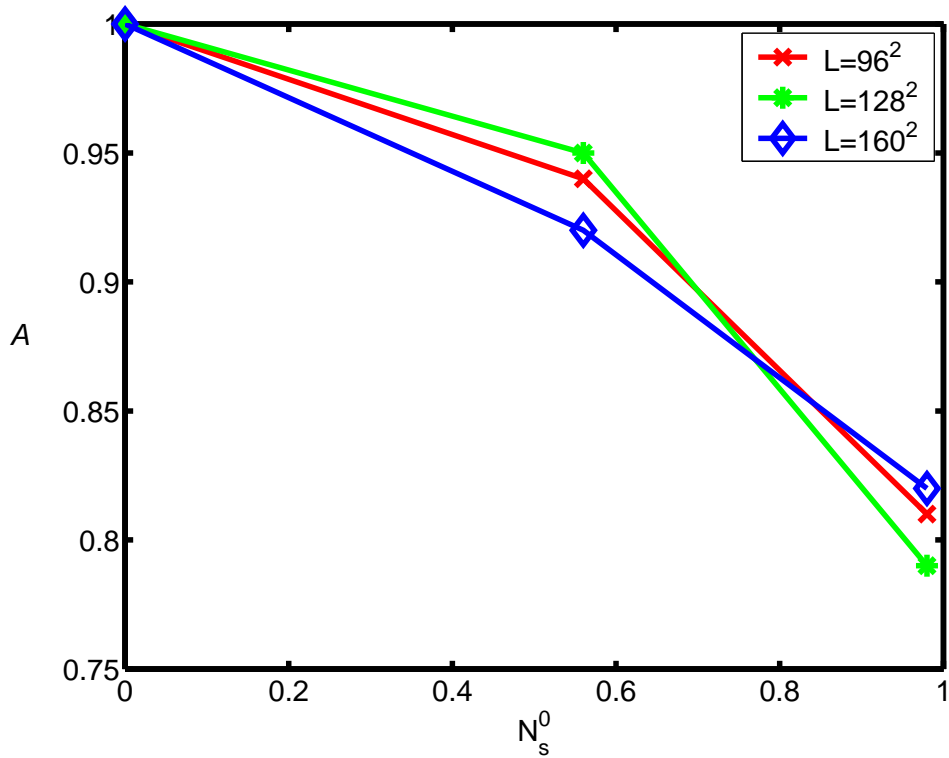}
\caption{A plot of $A$ versus $N_{s}^0$ for various system sizes
from our $2d$DNS studies.} \label{2dcfin}
\end{figure}

We now determine the scaling exponents $\chi_h$ and $z$. Due to the
very time consuming nature of the $2d$ simulations the quality of
our data for $2d$ are much poorer than those obtained from our $1d$
runs. Because of this difficulty we do not extract the roughness
exponent directly by plotting different correlation functions in the
steady state as functions of wavevector $k$. Instead we obtain the
ratio $\chi_h/z$ by plotting the width $W_{h}(t)$ as a function of
time $t$ in a log-log plot for $t \lesssim$ saturation time scale.
The slope yields the ratio $\chi_h/z$; we obtain $\chi_h/z=0.2\pm
0.1$ which is to be compared with our analytical result $1/5$,
obtained by means of SCMC and FRG calculations. At $d=2$ SCMC yields
$z=1.66$ which is close to Ref.~\cite{moore} whereas our DNS yields
$z\approx 1.60\pm 0.1$. Numerical studies of our type, performed on bigger
system sizes, should be able to
yield highly accurate values for the scaling
exponents which could be compared systematically with our SCMC/FRG
results and those of Ref.~\cite{moore}.
However, we refrain ourselves from making such detailed comparisons
due to the rather small system sizes we have worked with. We present
our result in Fig.\ref{slope} below for the system sizes $L=128^2$
and $L=160^2$ with $N_{s}^0=0.0$ and $N_{s}^0=0.9$ respectively.
Note that, within the accuracy of our numerical solutions, the ratio
$\chi_h/z$ does not depend upon the value of $N_{s}^0$, suggesting
that the scaling exponents are {\em independent} of symmetric
cross-correlations in agreement with our SCMC (Section
\ref{modecoupling}) and FRG (Section \ref{funcrg}) above.
\begin{figure}[h]
\includegraphics[width=8.0cm]{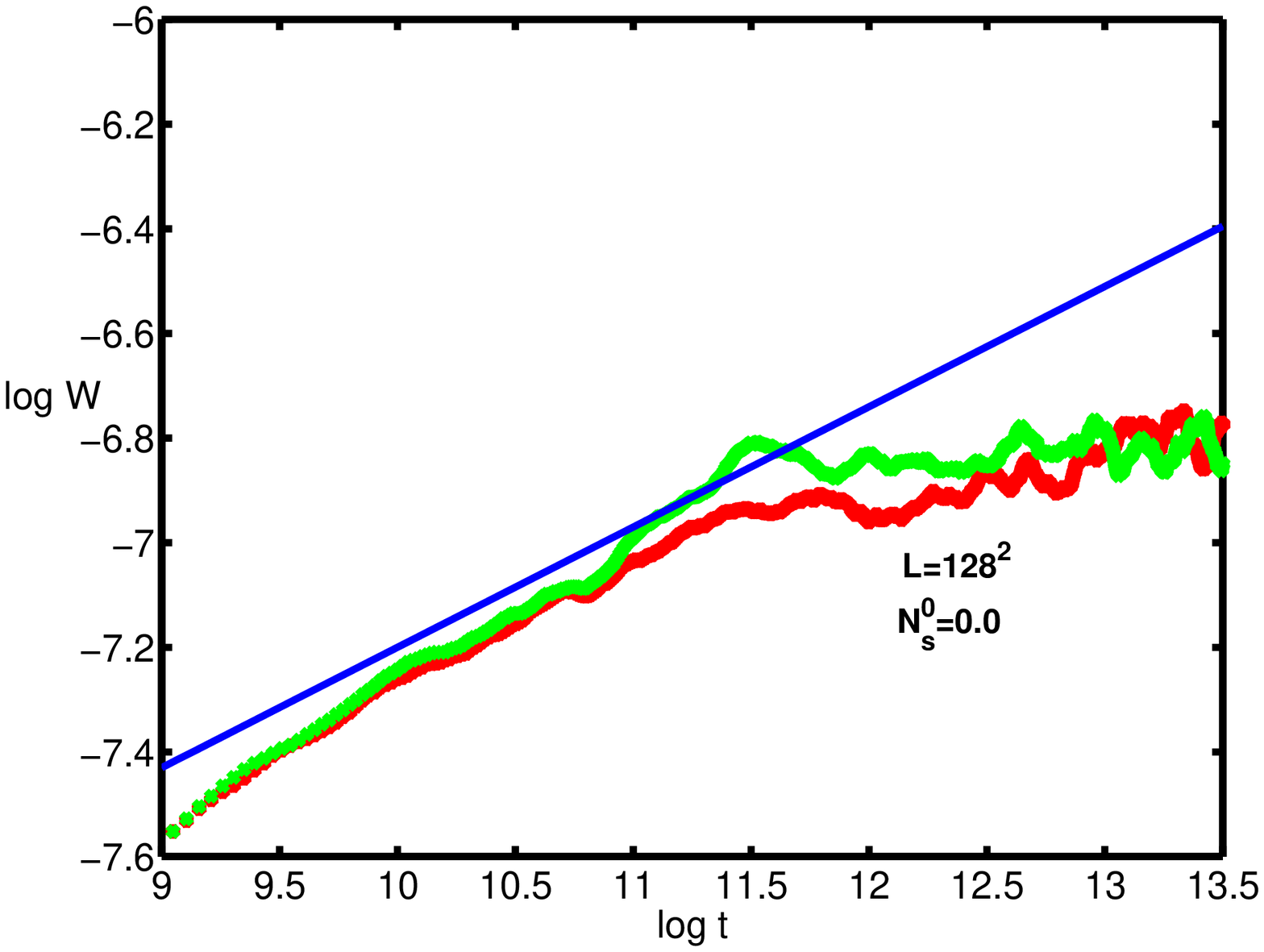}\hfill
\includegraphics[width=8.0cm]{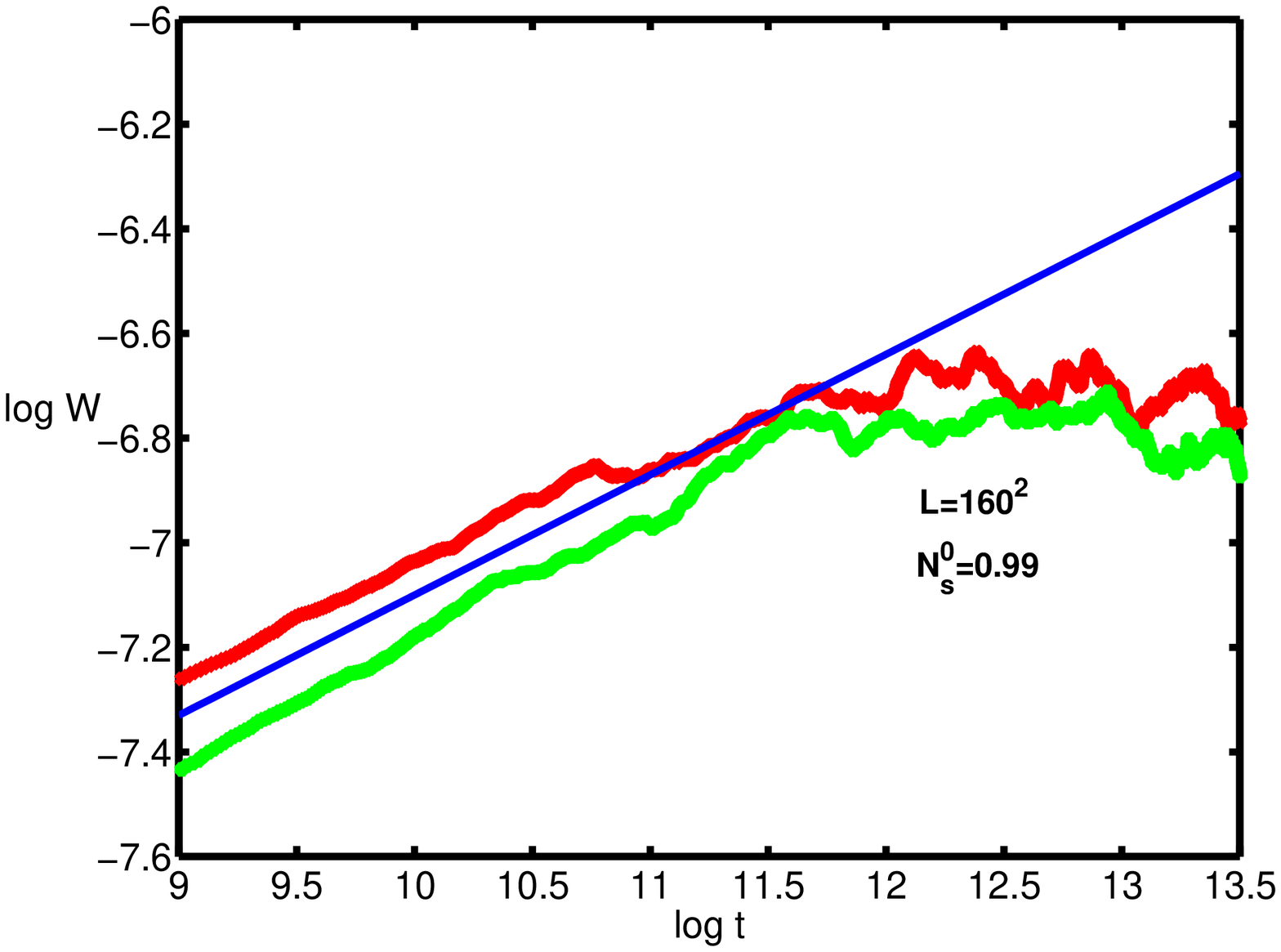}
\caption{Log-log plots of $C_{hh}(k,t=0)$ and $C_{\phi\phi}(k,t=0)$
versus $k$ from our $2d$DNS studies: (i) $L=128^2,\,N_{s}^0=0.0$,
(ii) $L=160^2,\,N_{s}^0=0.9$. The slope $\sim 0.2$ yields $\chi_h/z$
(see text).} \label{slope}
\end{figure}

We conclude this Section by summarizing our results. We performed
DNS of the model Eqs. (\ref{eq2kpz1}) and (\ref{eq2kpz2}) at
$d=1,\,2$ in the presence of the symmetric cross-correlations
parametrised by $N_{s}^0$. By tuning the bare amplitude $D_0$ we are
able to access the rough phase in $2d$. In both dimensions we find
that in the rough phase the ratio $D_b/D_u$ decreases monotonically
from unity as $N_{s}^0$ increases from zero. In $1d$ we find
$\chi_h=0.5$ very accurately which is in good agreement with the
analytically obtained results. In $2d$ we find $\chi_h/z=0.2\pm 0.1$
which is  to be compared with our analytical estimate of 1/5. Extensive DNS studies on
larger system sizes would be required to calculate the scaling
exponents with high accuracy.

\subsection{Monte Carlo simulations of the coupled lattice-gas models in one
dimension} \label{coupmc}

So far in the above we have studied the applications of several
methods, analytical as well numerical, on the continuum model
Equations (\ref{eq:burgers_1}) and (\ref{eq:burgers_2}) [or Eqs.
(\ref{eq2kpz1}) and (\ref{eq2kpz2})] and obtained several results
concerning universal properties of the statistical steady-state of
the model. Note that all techniques used above have been applied on
the {\em same} model. To complement our studies we discuss the
results from the lattice-gas model in this section which we
constructed [see Section~\ref{coup-lat}],  and compare with our
results already obtained above.

In this section we simulate our proposed  one-dimensional
lattice-gas models for the model Eqs. (\ref{2kpz}). By using
relations (\ref{lattmod-corr}) we calculate the ratios of
appropriate correlation functions in the steady state. We use Monte
Carlo methods to simulate our models. We extended the
Restricted-Solid-On-Solid (RSOS) algorithm \cite{rsos} and the
Newman-Bray (NB) algorithm \cite{ns} for the KPZ surface growth
phenomena to construct the coupled lattice-gas models in $1d$ for
Eqs. (\ref{2kpz}) with cross-correlations. In such models particles
are deposited randomly from above and settle on the already
deposited layer following certain growth rules which define the
models. One typically measures the widths of the height fluctuations
of the growing surfaces as functions of time $t$. Note that our
numerical works on the lattice-gas models are restricted to lattices
of modest sizes ($L=4096$ is the largest system size considered).
This is due to the fact that the numerical generation of noise
cross-correlations of the type we considered is not uncorrelated in
space in $1d$, rather it has a variance proportional to $1/x$. Generation of
such noises is a very time consuming process: we first generate the
noises in the real space. These are uncorrelated Gaussian random
noises without any cross-correlations. We then bring them to Fourier
space by Fourier transform. Inverse Fourier transforms of particular
linear combinations of them yield noises with finite
cross-correlations which we use in our studies. Although we have
used Fast Fourier Transforms, they are still computationally rather
time consuming and hence reduce the over all speed of the code. In
contrast, the more common lattice-gas studies on the KPZ equation do
not require Fourier Transforms of the noise (since they do not have
any noise cross-correlations) and hence are much faster. This allows
them to study up to much larger system sizes.

\subsubsection{A coupled lattice-gas model in one dimension with RSOS update rules:}
\label{rsos}  The restricted solid-on-solid (RSOS) update rules
involve selecting a site on the lattice randomly and permitting
growth by letting the height of the interface at the chosen site
increase by unity such that the height difference between the
selected site and the neighbouring sites does not become more than
unity \cite{rsos}. Our model involves two sublattices where the
height fields on each of the sublattices satisfy each of Eqs.
(\ref{2kpz}). In our coupled lattice model each sublattice is
evolved according to the RSOS update rule described above. The
selection of the random sites in the two sublattices may be
correlated. This correlation models the noise cross-correlations of
the model continuum Eqs. (\ref{eq:burgers_1}) and
(\ref{eq:burgers_2}) or (\ref{2kpz}).

We simulated our coupled lattice-gas model described above and
calculate the widths $W_h (t), \; W_{\phi}(t) $ of the growing
surfaces $h(x,t)$ and $\phi (x,t)$ as functions of time $t$. Similar
to the single-component KPZ equation, the plots have two distinct
parts - an initial growing part and a late time saturated part.
Below we present our results graphically: We plot $W_h(t)$ (red) and
$W_{\phi}(t)$ (green) versus $t$ for system size $1024$ in Fig.
(\ref{rsos1024}). Results from system sizes $L=2048,\,4096$ show
similar behavior without any systematic system size dependence.
\begin{figure}[h]
\includegraphics[width=8cm]{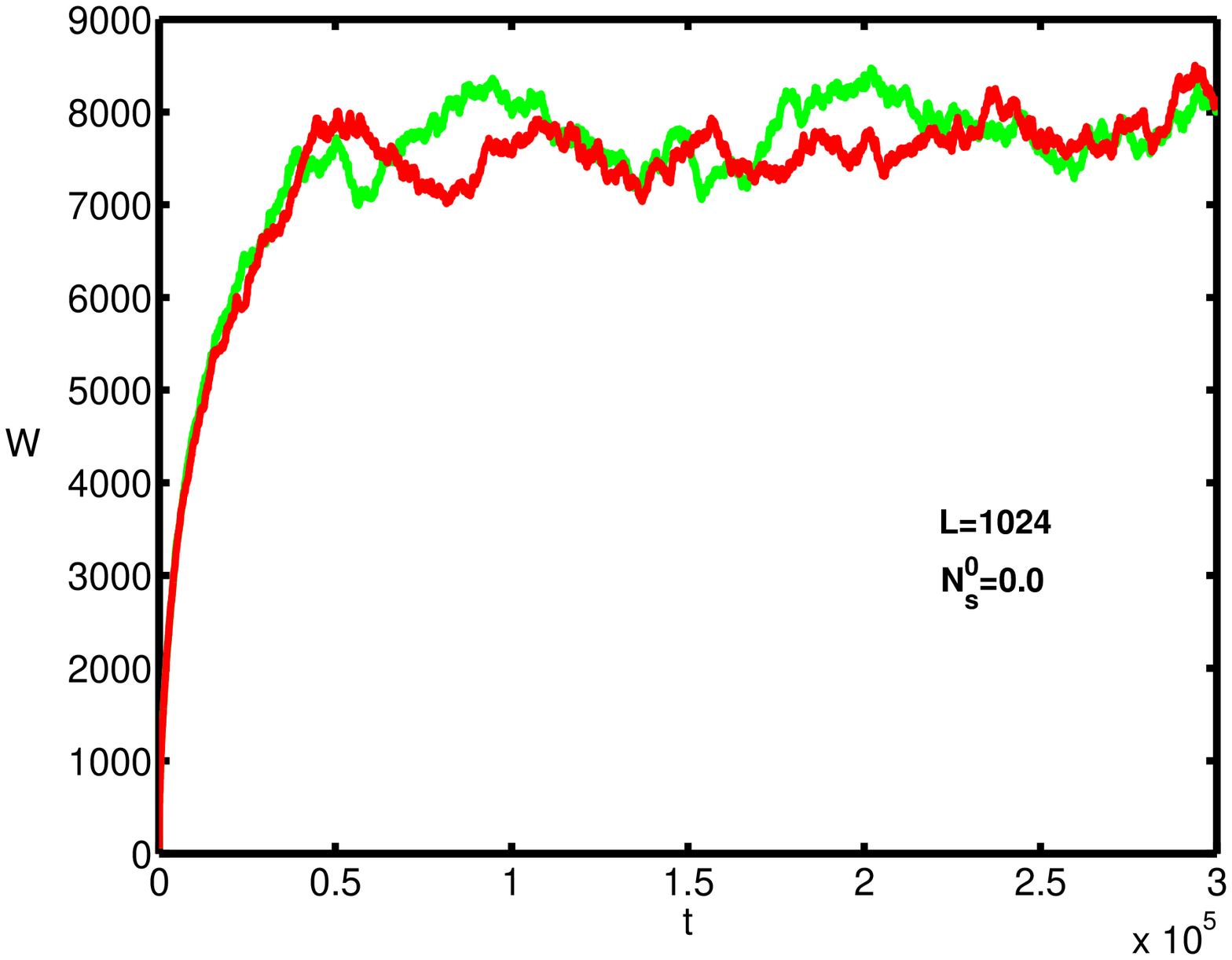}\hfill
\includegraphics[width=8cm]{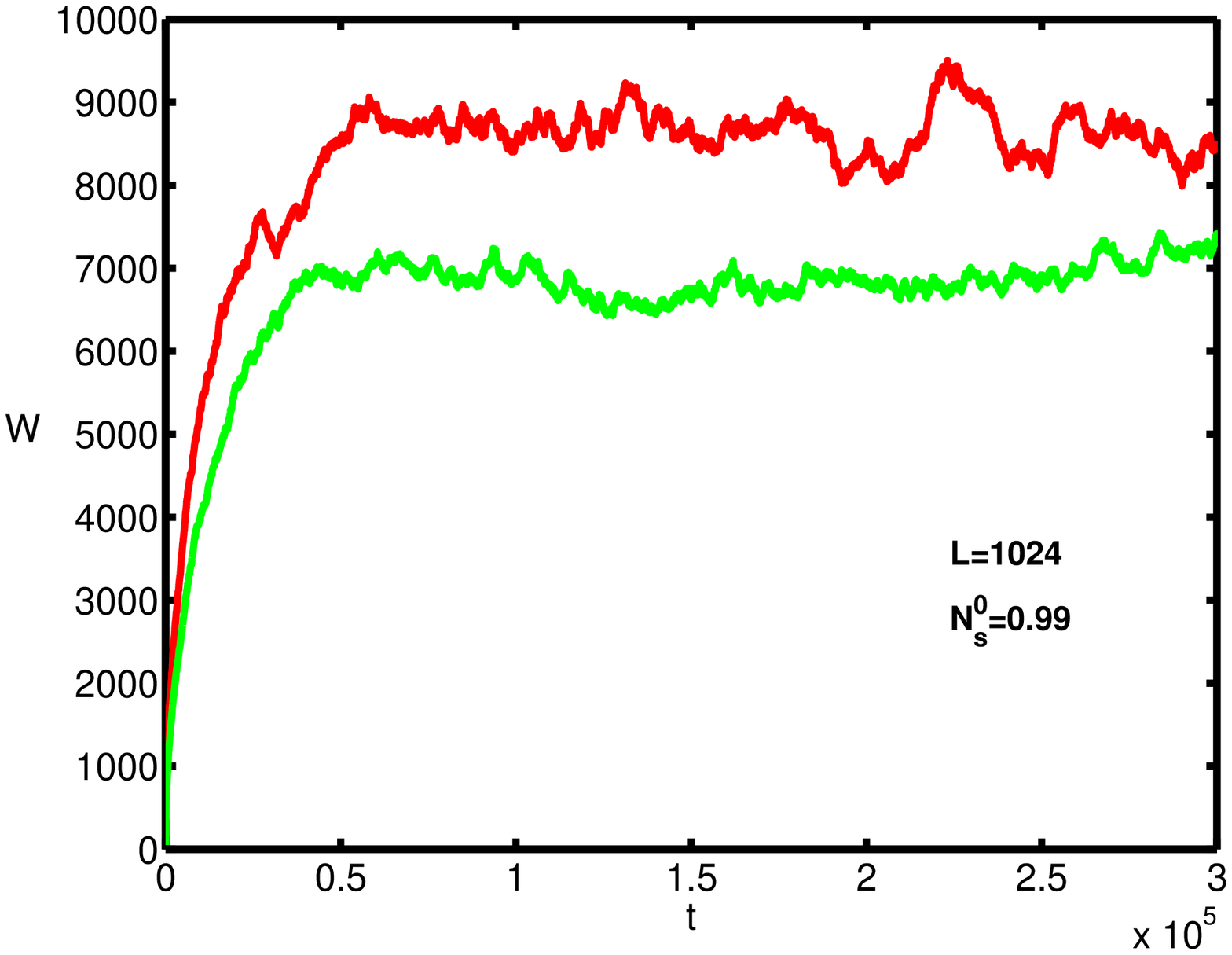}
\caption{Plots of $W_h(t)$ (red) and $W_{\phi(t)}$ (green) versus
$t$ for system size $L=1024$ for $N_{s}^0=0.0,\,0.9$  in our coupled
lattice-gas model based on the RSOS algorithm. Note the increasing
difference between $W_h$ and $W_{\phi}$ with increasing $N_{s}^0$.}
\label{rsos1024}
\end{figure}


In Fig.~\ref{rsosfinal} we present a plot of $A$ versus $N_{s}^0$
for the system sizes $L=1024,\,2048,\,4096$. As in our results
obtained analytically and DNS, $A$ decreases monotonically from
unity as $N_s^0$ increases from zero. We do not observe any
systematic system size dependence.
\begin{figure}[h]
\includegraphics[width=12cm]{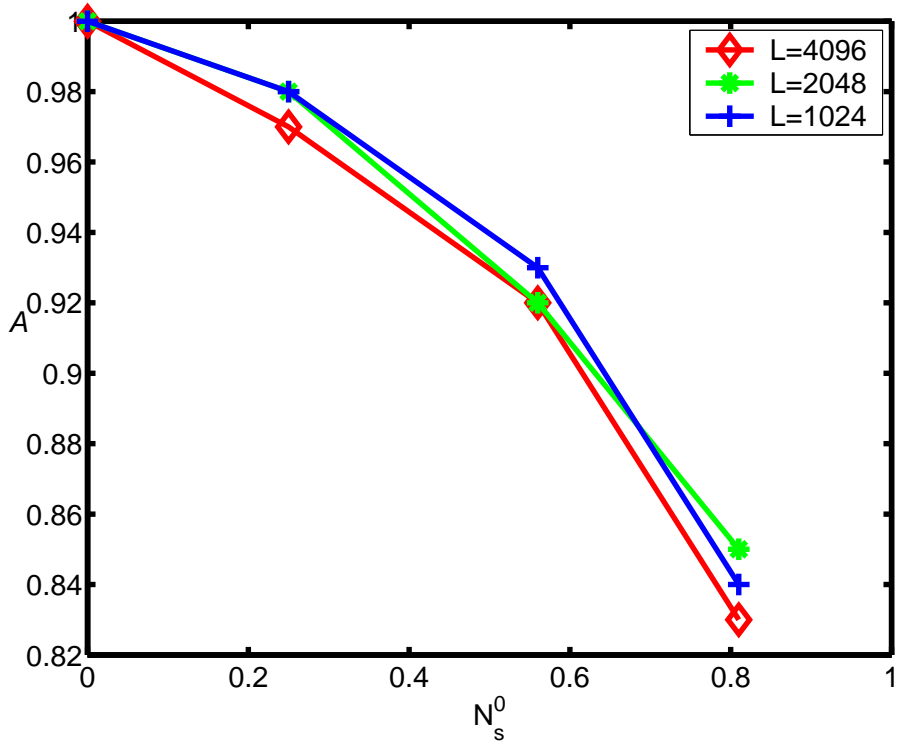}
\caption{A plot of $A$ versus $N_s^0$ for various system sizes
obtained from numerical simulations of our coupled lattice-gas
models with the RSOS algorithm.} \label{rsosfinal}
\end{figure}
Having shown the dependence of $A$ on $N_s^0$ we now estimate the
scaling exponents $\chi_h$ and $z$ to complete the discussions on
the universal properties of the lattice-gas model. Below we present
a log-log plot of the equal-time correlation function
$C_{hh}(k)\equiv \langle h(k,t)h(-k,t)\rangle$ in the steady state
as a function of the Fourier vector $k$ [Fig. (\ref{scal})]. We find
that slope in the scaling regime (small $k$) is very close to -2
corresponding to $\chi_h=1/2$ as predicted by analytical means
before.

\begin{figure}[h]
\includegraphics[width=5in]{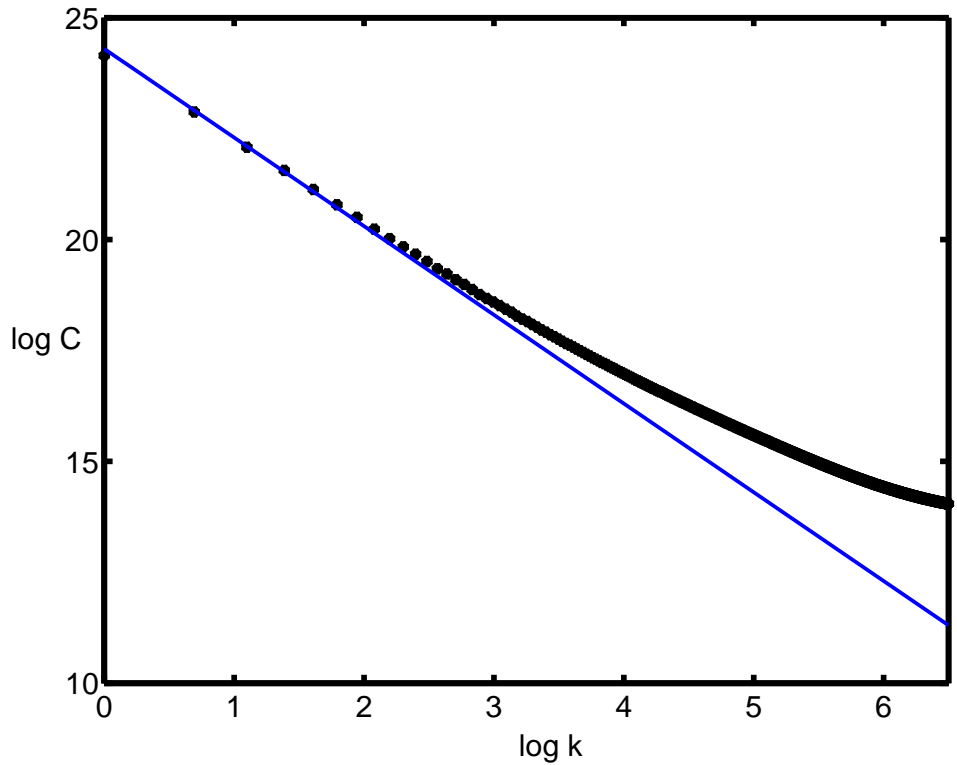}
\caption{A log-log plot of the correlation function $C_{hh}(k)$
versus $k$ for system size $L=2048$ from our coupled lattice-gas
model with the RSOS algorithm. The straight line with a slope of -2,
corresponding to $\chi_h=1/2$, represents the scaling regime of
$C_{hh}(k)$ for small $k$ (long wavelength limit).} \label{scal}
\end{figure}
An alternative way to obtain the scaling exponent $\chi_h$ is by
finding the dependence of the saturated value of the widths $W_h
(t)$ and $W_\phi (t)$ on the system size $L$. Since, after saturations in the steady
states $W_{h,\phi}\sim L^{2\chi_h}$ with
$L$ being the system size, we obtain $\chi_h$ from the slope of the
plots of logarithm of the saturated values of the widths versus
logarithm of the corresponding system sizes. We find $\chi_h=0.49\pm
0.05$ which is very close to the analytically obtained value,
Further, from the time-dependences of the widths $W_h(t)$ and
$W_\phi(t)$ the exponent-ratio $\chi_h/z$ can be obtained.
Fig.~\ref{skslope} shows a log-log plot of $W_{h}(t)$ and
$W_{\phi}(t)$ versus $t$ from a lattice size $L=16384$. For such a
large lattice size the saturation time is very large and hence we
show only the growing part of the curves. The slope yields the ratio
$\chi_h/z$. The red and green points refer to $W_{h}(t)$ and
$W_{\phi}(t)$, and the blue line a slope of $0.32$ indicating
$\chi_h/z=0.32$ which is very close to the analytically obtained
value of $1/3$. In short, therefore, the results from the
Monte-Carlo simulations of our coupled lattice-gas model in $1d$,
based on the RSOS algorithm, yield results in close agreement with
those obtained through analytical means and $1d$DNS.
\begin{figure}[h]
\includegraphics[width=5in]{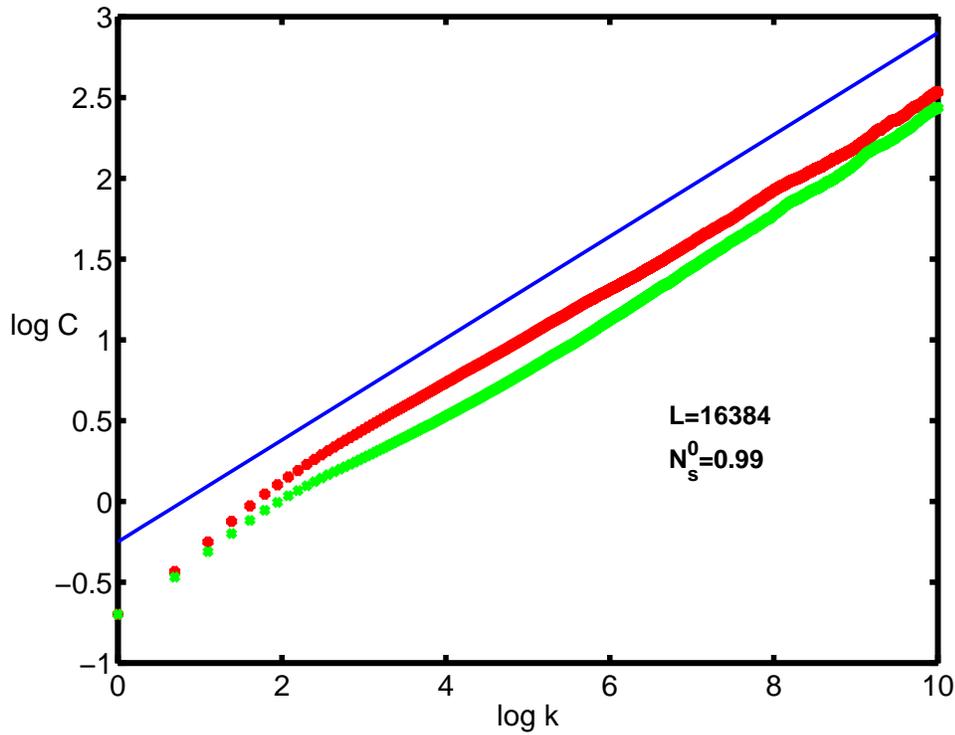}
\caption{A log-log plot of $W_{h}(t)$ (red) and $W_{\phi}(t)$
(green) versus $t$, obtained from our lattice-gas model with the
RSOS algorithm with system size $L=16384$. The black line indicates
a slope of 0.32 for easy eye estimation.} \label{skslope}
\end{figure}

\subsubsection{A coupled lattice-gas model in one dimension with the NB update rule:}
\label{ns}

As mentioned before, this model uses the mapping between the KPZ
surface growth problem and the equilibrium problem of a directed
polymer in a random medium (DPRM) which are connected by the
non-linear Cole-Hopf transformation $h_{1,2}({\bf x},t)= (2\nu/
\lambda)\ln Z_{1,2}$ leading to the Eqs.(\ref{colehopfeq}) for the
partition functions $Z_{1,2}$ for the two DPs. Further, one uses the
following update rules for $h_{1,2}$:
\begin{eqnarray}
\tilde{h}_i(t) &=& h_i(t) + \Delta^{1/2}\zeta_i(t),\nonumber \\
h_i(t+\Delta ) &=& \tilde h_i(t) + (\nu/\lambda) \ln \left[ 1+ (\Delta \nu/
a^2) \Sigma_{jnni}[e^{\lambda(\tilde{h}_j-\tilde {h}_i)/\nu} -1] \right],
\label{nsalgo}
\end{eqnarray}
where $\Delta$ and $a$ are the grid scales for time and space,
respectively, and $jnni$ indicates sites $j$ and $i$ being nearest
neighbors. Then, taking the strong coupling limit
$\lambda\rightarrow \infty$ one finally obtains
\begin{equation}
h_i(t+\Delta ) = {\rm max}_{jnni} (\tilde{h}_i,\{\tilde {h}_j\}).
\end{equation}
This is same as the zero-temperature DPRM algorithm, written in
terms of the fields $h_{1,2}$. We implement the above growth rule in
$1d$ with system sizes $L=1024, 2048, 4096$. Functions $W_{h}(t)$
and $W_{\phi}(t)$, as defined in Sec.\ref{rsos} are obtained by
taking appropriate linear combinations of the equal-time correlators
of $h_{1,2}$. We present our results graphically in Fig.
(\ref{sn1024}) for system size $L=1024$ below.

\begin{figure}[h]
\includegraphics[width=3.5in]{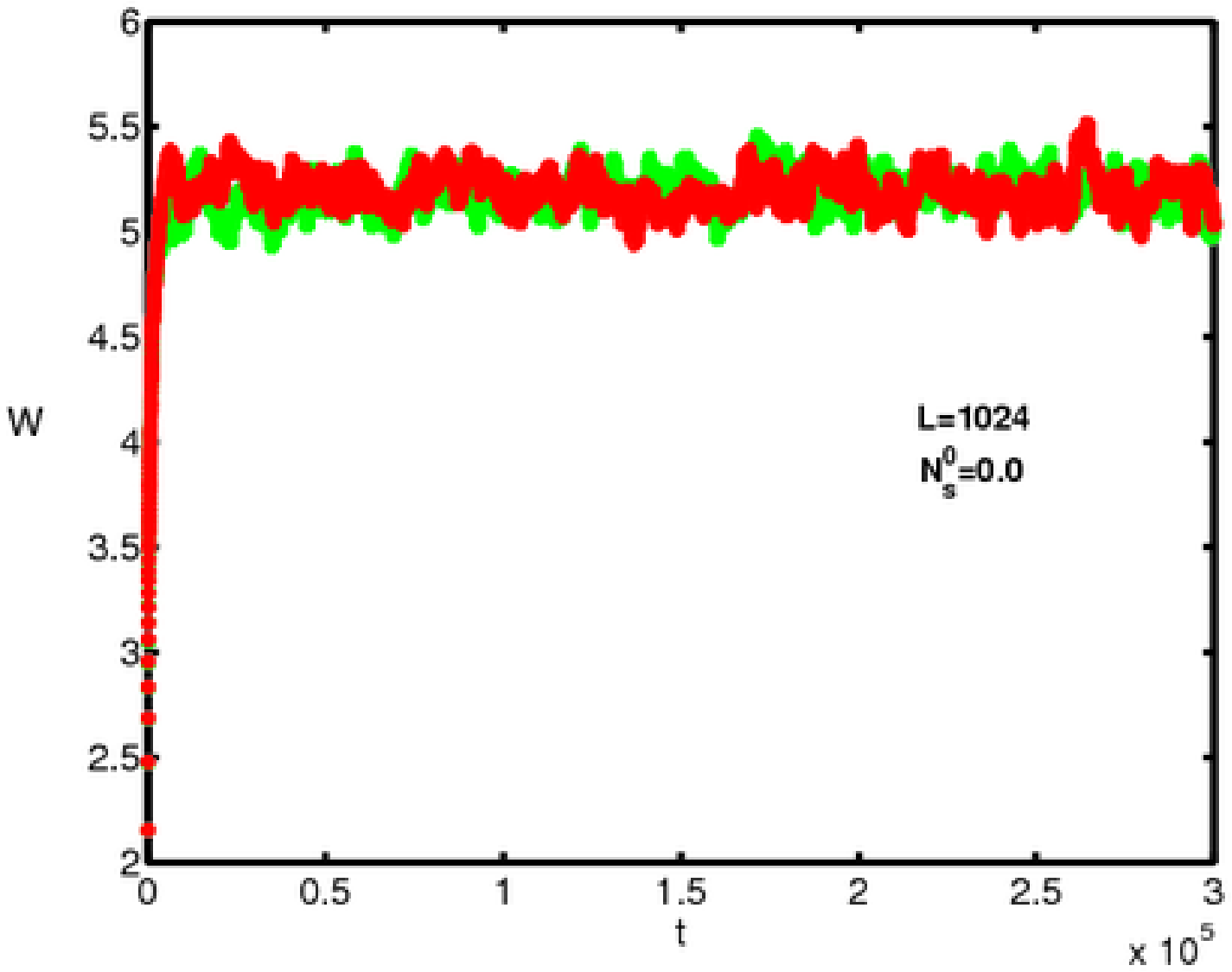}\hfill
\includegraphics[width=3.5in]{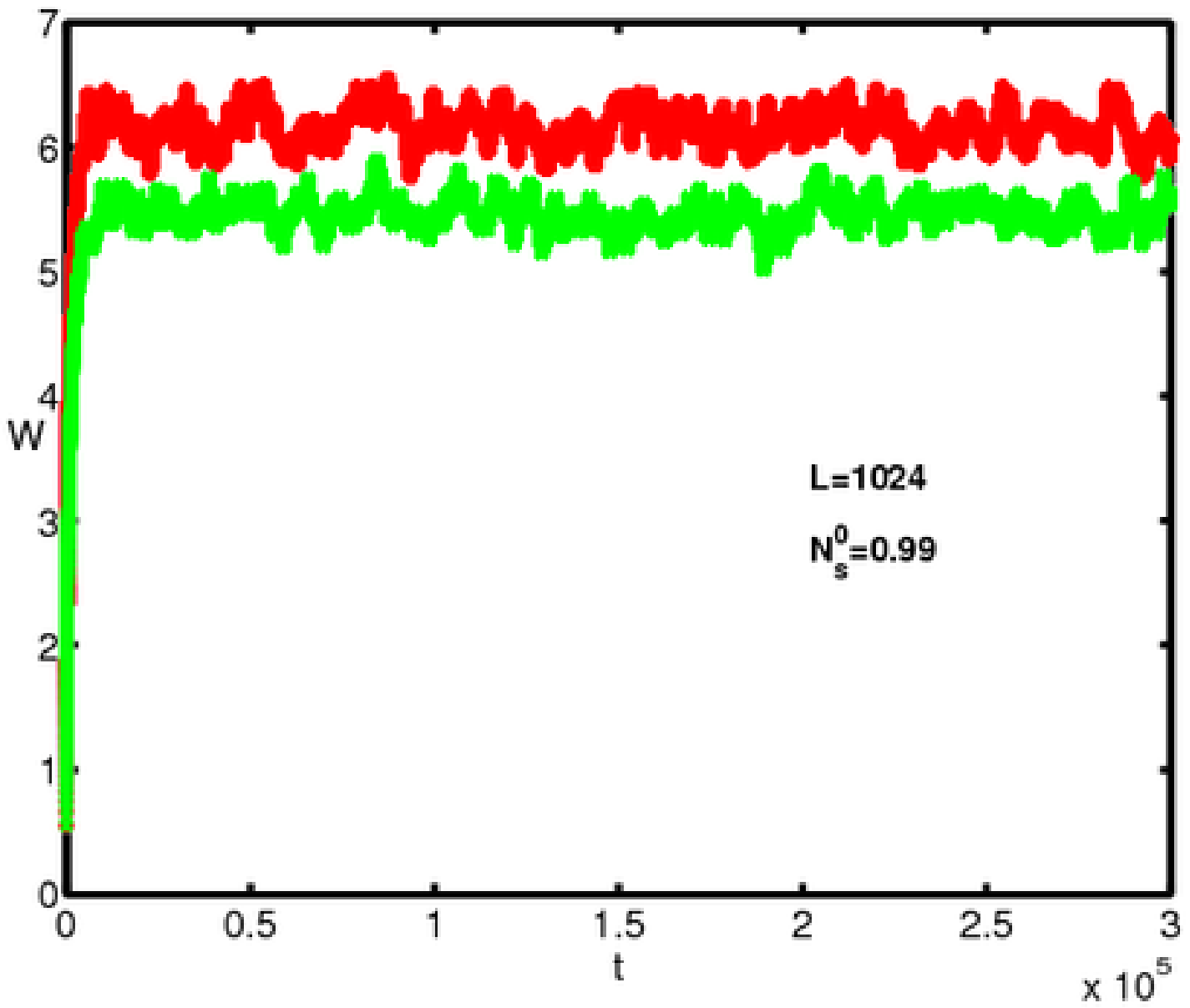}
\caption{Plots of $W_h(t)$ (red) and $W_{\phi}(t)$ versus $t$ for
system size $L=1024$ for various $N_s$ in our coupled lattice-gas
model based on the NB algorithm (see text). Note the increasing
difference between $W_{h}$ and $W_{\phi\phi}$ with
$N_{s}^0=0.0,\,0.55,\,0.9$.} \label{sn1024}
\end{figure}



From Fig.~(\ref{sn1024}) it is clear that after saturations, the
differences between $W_h$ and $W_{\phi}$, measured by $A$, increase
with $N_{s}^0$. Below in Fig.~(\ref{newmfinal}) we show a plot
depicting the variation of $A$ versus $N_{s}^0$ obtained from the
simulations of our coupled lattice-gas model with the NB update rule
in $1d$. As before, we find that $A$ decreases monotonically from
unity as $N_s^0$ increases from zero.
\begin{figure}[h]
\includegraphics[width=12cm]{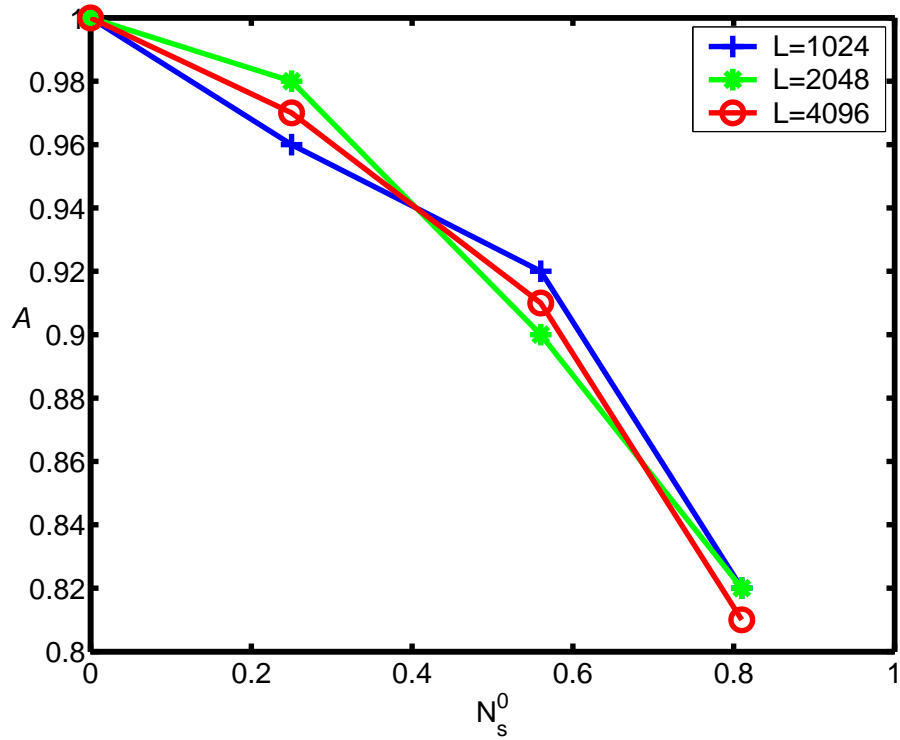}
\caption{A plot of $A$ versus $N_{s}^0$ obtained from numerical
solutions of our coupled lattice-gas model with the NB update rule
for system sizes $L=1024,\,2048,\,4096$ in $1d$.} \label{newmfinal}
\end{figure}
We further determine the scaling exponents $\chi_h$ and $z$ from our
numerical results, as we did above in Section~\ref{rsos}.
Fig.~(\ref{logsn}) shows a log-log plot of $W_{h}(t)$ and
$W_{\phi}(t)$ versus $t$ for the system size $L=4096$. We find that
the initial slope is $0.32\pm 0.02$ which is very close to the
analytically obtained $\chi_h/z=1/3$.
\begin{figure}[h]
\includegraphics[width=9cm]{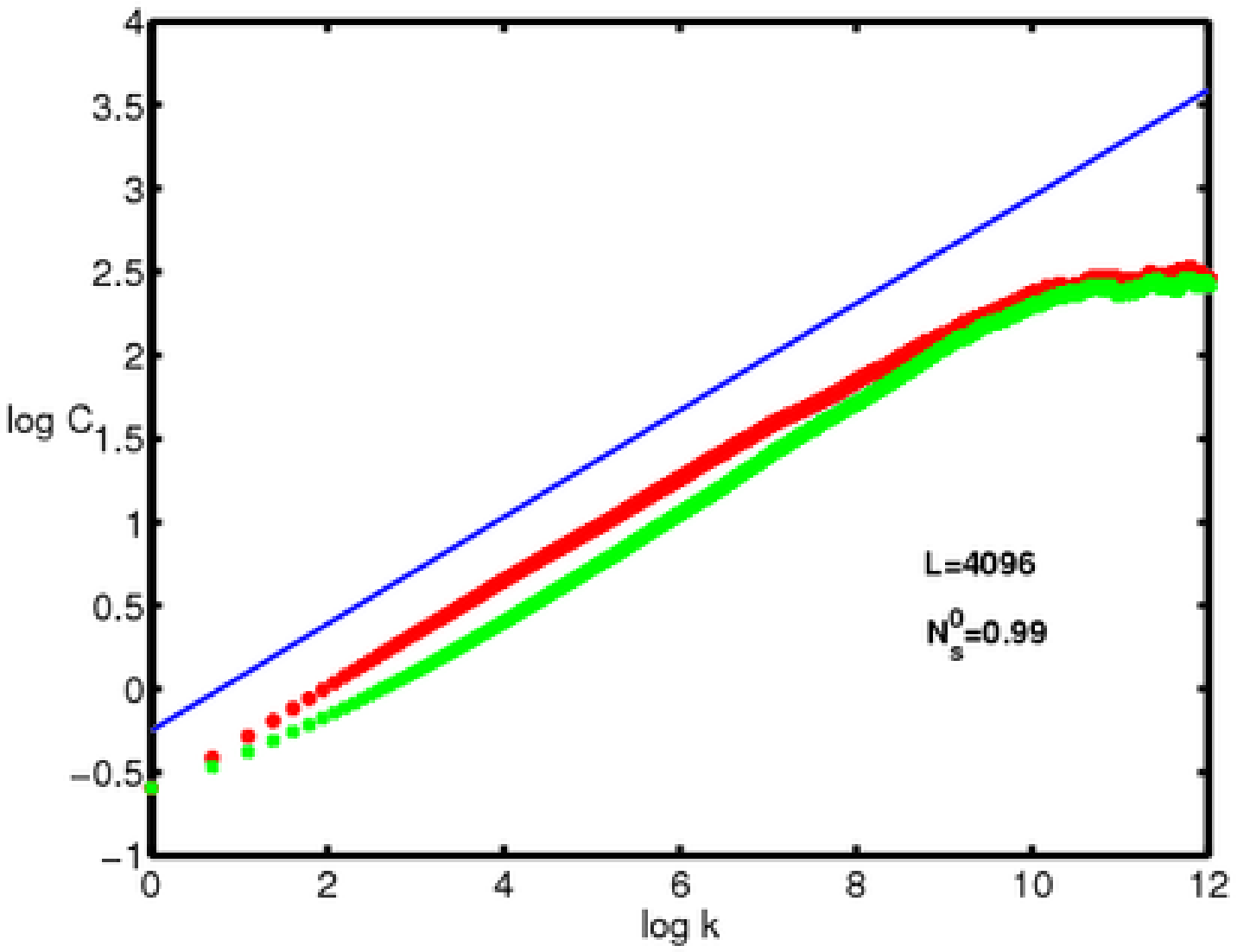}
\hfill\includegraphics[width=9cm]{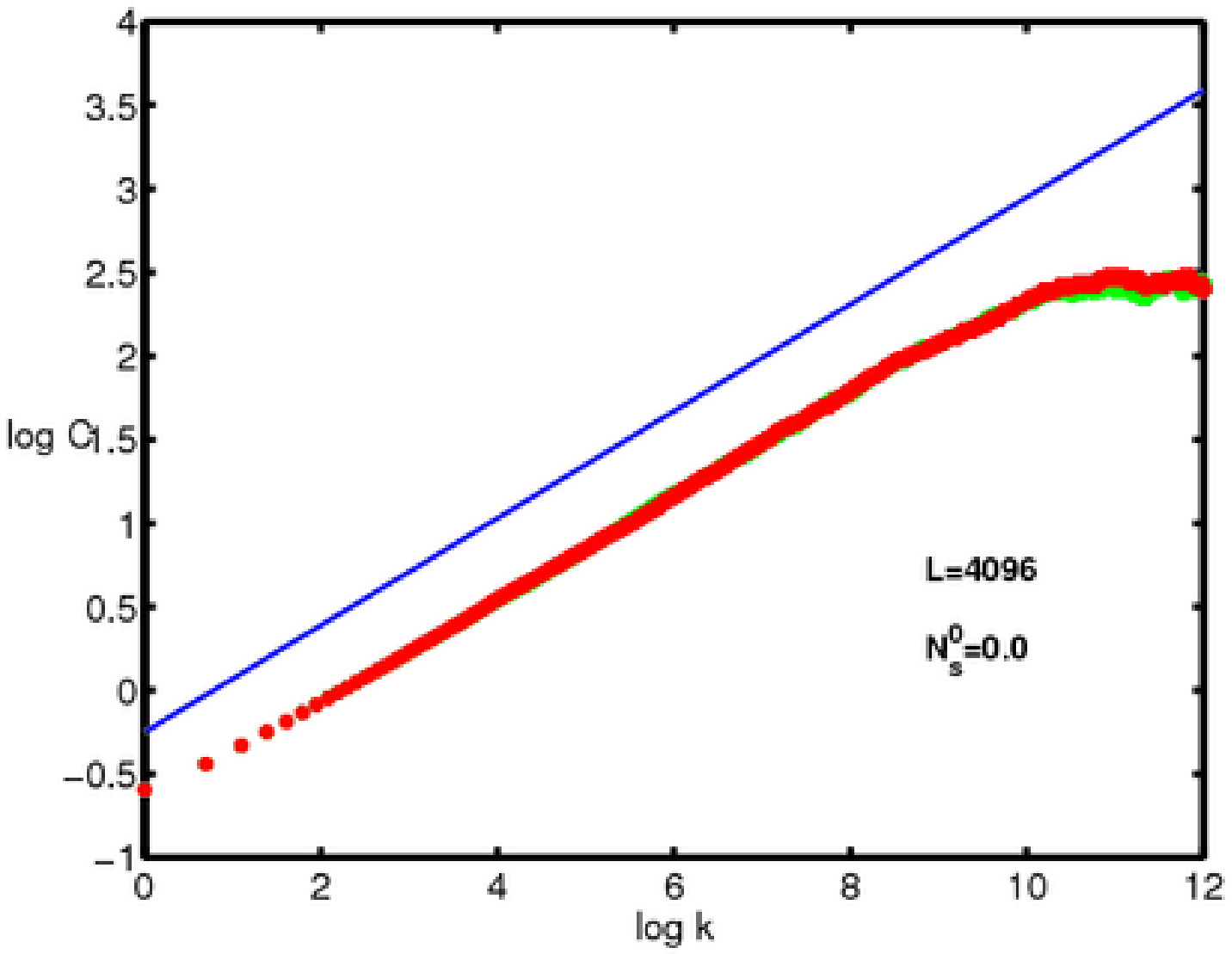} \caption{Log-log
plots of $W_{h}(t)$ (red) and $W_{\phi}(t)$ (green) versus $t$ for
system size $L=4096$ with $N_s^0=0.99$ (left) and $N_s^o=0.0$
(right) respectively. These are obtained from our lattice-gas model
with the NB algorithm The initial slopes are very close to the
analytically obtained value (see text).} \label{logsn}
\end{figure}
Furthermore,  the equal-time correlation function $C_{hh}(k,0)$ as a
function of $k$ is shown in a log-log plot [Fig. \ref{scal_ns}]. We
find that in the scaling regime (small $k$) $k^{-2}$ is an excellent
fit for $C_{hh}(k)$ yielding $\chi\simeq 1/2$. This, together with
the fact that $\chi_h/z\simeq 1/3$ give $z\simeq 3/2$, close to what
we estimated  by other means in $1d$ as reported above.
\begin{figure}[h]
\includegraphics[width=5in]{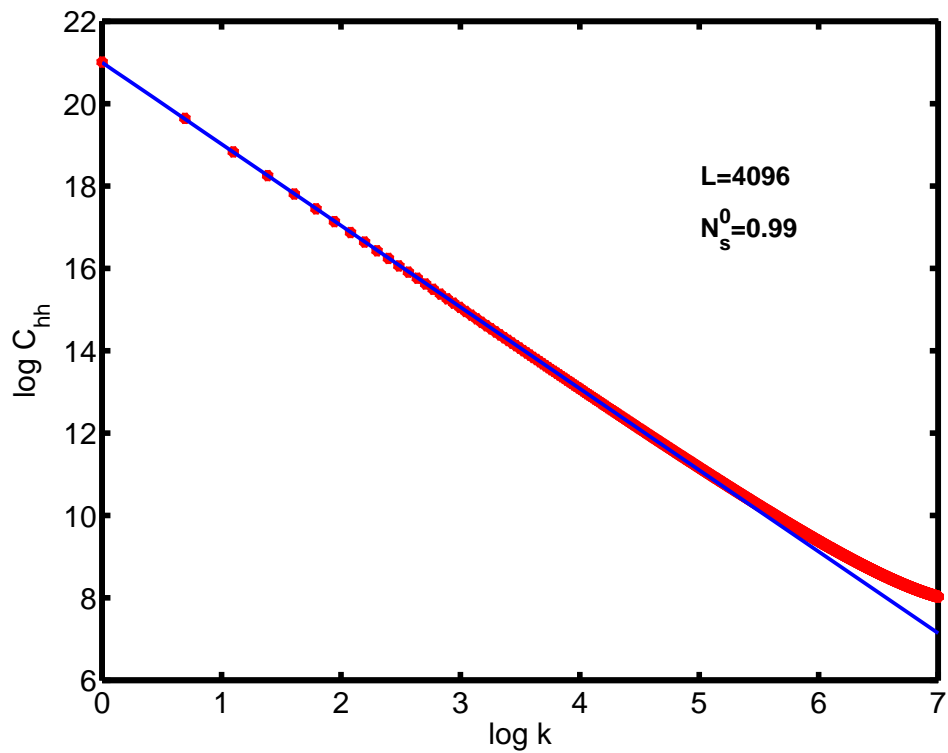}
\caption{A log-log plot of $C_{hh}(k)$ versus $k$ from simulations
of our coupled lattice-gas model with the NB algorithm (L=4096). The
black line is a guide to slope -2 which agrees very well with our
data in the scaling (small $k$) region.} \label{scal_ns}
\end{figure}

We close this section by summarizing our results from the
Monte-Carlo simulations of the two lattice-gas models proposed by
us. Both the models, based on the RSOS and the NB algorithms
respectively, exhibit similar dependence of the parameter $A$ over
$N_s^0$. They further yield the same scaling exponents. All these
are in close agreements with our analytical results obtained from
DRG/SCMC/FRG methods as well as $1d$DNS studies.

\subsection{Lattice-gas models and the continuum equations of motion:
Universality for systems out of equilibrium} \label{univ} In the
above we have obtained results on the scaling exponents and the
correlation functions in the statistical steady states from
different approaches. In particular we compare our results in $1d$
on the dependence of the amplitude ratio $A$ on the symmetric
cross-correlations $N_s^0$ from the numerical solutions (DNS) of the
continuum model Eqs. (\ref{eq:burgers_1}) and (\ref{eq:burgers_2}),
and the numerical simulations of the lattice-gas models based on the
RSOS and NB algorithms. Since in none of the cases we considered we
find any systematic system-size dependence, we take average of our
results from different system sizes for the same value of $N_s^0$
for a given type of study (DNS/NB/RSOS). We present our results in a
composite plot (\ref{compos}) below.
\begin{figure}[h]
\includegraphics[width=5in]{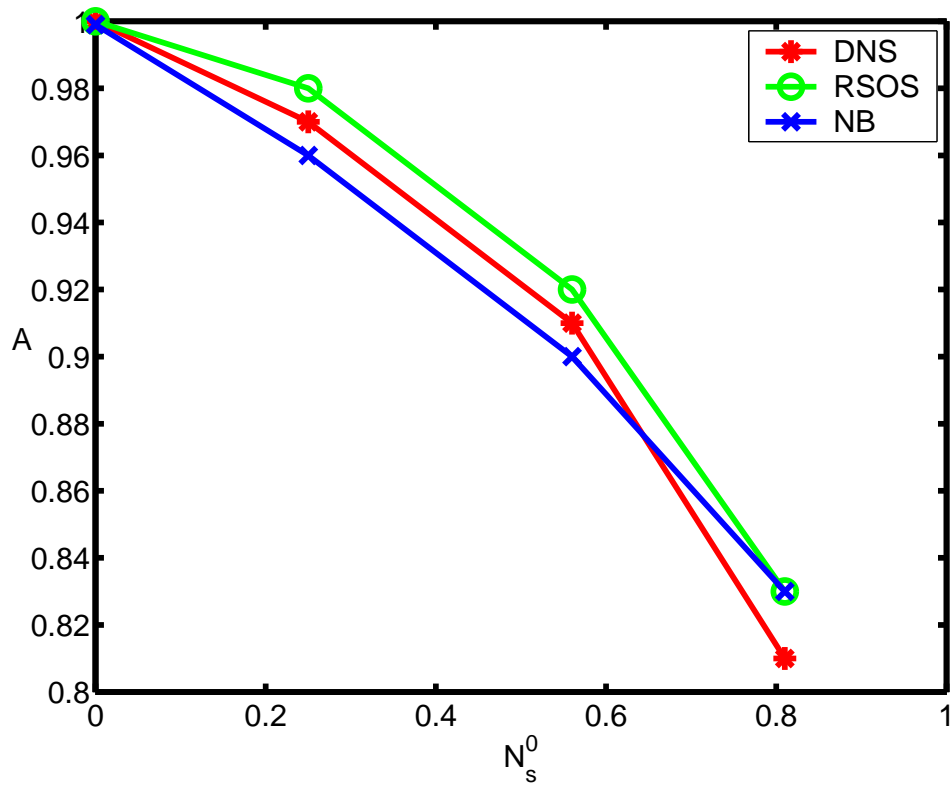}
\caption{A plot showing the variation of the amplitude ratio
$A$ as a function of the symmetric cross-correlations $N_{s}^0$
(see text) in $1d$ from RSOS (red x), NS (green *) and DNS (blue
diamond).} \label{compos}
\end{figure}
We find that the general trend that the ratio $A$ decreases from unity
when $N_s^0$ increases from zero persists in all the cases.
Moreover, all the cases yield values of the scaling exponents which
are very close to each other and in close agreement with the
analytically obtained values as well.

\section{Summary and outlook}
\label{conclu}

We have considered the statistical properties of the generalized
Burgers model (GBM) egiven by equations (\ref{eq:burgers_1}) and
(\ref{eq:burgers_2}), driven by conserved zero-mean Gaussian noise
sources with variances (\ref{noise1}, \ref{noise2}) and
(\ref{noise3}). We have undertaken a variety of analytical and
numerical investigations to uncover the universal properties of the
correlation functions in $d$ dimensions.

Upon employing analytical approaches like DRG, FRG and SCMC we have
calculated the scaling exponents and the amplitude ratios at $d=1$
and at $d \geq 2$. Each of these methods have a certain range of
applicability and also limitations. The DRG method yields results
only for the strong coupling phase in $d=1$ and the unstable
critical point (for $d>2$), for both symmetric and antisymmetric
cross-correlations. The FRG method works for the strong coupling
phases at $d=1$. For $d>2$ it is able to capture the physics in the
strong coupling phase, but only for symmetric cross-correlations.
The SCMC method is useful to study the strong coupling phases at
$d=1$ and $d>2$ for both symmetric and anti-symmetric
cross-correlations. Taking all the methods from the various
analytical methods together a coherent picture emerges. We broadly
find that symmetric cross-correlations affect the amplitude ratio
and the anti-symmetric cross-correlations affect the scaling
exponents. Since, in general, both symmetric and the anti-symmetric
cross-correlations are likely to be present, one may encounter
scaling exponents and amplitude ratios varying with the strength of
cross-correlations. In addition to our analytical studies, we have
performed extensive numerical work in $1d$. We have numerically
solved the continuum model equations using the pseudo-spectral
method in one and two dimensions, and numerically simulated two {\em
equivlent} lattice-gas models in $1d$. In all the cases we
numerically calculated the equal-time correlation functions of the
two height fields $h$ and $\phi$, and measured their spatial scaling
exponents and the amplitude ratios. Our numerical results on the
scaling exponents agree very well with those obtained analytically
for $d=1$. Further, our results on the dependence of the
amplitude-ratio on the strength of the cross-correlations from the
numerical studies agree broadly with those from the analytical
studies.

In $2d$ we have numerically solved the continuum model equations
(\ref{eq2kpz1}) and (\ref{eq2kpz2}) in the presence of symmetric
cross-correlations only. The amplitude ratios display dependences on
the cross-correlations similar to those obtained analytically. We
also obtained scaling exponents numerically. Due to the difficult
nature of the $2d$ simulations these are less accurate than their
$1d$ counterparts. Since we use a pseudo-spectral code, our method
is free from the controversial issue of real-space discretization
\cite{badkpz}. Owing to the
small system sizes and the high saturation time for the widths, the
quality of our data are less satisfactory than those obtained from
our $1d$ models. We are only able to measure the ratio $\chi_h/z$
from our numerical studies in $2d$. Nevertheless, the effects of
noise cross-correlations have qualitatively the same effects on the
amplitude ratio. Detail studies to yield the scaling exponents in
$2d$ with high accuracy will be presented elsewhere
\cite{abfreyunpub}.

Our results potentially have broad implications for natural systems
with coupled degrees of freedom. There are many such examples - MHD
turbulence, turbulence in a binary fluid mixture, biologically
relevant systems, e.g., coupled TASEPs (Totally Asymmetric Exclusion
Process) modeling transports in cellular systems, dynamics of motors
and microtubules, dynamics of visco-elastic active polar gels etc.
The general conclusions of our investigations, namely, the relevance
of the strength and the symmetry of the cross-correlations in
determining the long wavelength properties, are expected to be
visible in such systems as well, although detailed accounts are
likely to depend on the systems in question. To test our predictions
experimental measurements should include correlation functions and
effective transport coefficients. These are likely to be affected by
the presence of cross-correlations. In driven fluid-like turbulent
systems (MHD, binary fluid mixtures) cross-correlations can be
experimentally controlled by tuning the external (stochastic)
forces. In such systems one needs to measure correlation functions
of velocity and magnetic fields (in MHD) or velocity and
concentration gradient fields (in binary fluid mixtures) in the
scaling limit and obtain the scaling exponents characterizing the
correlation functions in the hydrodynamic limit as functions of the
amplitudes of the cross correlation functions.

\section{Acknowledgement}

A part of the work was done when one of the authors (AB) was working
at the Hahn-Meitner-Institut, Berlin as a Humboldt Fellow. AB wishes
to thank the Alexander von Humboldt Stiftung, Germany for partial
financial support. AB also thanks SERC (DST, India) for partial
financial support through a grant under the Fast Track Scheme
SR/FTP/PS-17/2006. Some of the computational works have been done
using the computational resources of Centre for Applied Mathematics
and Computer Science (CAMCS), SINP. AB thanks CAMCS for making such
facilities available for this work and for partial financial
support.

\section{Glossary}
\label{glos}For the convenience of the reader we provide below a
list of symbols for the functions and parameters which are used in
the bulk of the article.

\begin{itemize}
\item ${\bf u}({\bf x}, t),\;\; {\bf b}({\bf x}, t) \Rightarrow$
  Burgers  fields.
\item ${\bf z}^{\pm} = {\bf u\pm b} \Rightarrow$ \elss\;\; variables.
\item ${\bf u}=\nabla h,\,{\bf b}=\nabla\phi$. Fields $h$ and $\phi$
  are (non-conserved) height variables.
\item ${\bf z^{\pm}}=\nabla h_{1,2}$, $h_{1,2}$ are (non-conserved)
  height fields; $h_{1,2}$ also represent the free energies of two
  (identical) directed polymer in a random medium;
  $h_{1,2}=\frac{1}{2}(h\pm \phi)$.
\item $\nu_0,\,\mu_0\Rightarrow$ Bare
  (unrenormalized) viscosities for the fields $\bf u$ and $\bf b$;
$\nu,\,\mu\Rightarrow$ Renormalized fluid and magnetic
  viscosities.
\item $\lambda_1,\,\lambda_2,\,\lambda_3\Rightarrow$ coupling
  constants. They are set to unity in most of what follows.
\item $G_{u,b} ({\bf k}, \omega)\Rightarrow$ Renormalized
  propagators/response functions of the fields $\bf u$ and $\bf b$.
\item $\Sigma ({\bf k}, \omega) \Rightarrow$ Self-energy function of
  $\bf u$ and $\bf b$; $\Gamma\Rightarrow$ amplitude of $\Sigma ({\bf
    k},0),\,\Gamma=\nu=\mu$ in the present model.
\item $C^u_{ij},\,C^b_{ij},\,C^{\times}_{ij}\Rightarrow$ Auto
  correlation functions of $\bf u$ and $\bf b$, and cross correlation
  function respectively, $i,j\Rightarrow$ Cartesian indices.
\item $D_{u}^0,\,D_{b}^0\Rightarrow$ Bare amplitudes of the
  auto-correlations of the external Gaussian noise sources in the $\bf
  u$ and $\bf b$ equations [Eqs. (\ref{eq:burgers_1},\ref{eq:burgers_1})]; $D_u,\,D_b\Rightarrow$ corresponding
  renormalized ({\em scale-dependent}) amplitudes.
\item $D_{s}^0,\,D_{a}^0\Rightarrow$ Bare amplitudes of the symmetric
  and anti-symmetric parts of the noise cross-correlations in Eqs. (\ref{eq:burgers_1},\ref{eq:burgers_1});
  $D_s,\,D_a\Rightarrow$ corresponding renormalized amplitudes.
\item $P_m = \nu/\mu \Rightarrow$ Renormalized  Prandtl
  number.
\item $A = D_b/D_u\Rightarrow$ Dimensionless renormalized amplitude ratio.
\item $N_{s}^0= (D_{s}^0/D_{u}^0)^2,\,N_s = (D_s/D_u)^2$, $N_a =
(D_a/D_u)^2$.
\item $D_0= D_{u}^0+D_{b}^0,\;\hat D= D_{u}^0-D_{b}^0$.
\item $V_1, V_2\Rightarrow$ Random potentials embedding directed
  polymers (DP) A and B.
\item $Z_1, Z_2\Rightarrow$ Partition functions of DP A and B.
\item $R_1,R_2, \tilde R, \hat R \Rightarrow$ Different variances of
  the random potentials $V_1$ and $V_2$.
\item $\Gamma_0 = \left[\frac{R_{\times}({\bf x})}{R_1({\bf
      x})}\right]^2,\,\gamma= \frac {\hat R({\bf x})}{R_1({\bf x})}$. Since,
  $R_1({\bf x}),\,\hat R({\bf x}),\,R_{\times}({\bf x})$ have same
  scaling behavior in the hydrodynamic limit, $\gamma,\,\Gamma_0$
  are constants. To the lowest order $\Gamma_0 =
  N_s,\,\gamma=A$.
\item $G=\frac{\lambda^2 D_u}{\nu^3}\Rightarrow$ Dimensionless
  coupling constant in one-loop dynamic renormalization group
  calculations in the presence of symmetric cross-correlations;
  $G_a=\frac{\lambda^2D_u}{\nu^3}\Rightarrow$ the same for
  anti-symmetric cross-correlations.
\item $\tilde\Gamma\Rightarrow$ Vertex generating functional.
\item $\epsilon = 2-d$ where $d$ is the physical space dimension.
\item $\chi_u,\,\chi_b,\,z \Rightarrow$ Two roughness and dynamic
  scaling exponents respectively of the fields $\bf u$ and $\bf b$. In
  the model here, $\chi_u=\chi_b = \chi,\,\chi+z=1$.
\item$\chi_h,\,\chi_{\phi}\Rightarrow$ Roughness exponents of $h$ and
  $\phi$; $\chi_h=\chi_{\phi},\,\chi_h+z=2$.
\item $\zeta =1/z\Rightarrow $ scaling exponent characterizing the
  scaling of transverse fluctuations of directed polymer with its
  longitudinal length.
\item $T=1/\nu_0\Rightarrow$ Temperature of a DP in a random medium.
\end{itemize}

\section{Appendix}
\subsection{Symmetries and Ward identities}
\label{ward}

In this section we elucidate the continuous symmetries under which the
equations of motion (\ref{eq:burgers_1}) and (\ref{eq:burgers_2})
remain invariant. These allow us to construct exact relations
between different vertex functions which in turn impose strict
conditions on the renormalization of different parameters in the
model.

We begin by rewriting the Eqs.(\ref{eq:burgers_1}) and (\ref{eq:burgers_2}) by
expressing the Burgers $\bf b$-fields (whose mean value is always kept zero
here and below) as a sum of a constant vector and a space-time dependent part,
i.e., splitting the fields ${\bf b}({\bf x},t)\rightarrow {\bf b}({\bf
x},t)+{\bf B_0}/\lambda_2$ where ${\bf B_0}/\lambda_2$ is a constant vector:
\begin{equation}
{\partial {\bf u}\over \partial t}+  \nabla {\bf B_0\cdot b}+ \lambda_1\nabla
{u}^2+ \lambda_2{\nabla b^2} =\nu_0 \nabla^2 {\bf u} +{\bf f} \label{mhdu1}
\end{equation}
and
\begin{equation}
{\partial {\bf b}\over\partial t}+(\lambda_3/\lambda_2) \nabla {\bf
B_0\cdot u}+ \lambda_3{\bf \nabla (u\cdot b)}= \mu_0 \nabla^2 {\bf
b} + {\bf g}. \label{mhdb1}
\end{equation}
In this notation the mean Burgers $\bf b$-field is $\langle {\bf b}\rangle +
{\bf B_0}=0$.

The generating functional corresponding to Eqs.
(\ref{eq:burgers_1}) and (\ref{eq:burgers_2}) is written as
\cite{msr,janssen} $Z[{\bf j_1,\hat{j}_1, j_2, \hat{j}_2}]=\int
Du_iD\hat u_iDb_iD\hat b_i\exp[{\mathcal S}]$ \cite{martin}, where
the action functional $\mathcal S$ is given as
\begin{eqnarray}
{\mathcal S}[\Phi]&=&-\int \hat{b}_i\hat{b}_jD^b_{ij} -\int
\hat{u}_i\hat{u}_jD^u_{ij} -i\int \hat{u}_i\hat{b}_j{\Dcross}_s
-i\int \hat{u}_i\hat{b}_j \varepsilon_{ijp}k_pi{\Dcross}_a \nonumber
\\&& -i\int \hat{u}_i[\partial_t u_i +
\frac{\lambda_1}{2}\nabla u^2+\frac{\lambda_2}{2}\nabla b^2
+\nabla{\bf B_0\cdot b}-\nu_0\nabla^2u_i]\nonumber
\\&-&i \int \hat{b}_i [\partial_t b_i+ \lambda_3\nabla ({\bf u\cdot b})
+(\lambda_3/\lambda_2)\nabla{\bf B_0\cdot u}-\eta_0\nabla^2b_i].
\label{wardaction}
\end{eqnarray}

From the dynamic generating functional one obtains the vertex generating
functional $\Gamma$ \cite{martin} through
\begin{eqnarray}
\tilde\Gamma[u_i,\hat u_i,b_i,\hat b_i]= W[j_i,\hat j_i,l_i,\hat
l_i]-\int j_iu_i -\int \hat j_i\hat u_i-\int l_ib_i -\int \hat
l_i\hat b_i,
\end{eqnarray}
where $W=\ln Z$ is the generator of the connected diagrams and $\tilde \Gamma$ is the
generator of the one-particle irreducible diagrams \cite{zinn}.

The equations of motion (\ref{eq:burgers_1}) and (\ref{eq:burgers_2}) or the action
functional given by (\ref{wardaction}) are invariant under the following
transformations:
\begin{itemize}
\item The Galilean transformation (TI):
${\bf u}({\bf x},t)\rightarrow {\bf u} ({\bf x +u_0}t,t)+{\bf
u_0},\;\frac{\partial}{\partial t}\rightarrow \frac{\partial}{\partial t}-\lambda {\bf u_0}.\nabla,$ and
${\bf b}\rightarrow {\bf b}$ \cite{abjkb,fns,abepl} with
$\lambda_1=\lambda_3=\lambda$ in Eqs. (\ref{mhdu1}) and
(\ref{mhdb1}. This implies, as we show below, non-renormalization of $\lambda_1$
\cite{abjkb,fns,freykpz}.

\item The transformation TII: ${\bf B_0\rightarrow B_0}+\lambda_2\delta$,
${\bf b}({\bf x},t) \rightarrow {\bf b}({\bf x},t)- {\bf \delta}$, $\bf
u\rightarrow u$. Here the shift $\delta$ is a vector.
\end{itemize}

Let us first consider the invariance under the Galilean
transformation TI and the corresponding Ward identities. The
invariance of the vertex generating functional $\tilde\Gamma$ yields
\begin{eqnarray}
\tilde \delta\Gamma &=& {\bf u_0}\cdot\int_{\bf q}\int dt i\lambda
{\bf q} t[\frac{\delta\tilde\Gamma}{\delta u_j({\bf q},t)}u_j({\bf
q},t) + \frac{\delta\tilde\Gamma}{\delta \tilde u_j({\bf
q},t)}\tilde u_j({\bf q},t) \nonumber \\ &+& \frac{\delta\tilde
\Gamma}{\delta b_j({\bf q},t)}b_j ({\bf q},t) +
\frac{\delta\tilde\Gamma}{\delta \tilde b_j({\bf q},t)}\tilde
b_j({\bf q},t)]\nonumber \\& +& {\bf u_o}\cdot\int_{\bf q}\int dt \delta ({\bf q})
\frac{\delta\tilde\Gamma}{\delta {\bf u}({\bf q},t)}=0.
\label{wardver1}
\end{eqnarray}
By taking functional partial derivatives with respect to different
fields we obtain (in time space)
\begin{eqnarray}
\lambda (q_i't' +q_i''t'')\tilde \Gamma_{\tilde u_j u_k}&=& - \int
dt \tilde\Gamma_{\tilde u_j u_i u_k} ({\bf q'},t'; {\bf -q'},t'';
{\bf q},t)|_{{\bf q}=0},\\
\lambda (q_i't' +q_i''t'')\tilde \Gamma_{\tilde u_j \tilde u_k} &=&
- \int dt \tilde\Gamma_{\tilde u_j \tilde u_k u_i} ({\bf q'},t';
{\bf -q'},t''; {\bf q},t)|_{{\bf q}=0}, \\
\lambda (q_i't' +q_i''t'')\tilde \Gamma_{\tilde b_j b_k} &=& - \int
dt \tilde\Gamma_{\tilde b_j b_k u_i} ({\bf q'},t'; {\bf -q'},t'';
{\bf q},t)|_{{\bf q}=0},\\
\lambda (q_i't' +q_i''t'')\tilde \Gamma_{\tilde b_j \tilde b_k}&=& -
\int dt \tilde\Gamma_{\tilde b_j \tilde b_k u_i} ({\bf q'},t'; {\bf
-q'},t''; {\bf q},t)|_{{\bf q}=0}.
\end{eqnarray}
The above Ward identities may, equivalently, be written in the
frequency space as
\begin{eqnarray}
\lambda  q_i\frac{\partial}{\partial \omega} \tilde\Gamma_{\tilde
u_ju_k}({\bf q}, \omega; {-\bf q},-\omega) &=& \tilde\Gamma_{\tilde
u_ju_iu_k}({\bf q},\omega; {\bf -q},-\omega;
0,0),\\
\lambda  q_i\frac{\partial}{\partial \omega} \tilde\Gamma_{\tilde
u_j \tilde u_k}({\bf q}, \omega; {-\bf q},-\omega) &=&
\tilde\Gamma_{\tilde u_j\tilde u_ku_i}({\bf q},\omega; {\bf
-q},-\omega; 0,0),\\
\lambda  q_i\frac{\partial}{\partial \omega} \tilde\Gamma_{\tilde
b_jb_k}({\bf q}, \omega; {-\bf q},-\omega) &=& \tilde\Gamma_{\tilde
b_jb_ku_i}({\bf q},\omega; {\bf -q},-\omega;
0,0), \\
\lambda  q_i\frac{\partial}{\partial \omega} \tilde\Gamma_{\tilde
b_j \tilde b_k}({\bf q}, \omega; {-\bf q},-\omega) &=&
\tilde\Gamma_{\tilde b_j\tilde b_ku_i}({\bf q},\omega; {\bf
-q},-\omega; 0,0).
\end{eqnarray}
Further, as discussed in Ref.~\cite{freykpz}, the diffusive dynamics
of the model and the corresponding ${\bf q}$ dependences of the
vertices yield the {\em exact} relation
\begin{equation}
\tilde \Gamma_{\tilde u_i u_j} ({\bf q=0}, \omega)=
i\omega,\;\;\;\tilde \Gamma_{\tilde b_i b_j} ({\bf q=0},\omega) =
i\omega.
\end{equation}
To proceed further we first define the renormalization $Z$-factor
for a quantity $\kappa$ in the usual way: $Z_{\kappa}\kappa_R=
\kappa$ where the suffix $R$ refers to renormalized $\kappa$. These
then lead to
\begin{equation}
Z_uZ_{\tilde u}=1,\;\;\;\;Z_b Z_{\tilde b}=1.
\end{equation}
For the choice $Z_u=1$ (one of the renormalization $Z$-factors can
be set to unity without any loss of generality) we then obtain from the above Ward identities that
the non-linearity $\lambda$ does not renormalize.

We now discuss the symmetry TII. This symmetry has recently been
discussed in the context of a symmetric binary fluid mixture model
\cite{jstat}; see also \cite{absriram} for a similar symmetry in a
different physical context. The essence of the invariance under the
transformation TII is the following: We decompose the total
Burgers ${\bf b}$-field as a sum of a constant vector $\bf B_0$ and
a space-time dependent part ${\bf b}({\bf x},t)$ while writing down
the Eqs. (\ref{mhdu1}) and (\ref{mhdb1}) where $\bf B_0$ is
completely arbitrary. It should be noted that $\bf B_0$ is {\em not}
the mean $\bf b$-field, i.e., $\bf B_0$ is not the average of the
total $\bf b$-fields.
The latter, in our notations, is given by ${\bf B_0}/
\lambda_2 +\langle {\bf b}({\bf x},t)\rangle$ since ${\bf B_0}/
\lambda_2+{\bf b}$ is the total field. In fact, in our calculations
here and below we have set the mean magnetic field to zero which can
be ensured by adding appropriate counter terms in the action above.
We did not show them explicitly, but these are built in our
calculations below.

Under the transformation TII the mean Burgers $\bf b$-field ${\bf
B_0}/ \lambda_2 +\langle {\bf b}({\bf x},t)\rangle$ remains
unaffected. We argue that such a freedom to decompose the total
magnetic fields into two parts should hold for the renormalized
versions of Eqs.(\ref{mhdu1}) and (\ref{mhdb1}) as well. In
other words, in the renormalized versions of Eqs. (\ref{mhdu1}) and
(\ref{mhdb1}) one would be able to combine the constant and
space-time dependent parts of the Burgers-$\bf b$ field and express
them in terms of the total Burgers-$\bf b$ field. The resultant
renormalized equations of motion must be identical with those which
are obtained from the bare Eqs. (\ref{mhdu1}) and (\ref{mhdb1}) by
first writing them in terms of the total Burgers-$\bf b$ fields (by
combining $\bf B_0$ and $\bf b$) and then expressing them in terms
of the renormalized fields and parameters. This can happen only when
the second and the fifth terms of Eq.(\ref{mhdu1}) renormalize in
the same way. We arrive at the same conclusion below by using a more
formal language. Before closing this discussion we again emphasize
that here we consider a homogeneous and isotropic system, i.e.,
there is no mean Burgers-$\bf b$ field in the system. Furthermore,
to clarify the matter from a technical point of view we note that
the imposition of the condition of zero mean Burgers-$\bf b$ field
can be achieved by any of the two choices below:

\begin{itemize}
\item The choice $\langle {\bf b}({\bf x},t)\rangle = -{\bf B_0}\neq 0$,
such that the actual value of the mean Burgers-$\bf b$ field
$\langle {\bf b}({\bf x},t)\rangle + {\bf B_0}= 0$,
\item The choice $\langle {\bf b}({\bf x},t)\rangle =0= {\bf B_0}$. This
too, of course, maintains zero value for the mean Burgers-$\bf b$
field.
\end{itemize}
Note that the choices above are connected to each other by the
transformation TII of our manuscript. We now demand the physical
requirement that the renormalized equations of motion (and hence all
measurable quantities) would be the same, regardless of the choice
that one may make to obtain them. Such a requirement is nothing but
the invariance under the transformation TII. In particular, we work
with the second choice above, namely, $\langle {\bf b}({\bf
x},t)\rangle =0= {\bf B_0}$ in our perturbative calculations, which
is the most convenient one. However, the fact that our perturbative
RG is in agreement with the Ward identity originated from the
transformation TII, ensures that the results from our perturbative
RG are actually independent of the choices mentioned above.

To proceed further, we note that in the present problem one has
$Z_u=Z_{\hat u}=1$ \cite{freykpz,ronis}. Since the corresponding
vertex generating functional is also invariant under the
transformation TII we find
\begin{equation}
\lambda_2\frac{\delta\Gamma}{\delta
{B_0}_i}=\frac{\delta\Gamma}{\delta { b_i}({\bf q=0}, \omega=0)}.
\end{equation}
This leads to
\begin{equation}
\lambda_2\frac{\delta}{\delta {B_0}_i}\frac{\delta^2 \Gamma}{\delta
{b_j}({\bf k})\delta \hat{u_j}({\bf -k})}=\frac{\delta^3
\Gamma}{\delta {b_j}({\bf k})\delta \hat{u_j}({\bf -k}) \delta {
b_i}({\bf q=0})}. \label{ward2}
\end{equation}
Therefore, for the renormalized action to be invariant under the same
transformations, we must have
\begin{equation}
Z_{B_0}Z_b=Z_{\lambda_2}Z_b^2. \label{zfac}
\end{equation}

We assume the vector ${\bf B_0}$ to have components ${B_0}_{\mu}$ where $\mu$
is arbitrary. From the action functional (\ref{wardaction}) then,
\begin{equation}
  \frac{\delta}{\delta {B_0}_{\mu}}\frac{\delta^2
\Gamma}{\delta {b_j}({\bf k})\delta \hat{u_i}({\bf
-k})}={ik_{\mu}}\delta_{ij}.
\end{equation}
Therefore, its renormalization $Z$-factor must be unity leading to
$Z_{B_0}Z_b=1$. Due to the Ward identity (\ref{ward2}) the two sides of it
renormalize the same way and the $Z$-factor of the right hand side is
$Z_b^{-2}$ (we have used the fact that $Z_{\hat u}=1$). The left hand side has
an overall $Z$-factor given by
$Z_{\lambda_2}Z_{B_0}^{-1}Z_b^{-1}=Z_{\lambda_2}$ where we have used the
relation (\ref{zfac}). This yields,
\begin{equation}
Z_{\lambda_2}Z_b^2=1. \label{keyz}
\end{equation}
We, therefore, conclude that the coefficient of the effective vertex
constructed by using $\sqrt \lambda_2 \bf b$ does not renormalize.
Hence, $\lambda_2$ can be set to unity by treating all the
Burgers-$\bf b$ fields as the {\em effective} Burgers-$\bf b$ fields
$\sqrt \lambda_2 \bf b$. In fact, non-renormalization of $\lambda_2$
can be argued in a less formal way: Let us assume that under mode
eliminations and rescaling $\lambda_2\rightarrow \lambda_2/\Lambda$.
This scale factor of $1/\Lambda$ can now be absorbed by redefining
the units of the Burgers-$\bf b$ fields by ${\bf b}\rightarrow \sqrt
\Lambda {\bf b}$. Mathematically this would mean
$Z_{\lambda_2}Z_b^2=1$ - the same conclusion as in Eq.(\ref{keyz})
obtained by using a more formal way. Since the Eq.
(\ref{eq:burgers_2}) is linear, redefining (rescaling by a
multiplicative factor) $\bf b$-fields leaves it unchanged (see,
e.g., Ref.\cite{abepl}). Henceforth, all $\bf b$ fields here are to
be understood as effective fields with the coefficient of the
effective vertex being set to unity. It should be noted that under
the rescaling ${\bf b}\rightarrow \sqrt\Lambda {\bf b}$ the
stochastic force $\bf Q$ is also scaled by a factor $\sqrt \Lambda$.
However, this does not affect any of the conclusions as the
assignment of canonical dimensions to various fields and parameters
is done after absorbing $\lambda_2$ in the definition of $\bf b$
(or, by assigning zero canonical dimension to $\lambda_2$).

\subsection{Noise cross-correlations in the real space}
\label{cross1d}

We now establish the form of the noise cross-correlations in real
space in one dimension. Note that unlike the auto-correlation
functions cross-correlations are {\em not} squares of modulii of
functions; rather they are products of two different complex
functions in the Fourier space. Hence they can be positive or
negative, real or imaginary in the Fourier space. This, together
with the fact that noise cross-correlations are  real in the direct
space, and the properties of the fields $u$ and $b$ under parity
inversion yields that noise cross-correlations in $1d$ [see Eq.~(\ref{psi3})] are imaginary
and odd in Fourier wavevector $k$. In $1d$ it has the form $i\tilde
D k/|k|$. An inverse Fourier transform yields the form in real space
$\tilde D(x)$:
\begin{eqnarray}
\tilde D(x) &=& \int_{-\infty}^{\infty} dk e^{ikx} i \tilde D
\frac{k}{|k|} = i\tilde D \int_0^{\infty} dk e^{ikx} - i\tilde D
\int_{0}^{\infty}dk e^{-ikx} \nonumber \\
&=&-2\tilde D\int_0^{\infty} \sin (kx) dk.
\end{eqnarray}
The last integral above can be evaluated including a convergence
factor (equivalently by using contour integration along a closed
contour from the origin to $\infty$ along the real axis, then along
a circle at infinity anticlockwise up to an angle $\pi/2$ and
finally back to the origin along the imaginary axis). This yields
\begin{eqnarray}
&&  \tilde D(x)= -2\tilde D Lt_{\gamma \rightarrow 0}
  \int_0^{\infty} \sin (kx) e^{-\gamma k} dk \;\;\; (\gamma_0 >0)\;\;
  \nonumber \\ &=&-2\tilde D Lt_{\gamma_0\rightarrow 0} \Im \frac{1}{ix-\gamma_0} =
  -2\tilde D \frac{1}{x}.
\end{eqnarray}
Therefore, the cross-correlations have a form $1/x$ in real space in
$1d$. This has the same dimension as the $\delta$-function in $1d$,
but has a longer range than the $\delta$-function.

\subsection{Pseudo-spectral schemes for DNS}
\label{pseudo}

We would like to solve the stochastically driven equations
(\ref{eq:burgers_1}) and (\ref{eq:burgers_2}) in a one-dimensional
box. We use a {\em pseudo-spectral scheme} with {\em periodic
boundary conditions} (PBC). Pseudo-spectral schemes have a long
history in the numerical studies of fluid turbulence [see, e.g.,
Ref.~\cite{pseudo}]. The PBC are the simplest boundary conditions
ensuring that there are no surface effects on the bulk properties
that we propose to measure. They ensure that the fields $u$ and $b$
can be expanded in terms of {\em discrete} Fourier modes, labeled by
$k$. The highest $k$-mode will determine the smallest scale that can
be resolved in the simulations. In pseudo-spectral schemes spatial
derivatives are evaluated in the Fourier space, while products of
fields are evaluated in the real space. In terms of the Fourier
modes $u(k)$ and $b(k)$ the equations  (\ref{eq:burgers_1}) and
(\ref{eq:burgers_2}) take the form
\begin{equation}
  \frac{\partial { u}}{\partial t} +
  \frac{ik}{2}\sum_q u(q,t)u(k-q,t)+\frac{ik}{2}\sum b(q,t)b(k-q,t)
  =-\nu k^2 u(k,t)+f, \label{fouru}
\end{equation}

\begin{equation}
\frac{\partial {b}}{\partial t} +ik\sum_q u(q,t)b(k-q,t)=-\mu k^2 b(k,t)+g.
\label{fourb}
\end{equation}
We consider the case $\nu=\mu$. For time integration we use an Adams-Bashforth
scheme (step size $\delta t$)
\begin{eqnarray}
  \frac{u_k^{n+1}-u_k^n\exp (-\nu k^2\delta)}{\delta t}&=&\frac{3}{2}\Omega_k^n \exp
  (-\nu k^2\delta t)+\frac{3}{2}J_k^n \exp(-\nu k^2\delta
  t)-\nonumber \\
  \frac{1}{2}\Omega_k^{n-1} \exp(-2\nu k^2\delta t) &-& \frac{1}{2}J_k^{n-1} \exp(-2\nu
  k^2\delta t)+f_k^n\exp(-\nu k^2\delta t), \label{adam1}
\end{eqnarray}
and
\begin{eqnarray}
\frac{b_k^{n+1}-b_k^n\exp (-\nu k^2\delta)}{\delta
t}&=&\frac{3}{2}E_k^n \exp (-\nu k^2\delta t)-\frac{1}{2}E_k^{n-1}
\exp(-2\nu k^2\delta t)\nonumber \\ &+& g_k^n\exp(-\nu k^2\delta t).
\label{adam2}
\end{eqnarray}
Here, $\Omega=1/2 u^2,\;J=1/2 b^2,\;E=ub$ which are algebraic
products of functions in real space, superscripts $n$ etc refer to the $n$-th time-step.
Functions $\Omega,\,J$ and $E$ are evaluated
in real space which are then brought to Fourier space by Fourier
transforms. Note that, unlike in the commonly used Euler method of time integration,
here one solves the linear parts of the Eqs. of motion {\em exactly} and one requires
the solutions at steps $n$ and $n-1$ to solve at the $(n+1)$-th step.
Time evolutions of $u$ and $b$ takes place in Fourier
space. Fast Fourier Transforms (FFT) are used to perform Fourier
transforms. We, in particular, used FFTW ({\tt http://www.fftw.org})
routines for this purpose.

\paragraph{The choice of updating time step $\delta t$:-} For such nonlinear
equations as ours there is no analytic method available to choose
the updating time step $\delta t $ which ensures the stability of
the scheme. To get a measure of stability criteria we perform a
von-Neumann linear stability analysis on the linear parts of our
PDEs in $k$-space leading to the following condition on $\delta t$:
\begin{equation}
|1-\nu k^2 \delta t| > 1\Rightarrow 2>\nu k^2\delta t.
\end{equation}
For our run with $L=6144$ for a linear box size $2\pi$,
$k_{max}=3072$. We choose $\nu=5\times 10^{-9},\delta t=0.008$. Thus
$\nu k^2\delta t=0.004 \ll 2$. We start from random initial
conditions on the $\bf u$ and the $\bf b$ fields, allow it to reach
steady states (we monitor it by noting the time evolutions of the
 energies of the $u$ and $b$ fields, which should fluctuate about their
mean values in the statistical steady state). We then perform all
our measurements in the statistical steady state obtained.

\subsection{Generation of noises}
\label{noisegen} For our DNS studies in $1d$ we need to generate
stochastic forcings $f$ and $g$ with variances (in $k$-space)
\begin{eqnarray}
\langle f(k,t)f(-k,0)\rangle=2Dk^2\delta (t),\nonumber \\
\langle g(k,t)g(-k,0)\rangle=2Dk^2\delta (t),\nonumber \\
\langle f(k,t)g(-k,0)\rangle=2i\tilde Dk|k|\delta (t). \label{vari}
\end{eqnarray}
To obtain $f$ and $g$ we first generate two independent sets of
random numbers with Gaussian distributions with zero mean and of
variances $2(D+\tilde D)k^2$ and $2(D-\tilde D)k^2$. Appropriate
linear combinations of these random numbers are the stochastic noise
sources $f$ and $g$ above which have the specified variances
(\ref{vari}). To generate noises $\psi_1$ and $\psi_2$ in
Eq.~\ref{2kpz} with variances (\ref{psi1}-\ref{psi3}) we first
generate two independent sets of random numbers with Gaussian
distributions with zero mean and of variances $D_0 \pm \tilde D_0$.
Appropriate linear combinations of these random numbers yield
$\psi_1$ and $\psi_2$. In higher dimensions, stochastic forces can
be analogously generated.


\begin{thebibliography}{99}
\bibitem{foot1}There are, however, certain driven systems where in
  contrast to the present model total energies diverge with the
  system size. Physical examples include fluid and magnetohydrodynamic
  turbulence. Such systems are however described by Langevin equations
  driven by noise sources with variances divergent in the infra-red
  limit. See, e.g., Ref.\cite{abepl} and references therein.
\end{thebibliography}
\end{document}